\documentclass[12pt,preprint]{aastex}

\newcommand{\musec}{\mu{\rm s}}
\newcommand{\ergscm}{\,{\rm erg}\,{\rm sec}^{-1}\,{\rm cm}^{-2}\,}

\shorttitle{RXTE/PCA Calibration}
\shortauthors{Jahoda et al.}

\begin{document}

\title{Calibration of the Rossi X-ray Timing Explorer \\
Proportional Counter Array 
			}

\author{Keith Jahoda\altaffilmark{1}, 
Craig B. Markwardt\altaffilmark{1,2},
Yana Radeva\altaffilmark{2,3},
Arnold H. Rots\altaffilmark{4}, 
Michael J. Stark\altaffilmark{5}, 
Jean H. Swank\altaffilmark{1}, 
Tod E. Strohmayer\altaffilmark{1}, and William Zhang\altaffilmark{1}}
\altaffiltext{1}{GSFC/LHEA Code 662, Greenbelt, MD  20771}
\altaffiltext{2}{Univ. of MD, College Park, MD}
\altaffiltext{3}{Connecticut College, New London, CT}
\altaffiltext{4}{Harvard Smithsonian Center for Astrophysics, Cambridge, MA}
\altaffiltext{5}{Lafayette College, Easton, PA}

\email{keith.m.jahoda@nasa.gov}

\begin{abstract}
We present the calibration and background model for the Proportional Counter 
Array (PCA) aboard the Rossi X-ray Timing Explorer (RXTE).  The energy 
calibration is systematics limited below 10 keV with deviations from a 
power law fit to the Crab nebula plus pulsar less than $1\%$.
Unmodelled variations in the instrument background amount to less 
than 2\% of the observed background below 10 keV and less than 1\% between
10 and 20 keV.  Individual photon arrival times are accurate to $4.4 \musec$ 
at all times during the mission and to $2.5 \musec$ after 29 April 1997.  
The peak pointing direction of the five
collimators is known at few arcsec precision.

\end{abstract}


\section{Introduction}

The Proportional Counter Array (PCA) aboard the Rossi X-ray Timing Explorer 
(RXTE) consists of 5 large area proportional counter units (PCUs)
designed to perform 
observations of bright X-ray sources with
high timing and modest spectral resolution.
With a 1$^{\circ}$ collimator (FWHM) and well modelled background, 
the PCA is confusion (rather than systematics) limited at fluxes of $\approx 4 \times 10^{-12} \ergscm$
in the 2-10 keV band (approximately 0.3 count sec$^{-1}$ PCU$^{-1}$) and is 
capable of observing sources with fluxes up to 20,000 
counts sec$^{-1}$ PCU$^{-1}$.  
%
The PCA has unprecedented sensitivity to a wide variety of timing
phenomena due to the combination of several instrumental and operational
characteristics.  Instrumental characteristics (large area, 
micro-second time tagging capability, and a stable and predictable
background)
are combined with operational characteristics (rapid slewing, 
flexible planning, and
access to the entire celestial sphere further than 30 degrees from the sun)
provide unprecedented sensitivity to variability associated with
galactic compact objects and powerful active galactic nuclei.
The PCA has been uniquely able to study accreting milli-second pulsars
\citep{deepto03}
and the power spectra of Seyfert galaxies \citep{markowitz03} 
to give just two disparate
examples.  More inclusive views of the power of the PCA can be found
in the proceedings of ``X-ray Timing 2003" \citep{kaaret04} and the 
RXTE proposal to NASA's Senior Review \citep{swank04}.

This paper presents calibration information relevant to scientific users
of the RXTE PCA.  
The primary focus of this paper is the energy and background calibration of the PCA. 
The response matrix generator and default parameters are those included 
in the FTOOLS software package v6.0 release\footnote{http://heasarc.gsfc.nasa.gov/ftools/}.  
The background models presented\footnote{pca\_bkgd\_cmfaintl7\_eMv20031123.mdl and 
pca\_bkgd\_cmbrightvle\_eMv20030330.mdl.}
are available from the RXTE Guest Observer Facility.
We also summarize calibration information related to the deadtime, absolute
timing, pointing, and collimator field of view of the PCA.

The paper is organized as follows:  Section \ref{sec2}
gives an overview of the instrument and its operations. Section \ref{sec3}
describes the elements of the response matrix, including a description
of the parameters which are supplied to the response matrix generator
{\tt pcarmf}.  Section \ref{parfits} describes the use of on-orbit
data to determine the best values of the response matrix parameters
and presents the resulting fits to spectra from the Crab nebula and
the Iron line in Cas-A. Section \ref{bkgd} describes the method used
to construct the background model and describes the results.  Section
\ref{timing} describes the effects of deadtime in the PCA and how to
correct for deadtime.  Section \ref{absolute} describes the absolute
timing accuracy.  Section \ref{fov} documents the relative pointing
of the PCU collimators and decribes the model of the
PCA collimators.  References to specific tools or parameters are
to those included in the v6.0 release of the FTOOLS unless otherwise
noted.  Modest updates, particularly to calibrate the energy scale at
future times, may be expected
in future releases;  the most current values should be available via
the High Energy Astrophysics Science and Archival Research Center
(HEASARC) Calibration Data base.  An on-line version of this paper
and links to other published and internal notes relevant to PCA calibration
is being maintained.\footnote{
http://universe.gsfc.nasa.gov/xrays/programs/rxte/pca/pcapage.html
\label{cal_page}}

\section{Instrument Description}
\label{sec2}

\begin{figure}[t]
\includegraphics[width=1.0\columnwidth,angle=0.0]{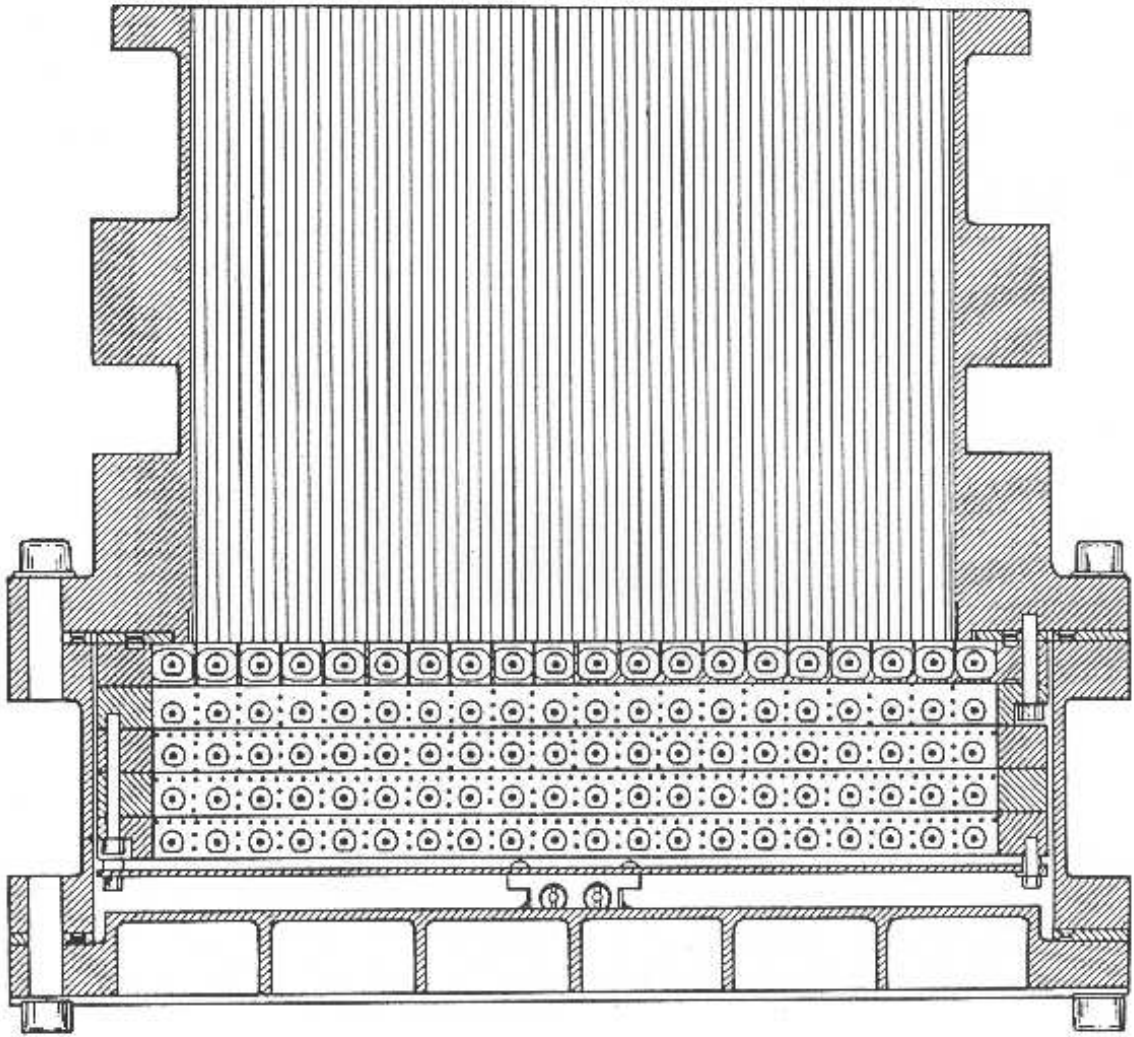}
\figcaption{Cross section view of one PCA detector.  The field of view is
defined by the collimator.  Below the collimator, there is a mylar window
above and below a propane veto volume.  The main xenon-filled detector volume
consists of 3 layers of signal anodes and a back layer of veto anondes.
Below the veto layer, there are two short anodes on either side of the
Am$^{241}$ calibration source.  Hatched areas in this figure represent 
the Aluminum counter body;  the Tin and Tantalum shielding is not shown.
\label{detxsec}}
\end{figure}

\begin{figure}[t]
\includegraphics[width=0.7\columnwidth,angle=0.0]{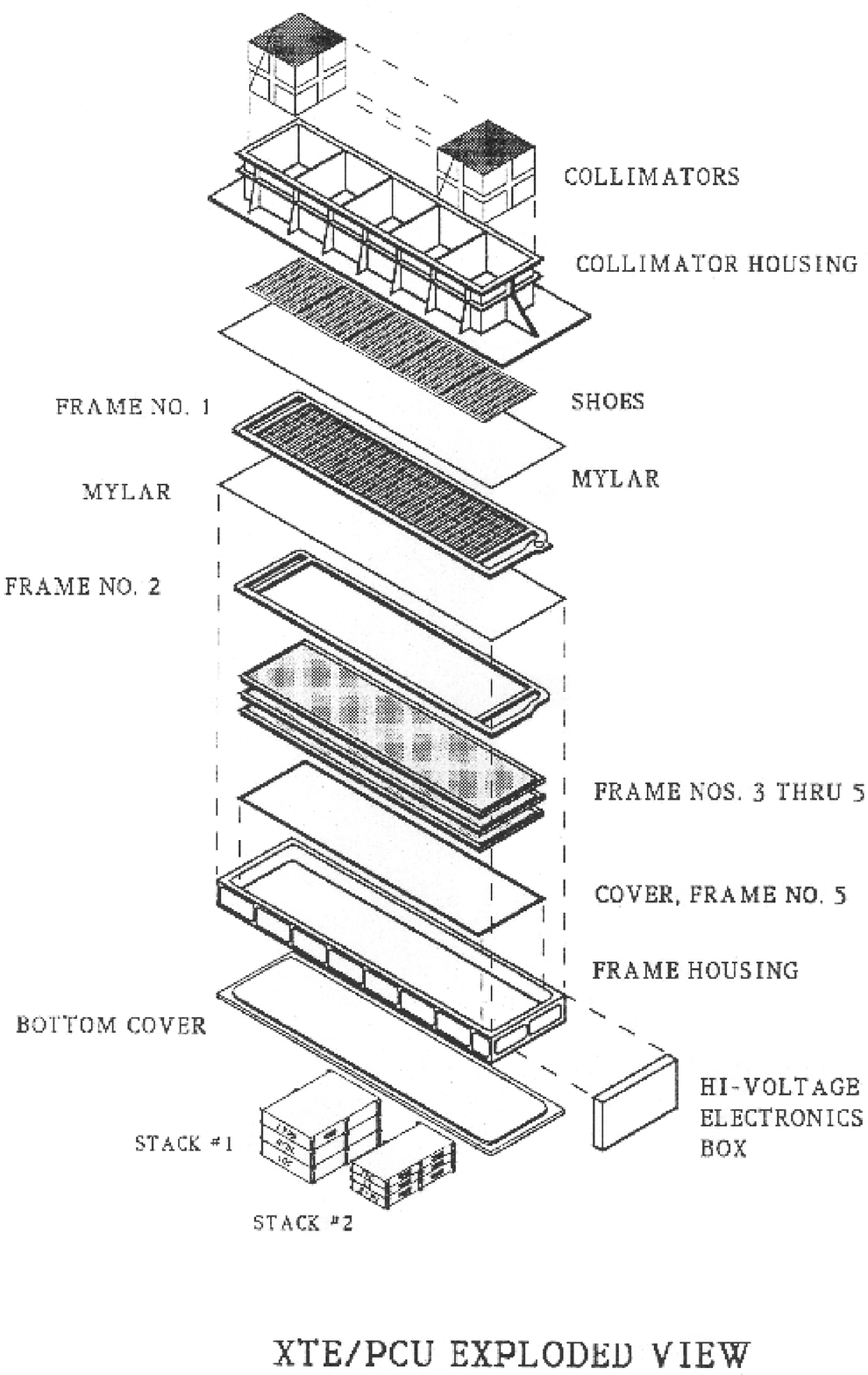}
\figcaption{Assembly  view of one PCA detector.  The field of view is
defined by the collimator.  The ``frames"  hold the anodes which
define the various detector layers.  The propane anodes are in frame
1;  the first Xenon signal layer is defined by frame 2;  frames 3 through 5
define the second and third Xenon layers and the Xenon veto-layer.  The
calibration source, not visible in this view, is mounted on the back
of the frame 5 cover.
\label{detassy}}
\end{figure}

\begin{figure}[t]
\includegraphics[width=0.7\columnwidth,angle=0.0]{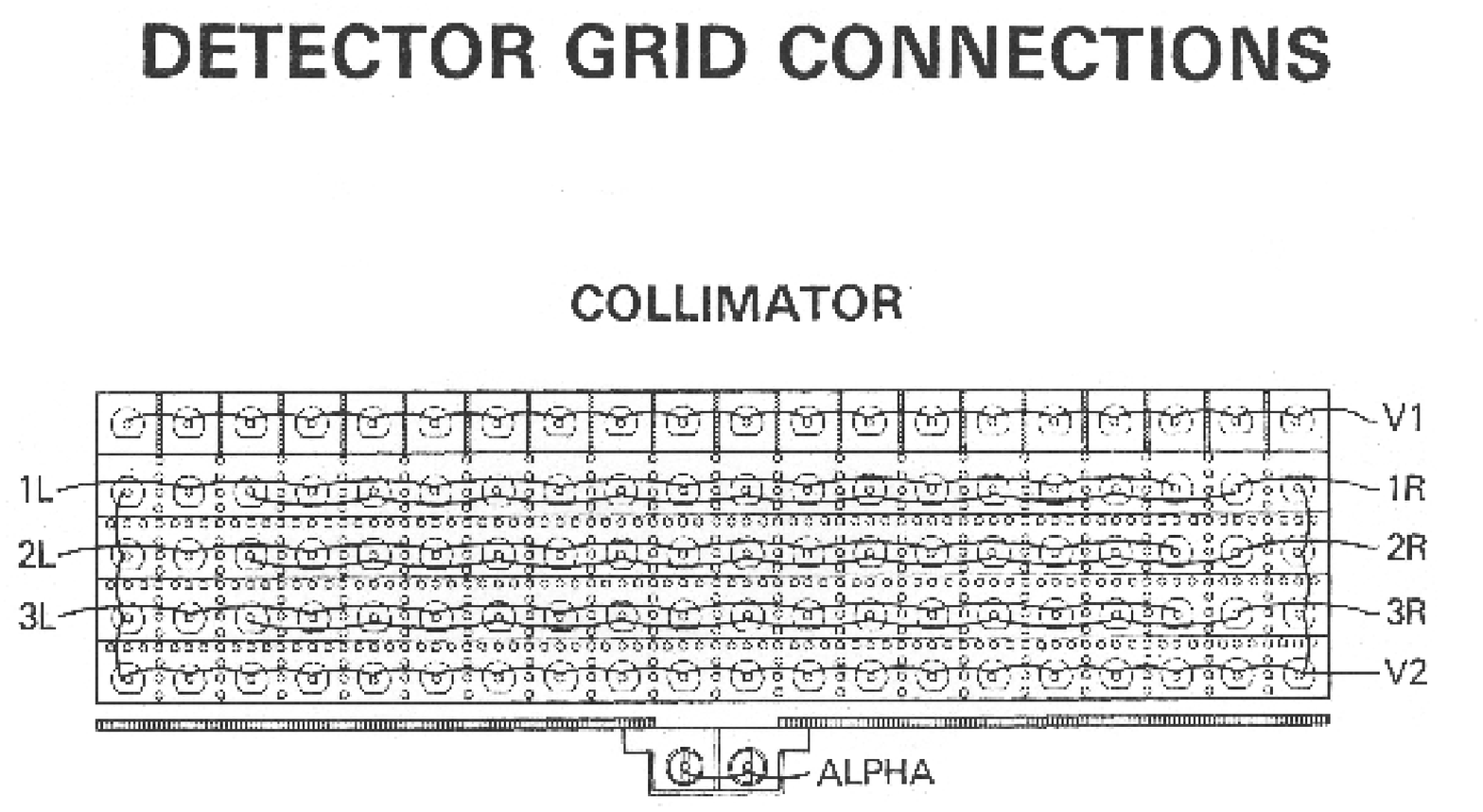}
\figcaption{Anode interconnections within a PCU.  The chain labelled
``V1" is usually referred to as the ``Propane" layer;  the chain
labelled ``V2" is usually referred to as the ``veto" layer or ``$V_x$"
chain.  Activity on either of these chains is treated as a coincidence
flag that inhibits further analysis of simultaneous events on the
main signal layers.
\label{anodes}}
\end{figure}

The PCA consisits of 5 nominally identical Proportional Counter Units (PCU).  
Each has a net geometric collecting area of $\sim 1600 \, {\rm cm}^2$.  
Construction, ground performance, and early inflight performance are described 
elsewhere \citep{gla94,zhang93,jah96}.
The proportional counters were designed, built,
and tested within the Goddard Space Flight Center (GSFC) Laboratory for 
High Energy Astrophysics.  
The essential features are visible in the detector cross section
(figure \ref{detxsec}) and assembly view (figure \ref{detassy}).
The detectors consist of a mechanical collimator with FWHM $\sim 1^{\circ}$, an aluminized mylar window,
a propane filled ``veto" volume, a second mylar window, and a Xenon filled 
main counter.  The Xenon volume is divided into cells of $\sim 1.3 \, {\rm cm} 
\times\,1.3 \, {\rm cm} \times 1\,{\rm m}$ by wire walls.  The detector bodies 
are constructed of Aluminum and surrounded by a graded shield consisting 
of a layer of Tin followed by a layer of Tantalum.
Nominal dimensions are given in table \ref{dimensions}.
There are three layers of Xenon cells, and each layer is divided in 
half by connecting alternate cells to either the ``right" or ``left" 
amplifier chain.  These six signal chains are referred to as 1L, 1R, 2L,
2R, 3L, and 3R.  The connections are shown in figure \ref{anodes}.
The digit denotes the layer while the letter denotes
``right" or ``left".  The division of each layer is significant
for data screening and background modelling.  The response matrices for 
each half of each layer are identical.   Scientific analyses generally 
combine data from the two halves.

There are 3 additional signal chains.  The anodes in the front volume
are connected together to form the ``Propane" chain;  events on this chain
are presented to the analog to digital coverter.  The outermost anode on
each layer and the entire fourth layer are connected together to form
a xenon veto ($V_x$) chain;  signals on this chain are not digitized but
are presented to the EDS as a two bit pulse height.  Finally, two short 
anodes on either side of the $Am^{241}$ source provide a flag which is
interpretted to mean activity on any of the other anodes is likely to be
a calibration event.  Of course, there is a chance of 50\% or more that the
calibration photon will go the wrong way or not be absorbed in the active volume.
It is also clear from the background spectra that some calibration events
are not flagged.

\placefigure{detxsec}

\placefigure{detassy}

Every event, whether due to background or a cosmic source, 
passes 19 bits over a serial interface to the Experiment Data System (EDS)
\footnote{The EDS was designed and built at the MIT Center for Space
Research.  A functional description is provided in the RXTE user guide at
http://heasarc.gsfc.nasa.gov/docs/xte/appendix\_f.html}.  
The information includes 8 bits of pulse height for events occuring on the
Xenon signal or propane veto layers, 2 bits of pulse height for 
events on the 
$V_x$ layer, 8 lower level discriminator bits (for the 6 signal chains, 
Propane chain, and the calibration flag) and a very large event flag.  
A `` good event" (i.e. an X-ray) is an event which triggers exactly one
discriminator;  the pulse height bits can be unambiguously associated with
that signal chain.  For coincidence events there is an ambiguity about
which signal was digitized, and these events are typically ignored.  The EDS
applies a time-tag and performs event selection and (multiple) data compressions.
Of particular interest, two standard
compression modes have been run throughout the entire mission.  Standard 1
provides light curves with 0.125 sec resolution and calibration spectra with full
pulse height information collected every 128 seconds.  Standard 2 provides pulse
height information for each layer of each detector with 16 second resolution and
29 rates which account for all the ``non X-ray" events.  Both Standard modes
count each event produced by the PCA exactly once.  Data used for in-flight 
calibration of the energy scale comes exclusively from the standard modes.
Scientific data in the telemetry stream identifies the detectors with a 0-4 scheme.
\footnote{Much of the engineering and housekeeping data uses a 1-5 numbering 
scheme.  All references to inidividual PCUs in this paper use the 0-4 convention.}

\subsection{Routine Operations}
\label{routine}
The RXTE spacecraft allows extremely flexible observations, often making more than 
20 discrete and scientifically motivated pointings within a day.  The data modes are 
selectable, allowing the compression which is required for bright objects 
(due to downlink bandwidth) to be performed in a user selected way.
More detail about the standard and user selectable data modes is contained in 
the RXTE user guide.  A brief description of
creating response matrices for user selected modes, using the ftool script {\tt pcarsp},
is given in the Appendix.
Throughout this paper we use a {\tt typewriter} font to refer to FTOOLS 
programs or scripts and
an {\it italic} font to refer to the associated parameters.

In addition to pointing direction and data compression mode, there are only two user
commandable parameters.  

Each detector is equipped with a High Rate Monitor (HRM) which disables the
high voltage when the total number of counts on any anode exceeds a preset
counting rate for three consecutive 8 second intervals.  The default is set
to 8000 count sec$^{-1}$.  This rate is occasionally adjusted upwards (typically to
24,000 count sec$^{-1}$) for observations
of extremely bright sources such as Sco X-1.
It is unlikely that the HRM would ever be tripped
if set much above 35,000 count sec$^{-1}$ as paralyzable deadtime losses prevent
any anode from registering more than about this counting rate.

Each detector is also equiped with a selectable Very Large Event window.
Events which saturate the pre-amplifiers cause ringing as the amplified signal
is restored to the baseline;  often the ringing can cause false events to be
pushed through the system.  The timing of these events is dependent on the
actual pulse height of the saturating event.  Each detector has 4 commandable
windows, set nominally to $20$, $60$, $150$, and $500 \, \musec$.  The pre-launch
intention was to allow ``extra clean data" for very faint sources and ``high
throughput" data for very bright sources.  The size of the VLE window 
affects the shape of the power spectrum (section \ref{deadtimeps}).  
Early observations
of Cyg X-1 with the VLE window set to the smallest value demonstrated
a failure associated with this window on one of the analog chains; 
use of the shortest VLE window has been discontinued.  The third window
is the default;
observations of some bright sources have been conducted with the 
second window.  The calibrated values of these two windows are
$70$ and $170 \, \musec$.

The High Voltage is commandable in discrete steps, and has been changed
by the instrument team 3 times during the mission,  resulting in 
discontinuities in the energy response.
In addition, the propane layer in PCU 0 lost pressure in the spring of 2000, 
in a manner consistent with a micro-meteorite induced pinhole.
We use the term epoch to describe the periods between these discontinuous 
changes in the instrument response.  Response changes within an epoch
are gradual and can be described by parameterizations with a small time dependence.
Definitions of the epochs are given in Table \ref{hv_epochs}.  Epoch 3 is divided
into 3A and 3B as the time dependence of the background model changed
significantly within epoch 3;  the boundary is near the moment when orbital
decay became important as discussed in section \ref{bkgd}.

Three of the PCUs became subject to periodic discharge after launch.  
The discharge is primarily evident in an increased rate on one signal
chain, and in extreme cases by a reduced high voltage.  (High rates 
can reduce the voltage by drawing current from the high voltage supply.
There is a large resistor between the supply and the input to the pre-amplifier.)
This behaviour began in March 1996 for PCUs 3 and 4 and in March 1999 
for PCU 1.  To minimize the occurence of discharge 
several operational steps were taken.  
First, detectors that break down are now cycled off and on and are not 
used when the scientific objectives do not require large instantaneous area.  
Second, the high voltage has been lowered.  Finally, the spacecraft roll 
angle has been changed slightly to increase the solar heating
of the PCA and to operate at a slighty warmer temperature.  While discharge
still occurs, the EDS detects these occurrences
by comparing the first moment of the pulse height distributions from the two
anode chains associated with each layer.  When mismatched moments are detected,
the EDS sets a status signal detectable in 
the satellite Telemetry Status Monitoring (TSM) system 
which causes the high voltage to be turned off.  The EDS calculation is performed
every 16 sec;  the TSM requires 5 consecutive readings (separated by 8 sec), so
periods of discharge are limited to less than one minute;  resting the 
detectors for a few hours allows them to be turned on again.
In mid-2004, the duty cycles (not counting the fraction of the orbit
when all detectors
are off due to passage through the SAA, which amounts to $\sim 20\%$)
of PCUs 0-4 were $\sim$ 1.0, 0.2, 1.0, 0.6, and 0.2.

The high voltage was lowered in March 1996 and again in April 1996 as
new operating procedures were installed after the first breakdowns of
PCU 3 and 4.  The high voltage was lowered again in April 1999 after the
first breakdown in PCU 1.  The times are given in Table~\ref{hv_epochs} along
with the high voltage settings (pre-launch measurements) within the Xenon detectors.

There was a brief period in March 1999 when the high voltage fluctuated
from the desired setting (level 4) to the epoch 3 setting (level 5) due
to an error in the high voltage commanding surrounding the South Atlantic
Anomaly.  The archived PCA housekeeping data correctly records the instantaneous
high voltage state.

\section{Response Matrix Overview}
\label{sec3}
   The response matrix provides the information about the probability that
an incident photon of a particular energy will be observed in a particular
instrument channel.  The PCA response matrices are generated within the standard
FTOOLS environment \citep{blackburn} 
and are designed for use with the XSPEC convolution and spectral model fitting 
\notetoeditor{this is the second footnote reference to ftools;  can it just 
point back to the previous one?}
program \citep{arnaud96}.
A satisfactory matrix must account for 
\begin{itemize}
\item The energy to mean channel relationship
\item the quantum efficiency as a function of energy
\item the spectral redistribution within the detector
\end{itemize}

We describe below the model as implemented by the program {\tt pcarmf} as 
released in FTOOLS v5.3 in November 2003.\footnote{Neither the 
response generator nor the default parameters were updated in 
the v6.0 release of April 2005.  Future updates will be made as required.}
The ftool {\tt pcarmf} produces a 256 channel response matrix for a single detector
or detector layer.  The appendix describes the additional steps needed
to produce a response matrix any of the rebinned and compressed
modes commonly used by the PCA and found in the RXTE archive.

\subsection{Energy to channel model}
The energy to channel relationship in proportional counters is non-monotonic in
the neighborhood of atomic absorption edges \citep{jah88,bav95,tsunemi93}.  
The mean ionization state of an atom which absorbs a photon just above 
an atomic absorption edge
is greater than the mean ionization state if the photon is just below the edge.
Because more energy goes into potential energy assoicated with the absorbing
atom, the photo-electron (and other electrons ejected from the
atom by Auger or shake-off processes) have less kinetic energy
and the mean number of
electrons produced in the absorbing gas is smaller.  
Between edges, the energy available for ionizing the gas
increases smoothly as does the mean number of electrons created as 
the final result of the photo-electric absorption \citep{santos93,santos94,
borges03}.  

While the voltage pulse in a ``proportional counter" is
only approximately proportional to the incident photon energy, the pulse is 
proportional to the number of electrons produced in the absorption region.
The observed pulse heights are just the number
of electrons produced in the gas multiplied by the electronic gain
from both the proportional counter itself and the amplifier chains.
The response matrix provides the probabilities that an X-ray of a
particular energy ($E_{\gamma}$) will be observed in each of the 
detector pulse height channels (numbered 0 through 255).  When
analyzing or plotting the data, it is often convenient to label
the channels with energies.  Plotting pulse height spectra as
a function of energy
is facilitated by a single valued channel to apparent energy relationship.
We define a second scale, $E_p$, proportional to the 
number of electrons produced and normalized to be approximately 
equal to energy by taking advantage of the detailed predictions
about the number of electrons produced by photo-electric
absorption in Xenon \citep{dias93,dias97}.  A cut through the response matrix
gives the probabilities that a photon of energy $E_{\gamma}$ will 
be observed at apparent energy $E_p$.  The matrix captures both the
non-linearities in the this relationship and the instrumental
redistribution.  The XSPEC spectral convolution program
provides the option of plotting or data selection
using the $E_p$ or channel scale.

\begin{figure}[t]
\includegraphics[width=0.8\columnwidth,angle=270.0]{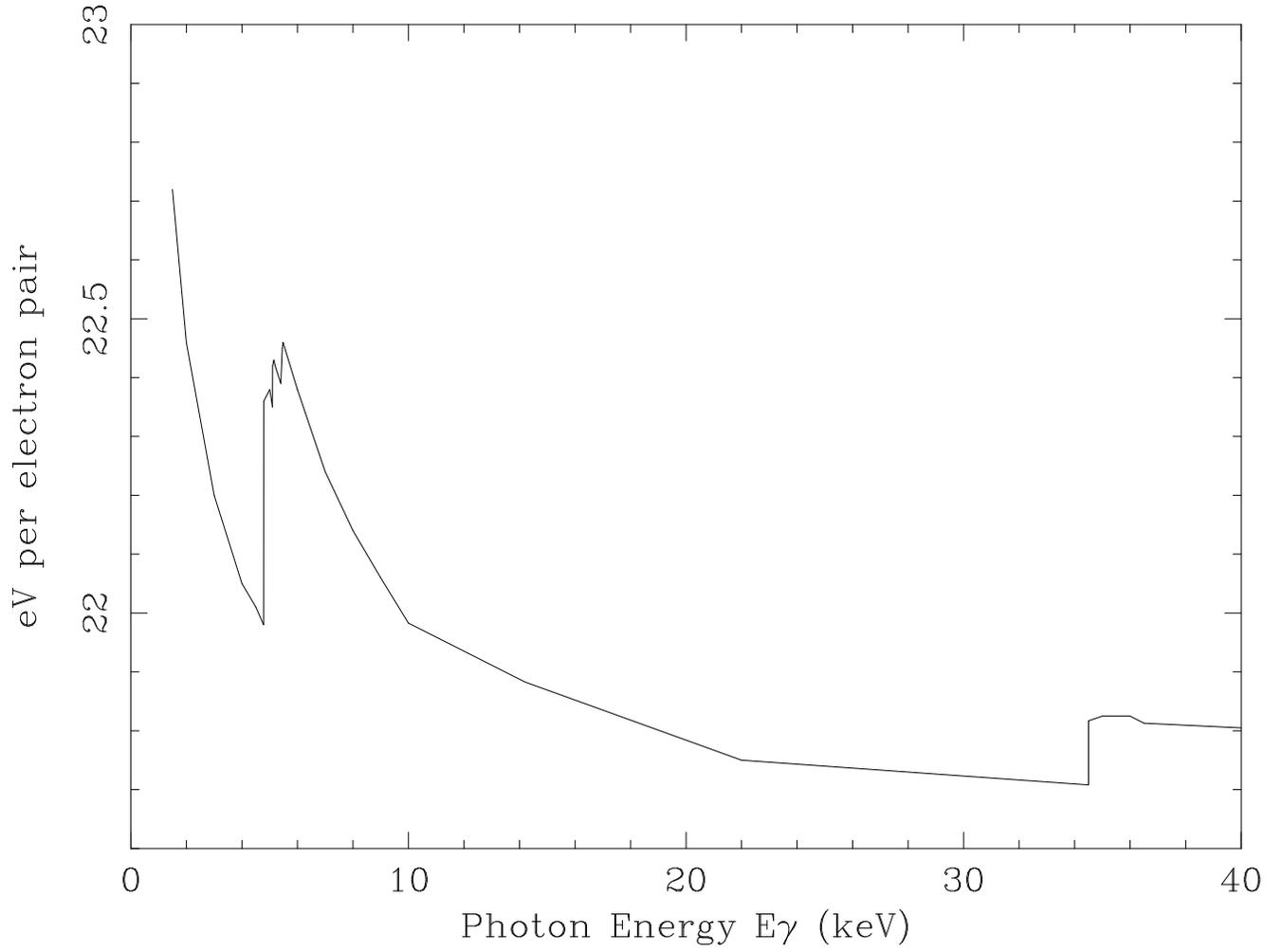}
\figcaption{Average number of eV required to create on electron-ion pair 
in Xenon as a function of incident photon energy.  Data is from \citet{dias93,dias97}.
\label{wxe}}
\end{figure}

The average number of eV ($w$) required to produce one ionization electron 
in Xe is shown as a function of
photon energy ($E_{\gamma}$) in figure \ref{wxe}.
The values of $w(E_{\gamma})$ 
are near 22 eV and we define
\begin{equation}
\label{ep}
E_p = {22.0 \over { w(E_{\gamma})} } E_{\gamma}.
\end{equation}
With this definition there are discontinuities of 76.4, 10.9, and 12.6 eV 
in $E_p$ at the Xenon $L_1$, $L_2$,
and $L_3$ edges with $E_{\gamma} = 4.78, 5.10, {\rm and}\, 5.45$ keV
and a discontinuity of 173 eV at the Xe $K$ edge at 34.5 keV.

\placefigure{wxe}

We find that the PCA data are better fit with smaller jumps, and introduce a 
parameter $f$ which lessens the difference between $E_p$ and $E_{\gamma}$ 
and reduces the size of the jumps.  For values $0 \le f \le 1$ we define
\begin{equation}
\label{wf}
 w_f(E_{\gamma}) = ( w(E_{\gamma}) -22.0) f + 22.0  .
\end{equation}
The value of the parameter $f$ is optimized in our fitting procedure.
Our best fit value for $f$ is 0.4, resulting in a total jump of 40 eV
summed across the three edges.

Our response matrix has an ``instantaneous" quadratic relationship between 
channel and $E_p$.  Within each high voltage epoch we fit a model where 
\begin{equation}
ch(E,T) = A + B E_p + C_o E_p^2.
\label{e2c}
\end{equation}
The constant and linear coefficients are time dependent:
\begin{equation}
A = A_0 + A_1 \Delta T + A_2 (\Delta T)^2 
\end{equation}
and
\begin{equation}
\label{e2c2}
B = B_0 + B_1 \Delta T + B_2 (\Delta T)^2  E_p.
\end{equation}
$\Delta T$ is the time in days between $T$ and a reference date $T_0$.
The cause of the time dependence is not known, but may reflect a slow
change in purity or pressure of the gas or other changes within the amplifier
chain.

\subsection{Quantum Efficiency}
To model the quantum efficiency, each PCU is treated as a series of parallel slabs
of material.  No account is made of possible bowing of the front entrance
aperture on any scale.  
All quantum efficiency parameters, specified in {\tt pcarmf.par} and documented
in table \ref{v5.3_pars}, are reported with units
of gm cm$^{-2}$:

We specify for each detector 
\begin{itemize}
\item Xe$_{l(m)}$, the amount of Xenon in detector layer $m$; $m$ runs 
from 1 to 3.  1 represents the top layer and 3 the innermost layer as
seen from the collimator (figure \ref{anodes});
\item Xe$_{pr}$, the amount of Xenon in the propane
layer on 1997 Dec 20;
\item $d({\rm Xe}_{pr})/dt$, the rate of change (per day) of Xenon in the propane layer;
\item Xe$_{dl}$, the thickness of a dead layer of Xenon between adjacent layers;
\item $Mylar$, the total thickness of the two Mylar windows;

\item $Aluminum$ The total amount of aluminization on the four sides of the two windows;
\item $Propane$ The amount of propane in the first gas volume
\end{itemize}

The dead layer, as modelled here, assumes that there is a small region where events
will be self vetoed as part of the electron cloud is collected in one anode volume
and part in the adjacent volume.  In detail, such a region probably has a soft
edge, though our data cannot distinguish this.
Additionally, the model requires that PCU 0, despite the loss of pressure 
in the propane layer, requires residual Xenon in the propane layer in epoch 5.  
The best parameterization requires half as much Xenon as before the loss of 
pressure.  This model is unsatisfactory,
and the fits are less good for PCU 0 in epoch 5.  
There could be many causes, including bowing
of the internal window which now supports a 1 atm pressure differential.

The Xenon absorption cross sections
are from \citet{mcmaster}.
All other absorption cross sections are derived from \citet{henke93}.  

\begin{figure}[t]
\includegraphics[width=0.8\columnwidth,angle=270.0]{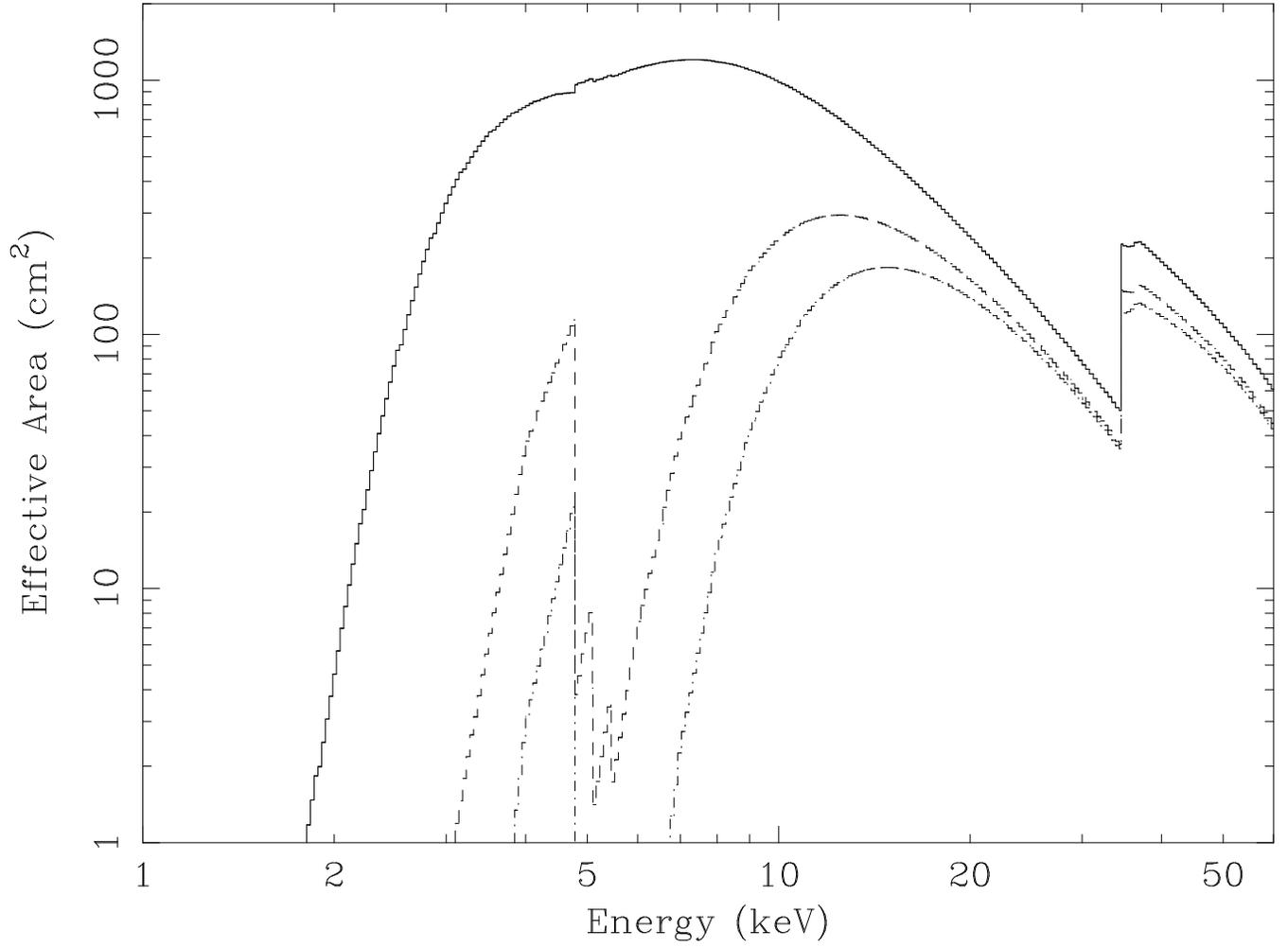}
\figcaption{Quantum efficiency for layers 1 (solid line), 2 (dashed), and 3 
(dot-dash) of PCU 2 for 13 January 2002.
\label{qeff}}
\end{figure}

Figure \ref{qeff} shows the modelled quantum efficiency for
PCU 2, layers 1, 2, 3 on January 13, 2002.  The peak quantum efficiency 
has decreased by $\sim 1\%$ per year over the lifetime of the mission 
due to the diffusion of Xenon into the front veto volume.

\placefigure{qeff}

\subsection{Redistribution matrix}
\label{redistribution}
Our model accounts for the intrinsic resolution of the proportional counter, K and L escape
peaks, losses due to finite electron track length (which causes events to be 
self-vetoed), and losses due to partial charge collection (which causes events
to show up in a low energy tail).  These contributions are described in more detail
below.

Proportional counter resolution is typically limited at high energies by fano factor
statistics.  At low energies there may be a significant component related to detector non-uniformities and
readout electronics \citep{fraser89}.  The resolution 
($\Delta E/E$, where $\Delta E$ is the FWHM) is $\sim 0.17$ at 6 keV and $\sim 0.08$ at 22 keV,
measured with ground calibration sources.
We model the resolution (FWHM) in channel space as
\begin{equation}
{\Delta ch} = (\sqrt{aE + b}) B
\label{energy_res}
\end{equation}
where $B$ is the slope term from equation \ref{e2c}.  

The Xenon L-escape peak is not prominent.  The 4.1 keV escape photon has a 
mean free path of 0.5 cm in 1 atm of Xenon.  Escape is only possible 
from the first layer.  For the inner layers (and the front layer) 
there is a chance that the escape photon will be re-absorbed in an
adjacent layer.  The resulting coincident signals will affect the overall efficiency
as the EDS will not recognize this as a good event.  The
model allows an efficiency correction for energies above the Xenon L-edges and below 
the K edge, though our parameterization sets this self vetoed fraction to 0.
The overall normalization is adjusted by the values of geometric area
which parameterize the tool {\tt xpcaarf}.

\begin{figure}[t]
\includegraphics[width=0.8\columnwidth,angle=270.0]{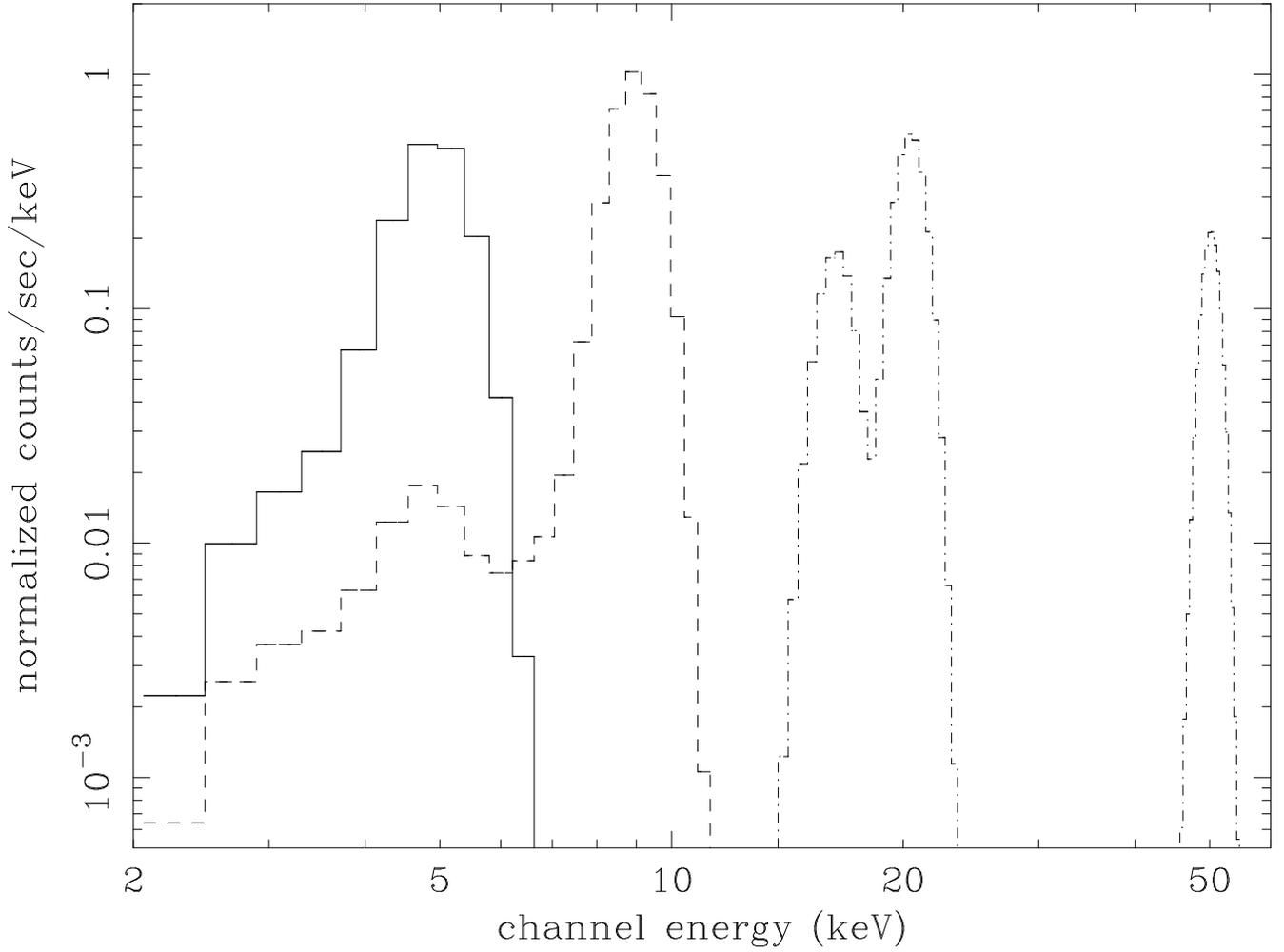}
\figcaption{The model response is shown for 3 narrow input lines:  at 5 keV where partial
charge collection is important (solid line), at 9 keV where there is a small L-escape peak (dashed), and at
50 keV where there are prominent K-escape peaks (dot-dash).  
\label{3line}}
\end{figure}

Above the K-edge, only the K-escape peaks are modelled, and these are more prominent 
than the main photo-peak.  Detailed prediction of the amplitudes would 
require a complex integration over the detector geometry, and
might be slightly different for the three detector layers.  We 
assume that the escape fraction is the same for each layer.
Initial estimates of the escape fractions come from laboratory
data obtained at the Brookhaven National Laboratory National 
Synchrotron Light Source
\footnote{ The Brookhaven National Laboratory
National Synchrotron Light Source
is supported by the U.S. Department of Energy.  
At the time of our visit, the  Contract Number was DE-AC02-76CH00016.}
where small areas of one of the detector were illuminated by monochromatic beams. 
Our model allows a correction for self vetoing events when the escape photon is
absorbed in a different anode volume.  Our estimate is that 0.09 of
the events above the K-edge are vetoed by the reabsorption of an
escape photon in a different signal volume.  The ratio of events in
the photo-peak, the $K_{\alpha}$ escape peak, and the $K_{\beta}$
escape peak is 0.29 / 0.55 / 0.16.  The redistribution function
for a 50 keV input is illustrated in
figure \ref{3line}.

Events can be lost if the photo-electron travels into a second anode volume. 
\Citet{youngen94} demonstrated the importance of this effect for the 
HEAO-1 A2 High Energy Detector \citep{roth79} which has a similar internal 
geometry to the PCA.  The Youngen model is a Monte Carlo,
making it difficult to use the results directly.  We model the unvetoed 
fraction as
\begin{equation}
f_{unveto} = 1.0 - A_o g(layer) ({E_{electron} \over E_{point}})^\delta
\label{e_tracking}
\end{equation}
where $E_{electron}$ is the energy of the initial photo-electron, 
$A_o$ is an amplitude, $E_{point}$ is an empirical parameter that
is approximately the energy where this effect becomes important, and
the factor $g(layer)$ (= 1, 1.33, 1.33 for the first, second, 
and third layers) accounts
for the fact that the inner detection cells have more edges which 
the electron might cross.  
Because the exponent $\delta$ is large ($\sim 3$), the fraction lost is
near constant below $E_{point}$ and rises quickly above this energy.

We also model partial charge collection.  For X-rays which are absorbed near the 
edge of the active region, there is a competition between diffusion and drift 
towards the high field region where charge multiplication occurs.  The effect 
has been modelled at low energies \citep{inoue78} and parameterized
in terms of the ratio $\kappa = {D \over {w \lambda}}$ where $D$ is the 
diffusion coefficient, $w$ is the drift velocity, and $\lambda$ is the 
mean absorption depth.  For an initial cloud of $N_o$ electrons,
the number which reach the amplification region is
\begin{equation}
f(N)dN = \kappa ( 1 - { N \over N_o }) ^{\kappa - 1} dN.
\label{pcc}
\end{equation}
\citet{jah88} used this equation directly for an Argon based counter while 
\citet{inoue78} found it necessary
to introduce an ad-hoc factor of $\sim 3$ for a Xenon based counter.  
We have chosen to treat the ratio
of $D$ and $w$ as a free parameter which we fit.   
Equation \ref{pcc} was derived for
energies where the photons are absorbed near the entrance window.
However, this model predicts very small
losses for energies where this condition is not satisfied, so we apply this 
model at all energies.

Figure \ref{3line} shows the redistribution function for lines with 
energies at 5, 9, and 50 keV to illustrate the magnitude of the 
incomplete charge collection, L-escape, and K-escape peaks.
The normalization of the lines is arbitrary and selected for illustrative purposes.
 
\placefigure{3line}

\subsection{Detailed construction of the matrix}
\label{details}
This section describes the operation of the PCA response generator {\tt pcarmf};  
default values of the parameters and their mnemonics are listed in table 
\ref{v5.3_pars} and described below.

The matrix is constructed to have 296 energy bins in $E_{\gamma}$, spaced logarithmically
from 1.5 to 80.0 keV.  Extra bin boundaries are inserted at the three 
Xenon L edges and the Xenon K edge, so there are a total of 300 energy bins.
The edge energies are read as the parameters
{\it xeL3edge, xeL2edge, xeL1edge} and {\it xeKedge}.
The matrix maps monochromatic input from the mean energy of each bin to
256 detector channels, indexed from 0 to 255 to match the channel-ids reported by 
the PCA to the EDS.  

The response matrix generator works on one detector and one layer at
a time.  The detector is specified by {\it pcuid} with allowed values
ranging from 0 to 4.  The layer is specified by {\it lld} with allowed
values of 3, 12, 48, 63, and 64 for the first, second, third, summed Xenon,
and propane\footnote
{The propane layer is poorly calibrated as we have no in flight
energy scale diagnostics.} layers respectively.
The date is specified by {\it cdate}, in yyyy-mm-dd
format.

A matrix corresponding to the selected detector, layer, and date is generated
with the following steps.  
For each energy in the matrix we
\begin{itemize}

\item Determine the overall quantum efficiency.  

\item Calculate $E_p$ (equations \ref{ep} and \ref{wf});  the parameter $f$ is 
specified by {\it w\_xe\_fact}.
\footnote{This is {\it e2c\_model} ``3";
{\tt pcarmf} is capable of handling an alternate energy to channel 
model (model ``2") where 
$E_p = E_{\gamma} - \Sigma(\Delta_{edge})$. 
The $\Delta_{edge}$ are defined so that $E_p = E_{\gamma}$ between the 
Xenon L and K edges and
$E_p$ is greater (less) than $E_{\gamma}$ for energies below the L edges 
(above the K edge).  The size of the jumps, in keV, is given by 
{\it DeltaE\_L$n$} and {\it DeltaE\_K}
where $n$ = 1,2,3.  This model has a strictly linear $E_p$ vs $E_{\gamma}$
relationship between the edges.  The energy to channel relationship 
has not been provided for this model.}

\item Calculate the mean channel ($ch_{mean}$).
The coefficients in equation \ref{e2c} are stored in pca\_e2c\_e05v03.fits 
found in the \$LHEA\_DATA directory which is part of the FTOOLS v6.0 
package.  The latest version of
this file is also available from the HEASARC calibration data base CALDB.

\item Determine whether the photon energy is below the Xenon L3 edge (parameter {\it xeL3edge}) where no escape peak is possible,
between the L3 and K edge (parameter {\it xeKedge}) where an L-escape photon is 
possible, or above the K edge where
both the Xe K$_{\alpha}$ and K$_{\beta}$ escape photons are possible.

\item For energies where escape peaks are possible, calculate the mean 
channel of the escape peak.  
Monte Carlo and experimental results \citep{dias96} indicate that
the mean channel associated with an escape peak is not in the same location
as the photo-peak for an X-ray with energy equal to $E_{\gamma} -
E_{esc}$.  The offsets can be as large as 70 eV and
are quantified in terms of a number of electrons;  
the offsets are parameterized as {\it delta\_el\_L, delta\_el\_Ka} and 
{\it delta\_el\_Kb} for the L, K$_{\alpha}$, and
K$_{\beta}$ escape peaks respectively.  The default values, taken from
\citet{dias96} are 3.9, -2.26, and 3.84 electrons respectively.

\item For energies where K escape peaks are possible, the fraction of events 
in the escape peak are given by {\it EscFracKa} and {\it EscFracKb} for 
the K$_{\alpha}$ and K$_{\beta}$ peaks respectively.
The fractions are modelled to be the same for each layer in the detector.  
When K escape peaks are possible,
the L escape fraction is modelled to be 0.

\item  For energies where the L escape peak is possible, the fraction of 
events in the escape peak is given by {\it EscFracLM} where {\it M} 
is 1, 2, or 3 corresponding to the layer that
the photon is absorbed in.  The default parameterization is no L escape
peak for photons absorbed in the second or third layer, and $1\%$ for 
photons absorbed in the first layer.

\item For the main photo-peak, evaluate the resolution.  
The coefficients $a$ and $b$ in equation \ref{energy_res} are given 
by {\it resp1\_N} and {\it resp2\_N} where $N$ identifies the PCU.  
These coefficients have not been separately fit by PCU.  The resolution is assumed
to be constant in in keV;  the coefficient $B$ converts the resolution
in keV to the (time dependent) value in channels.

\item Correct the quantum efficiency for the effects of electron tracking 
(equation \ref{e_tracking}) where the coefficients $A_o$, $E_{point}$, 
and $\delta$ are given by {\it track\_coeff, epoint} and {\it track\_exp}.

\item Distribute the corrected quantum efficiency, further corrected for the 
escape fractions, in a gaussian centered at $ch_{mean}$ and with $\sigma =
\Delta ch / 2.35$

\item  Treat the escape peaks similarly.  The resolution and electron tracking 
are calculated for $E_{\gamma} - E_{esc}$, and
the resulting peaks are added to the 256 channel spectrum.

\item Correct the entire 256 channel spectrum for partial charge collection 
(equation \ref{pcc}).  As noted above, this is certainly unphysical for 
large energies, but the effect is small here, and the model is plausibly correct
at the energies where the effect is important.

\end{itemize}

\section{Calibrating the parameters}
\label{parfits}
\subsection{Calibration Sources and Targets}
Data suitable for parameterizing the energy scale, and monitoring variations, comes from
three regularly observed sources:  the internal calibration source provides continuous
calibration lines at 6 energies from 13 to 60 keV;  approximately annual observations
of Cas A provide a strong and well measured Iron line at 6.59 keV; and regular monitoring
observations of the Crab nebula provide an opportunity to measure the location of the
Xenon L-edges near 5 keV.  Almost all of the ground calibration data was
obtained with High Voltage settings that correspond to Epoch 1.  In addition
no ground data was obtained which illuminated an entire detector.  Ground
calibration thus verified basic operation of each detector and the 
general features of the response.  The detailed response is parameterized
with flight measurements, which also allows a correction for a small
time dependence.

\subsubsection{Internal Calibration Source}
\begin{figure}[t]
\includegraphics[width=0.8\columnwidth,angle=270.0]{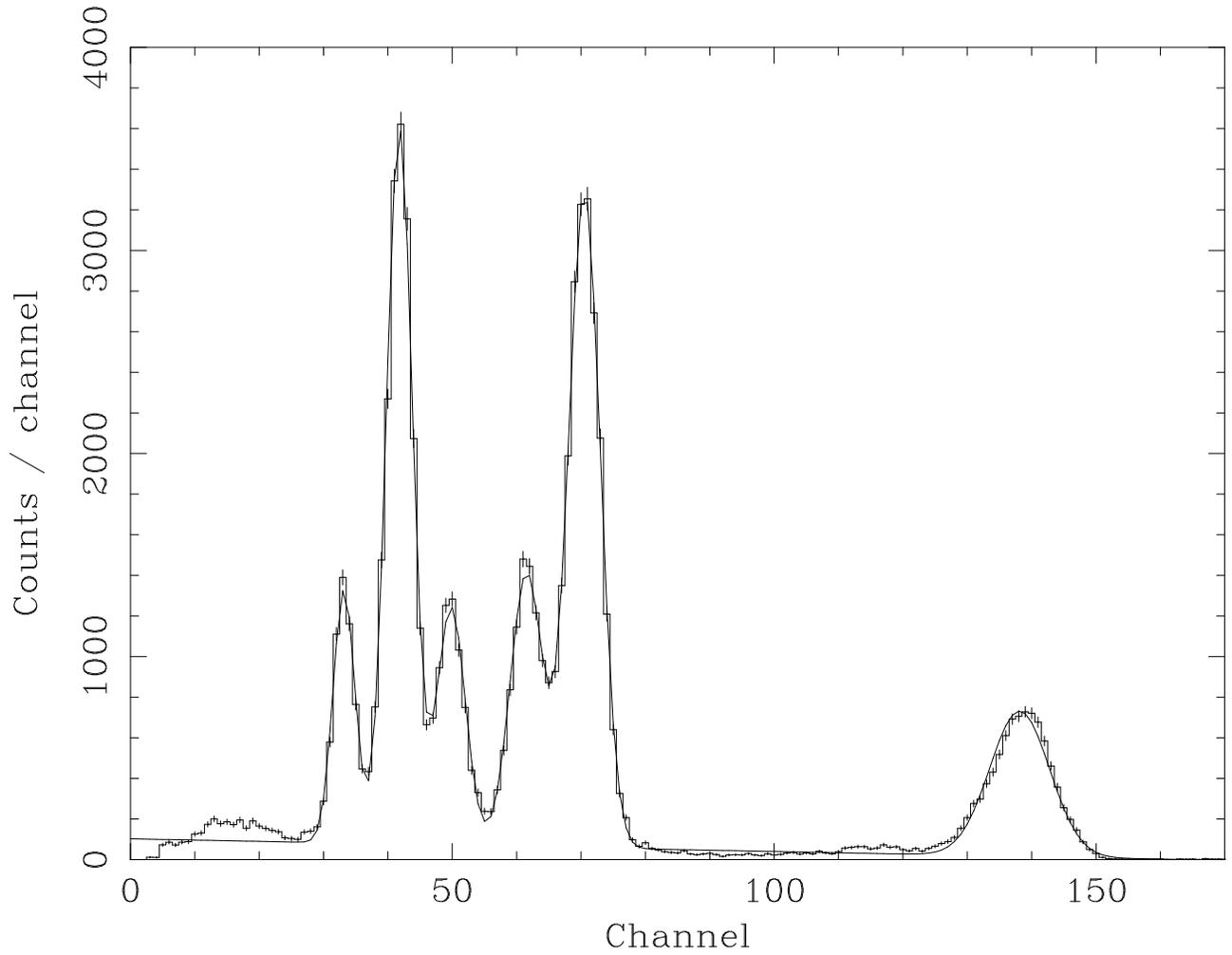}
\figcaption{Data and gaussian fits to the Am$^{241}$ calibration line collected in 
PCU 2 during sky background pointings in September 2000.  The 60 keV peak
is slightly assymetric as the Compton scattering cross section has become
noticable (1.5\% of the photo-electric cross section).  We have ignored this
effect.
\label{cal_spec}}
\end{figure}

Each PCU contains a small Am$^{241}$ source which provides a continuous source of
tagged calibration lines with energies between 13 and 60 keV \citep{zhang93}.  
The energies and notes about each line are listed in table \ref{line_energies}.    
Calibration spectra are telemetered with the full 256 channel resolution of the PCA, so we perform fits directly in channel space.  A sum of 6 gaussians with a small 
constant and linear term provides an excellent fit.  An example is shown in 
figure \ref{cal_spec}.  The 60 keV line is imperfectly modelled by a Gaussian as
the Compton cross section is beginning to become significant with respect to
the photo-electric cross section.  We have ignored this effect.

We collected calibration data from dedicated sky background pointings within each 
calender month of the mission (for the period before November 1996 when dedicated 
background observations began, we have used observations of faint sources).  
During months when the high voltage changed, we collected 2 distinct spectra.  
This procedure provides data which is systematics (i.e. not statistics) limited.  
For instance, the line near 26 keV is consistently broader than the others as 
it is a blend of a nuclear line from the Am$^{241}$ source and the K-$\alpha$ 
escape peak from the 60 keV line.

\subsubsection{Cassiopeia A}
\label{casa}
\begin{figure}[h]
\includegraphics[width=0.8\columnwidth,angle=270.0]{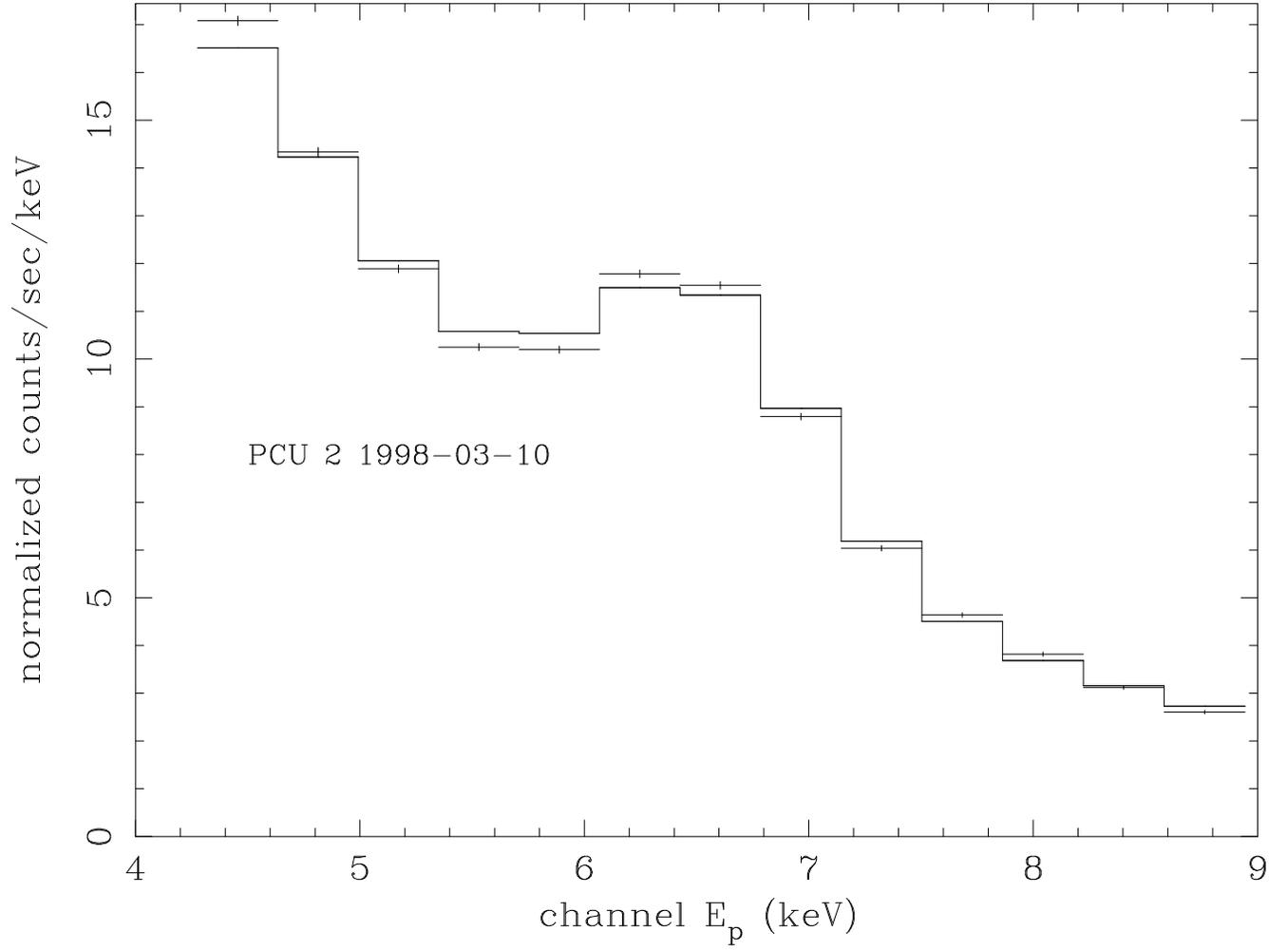}
\figcaption{Power-law  plus gaussian fit to the Cas-A Fe line.  
\label{casa_fit}}
\end{figure}

The supernova remnant Cassiopeia A has a bright, strong, Iron line easily visible 
in the PCA count spectra.  Cas-A has been observed approximately annually as a
calibration source, and we have fit a model with a power law continuum and a gaussian
line to the data between 4 and 9 keV (figure \ref{casa_fit}).  Over this band
pass, this simple model provides an acceptable fit to the data;  over a broader band
the models become quite complex \citep{allen98}.  The fit energy centroid can be 
unambiguously converted to a mean channel.  We assume that the actual average
energy of 
the Fe line complex is 6.59 keV, consistent with {\it Ginga} \citep{ginga92} 
and ASCA \citep{holt94} results.  The mean channel can be accurately determined
without considering the line complex at 8.1 keV observed by \citet{pravdo79} and
\citet{bleeker2001}.

\placefigure{casa_fit}

While there are relatively few pointings at Cas-A, and this line thus has limited 
weight in the channel to energy fits, it does provide a valuable check on the 
accuracy of the energy scale in the Fe line vicinity.  Results are presented in 
section \ref{resultsf}.   While calibration near the Iron line is unavailable for 
the second and third Xenon layers,  there is also little significant signal 
from cosmic sources here.  

\subsubsection{Xenon L-edge}
Each detector has a front layer filled with propane;  a small amount of Xenon is 
also present, due to diffusion either through the window or the o-ring seal 
that separates the two volumes.  Although the presence of Xenon in this layer 
does reduce the efficiency in the active volume, it also provides a 
calibration opportunity.  The regular monitoring of the Crab nebula plus 
pulsar allows us to measure and monitor the energy calibration near the 
Xenon L-edge at 4.78 keV.

The response matrices with default parameters account
for the Xenon in the propane layer, however we 
can construct a matrix which artificially sets the amount of Xenon to zero.  
The Crab continuum spectrum, analyzed with this incorrect matrix, 
requires an absorption edge to mimic the unaccounted for Xenon.
Fits to this edge can be interpretted in terms of the energy scale.

\begin{figure}[t]
\includegraphics[width=0.8\columnwidth,angle=270.0]{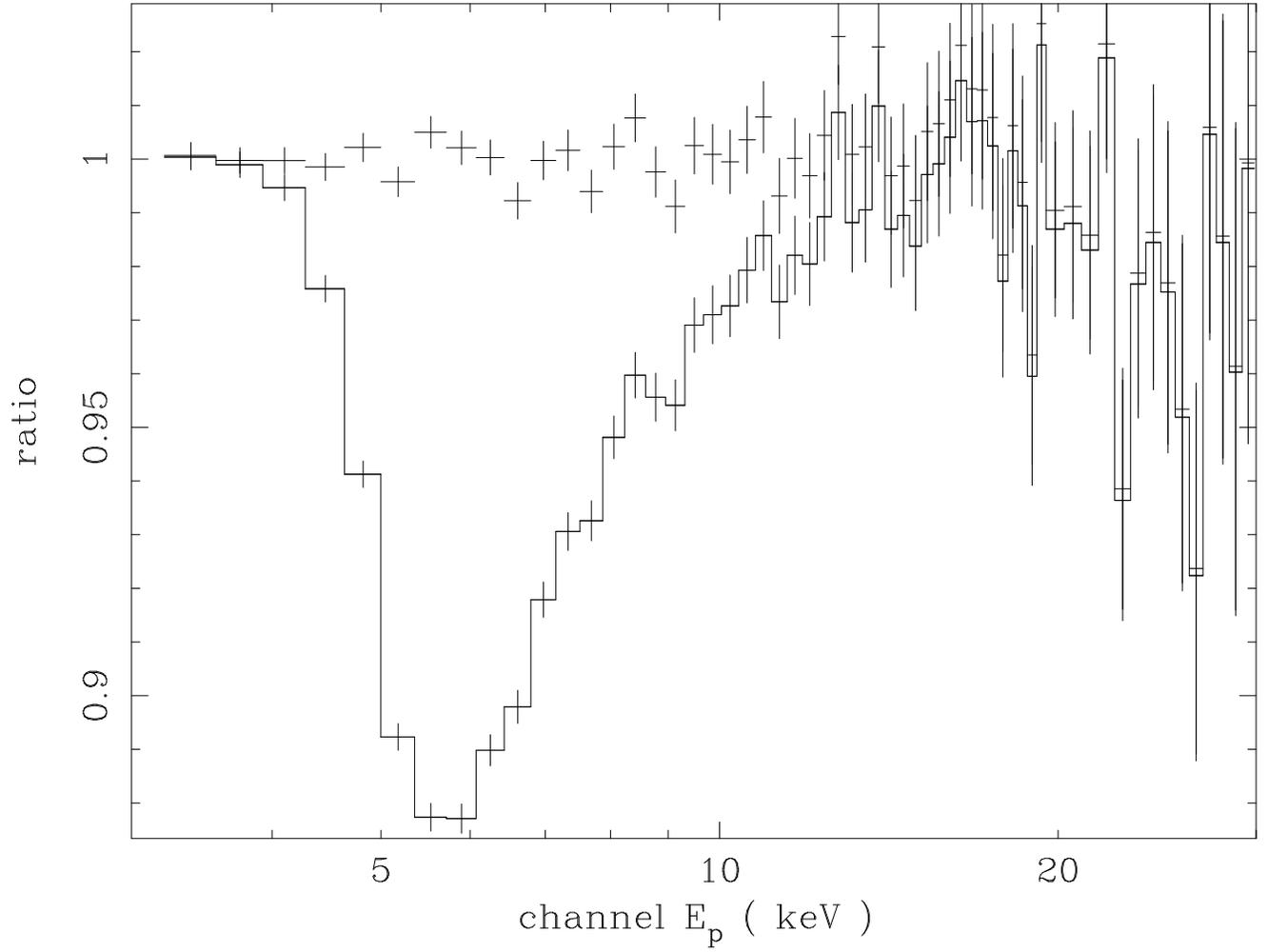}
\figcaption[edge_3tauv2_p2l1.ps]{Ratio of model to data for Crab fit with a Xenon-free propane layer and the edge model described 
in the text (data points) and for the same continuum model with the optical depth set to 0 (line).  The data is from 1997-12-20, 
PCU 2, layer 1.
\label{edge_fig1}}
\end{figure}

\placefigure{edge_fig1}

We perform fits in energy space, and convert the fit energy back to channels.  
The resulting set of date/channel pairs for 4.78 keV are fed back into the 
procedure that determines the energy to channel relationship.  
\footnote{Fitting in energy space allows the use of the 
XSPEC convolution code as well as the pre-defined absorption models.}

\begin{figure}[t]
\includegraphics[width=0.8\columnwidth,angle=270.0]{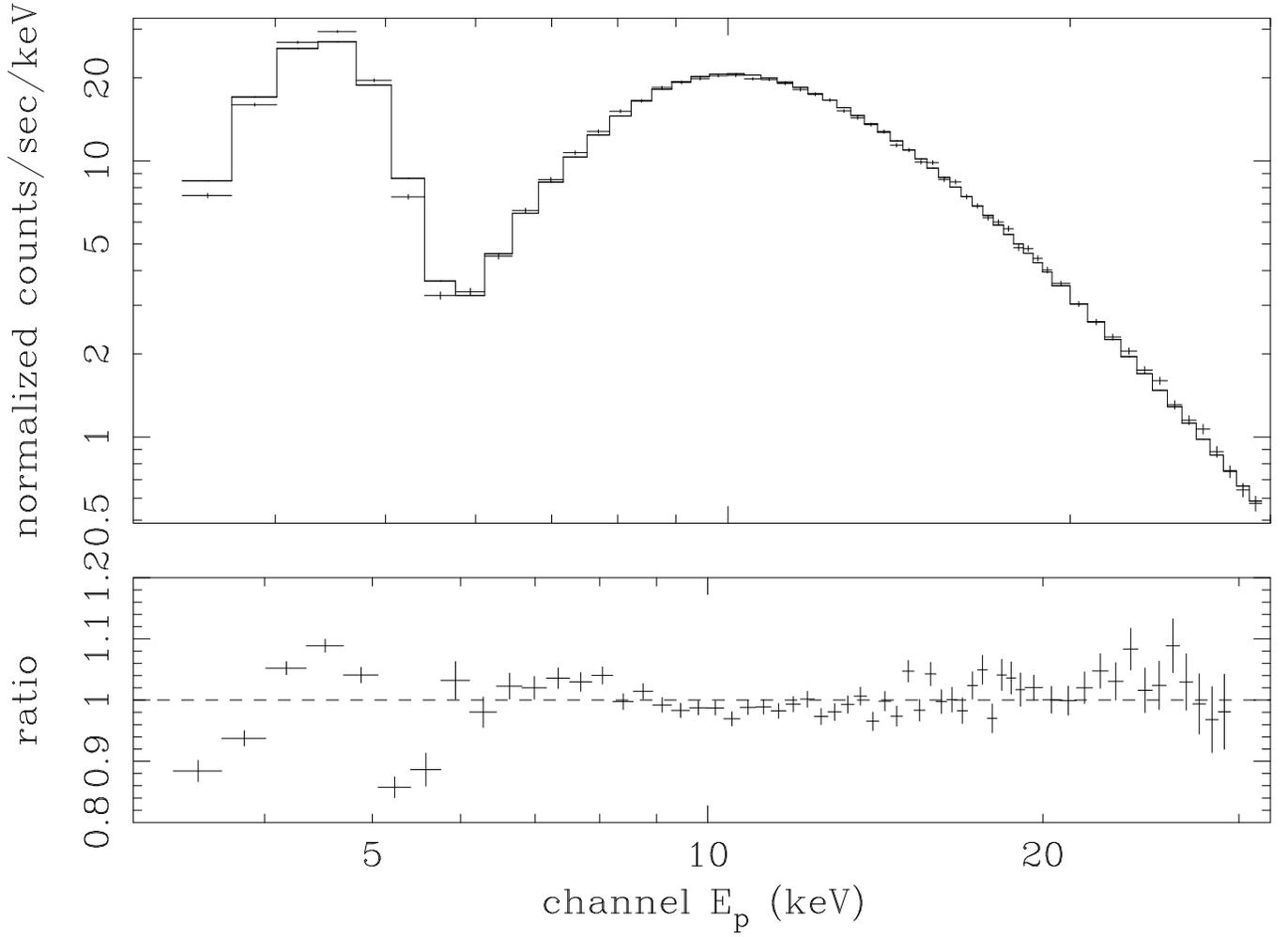}
\figcaption{Data and model for PCU 2, second layer, with a matrix that sets the total Xenon in layer 1 and the propane layer to 0.  The primary
edge has an optical depth of $\sim 4$.\label{edge_fig2}}
\end{figure}

The Xenon L-``edge" has structure on a scale that is fine with respect to the 
energy resolution of the PCA.  There are 3 edges with energies at 4.78, 5.10, 
and 5.45 keV.  With proportional counter resolution these cannot be fit 
simultaneously.   We use a model with a power-law, 3 edges, and 
``interstellar absorption" with variable elemental abundances;  the 
absorption and edges mimic the absorption due to the Xenon in the propane layer.  
The energies of the second and third edges are fixed at 1.07 and 1.18 times 
the energy of the first edge, and the optical depth is fixed at 0.44
and 0.18 times the optical depth of the first edge \citep{henke93}.  
The absorption uses the XSPEC {\tt varabs} model, itself based on the 
cross-sections tabulated by \citet{BCM}.  We keep the relative abundances of
H, He, C, N, O, and Al fixed with respect to each other.  The abundance of 
Fe is allowed to vary.  The abundances of all other elements are fixed at 
zero.  This description is not intended to be physical, but rather to 
produce a fit with small residuals;  examples are shown for PCU 2,
layers 1 and 2, in figures \ref{edge_fig1} and \ref{edge_fig2}.

\placefigure{edge_fig2}

The edges are quite precisely fit with typical $3 \sigma$ errors less 
than $\pm 0.07$ keV for the first layer.  
We include over 100 observations in our determination of the energy to 
channel law, and the average is extremely well determined.  
Systematics associated with the model are thus quite important, and
it remains possible that a different parameterization of the edge
could provide a matrix with smaller residuals in the vicinity of the 
L-edge (see section 
\ref{residualsf}).  

\subsection{Energy to channel relationship}
\begin{figure}[t]
\includegraphics[width=0.8\columnwidth,angle=270.0]{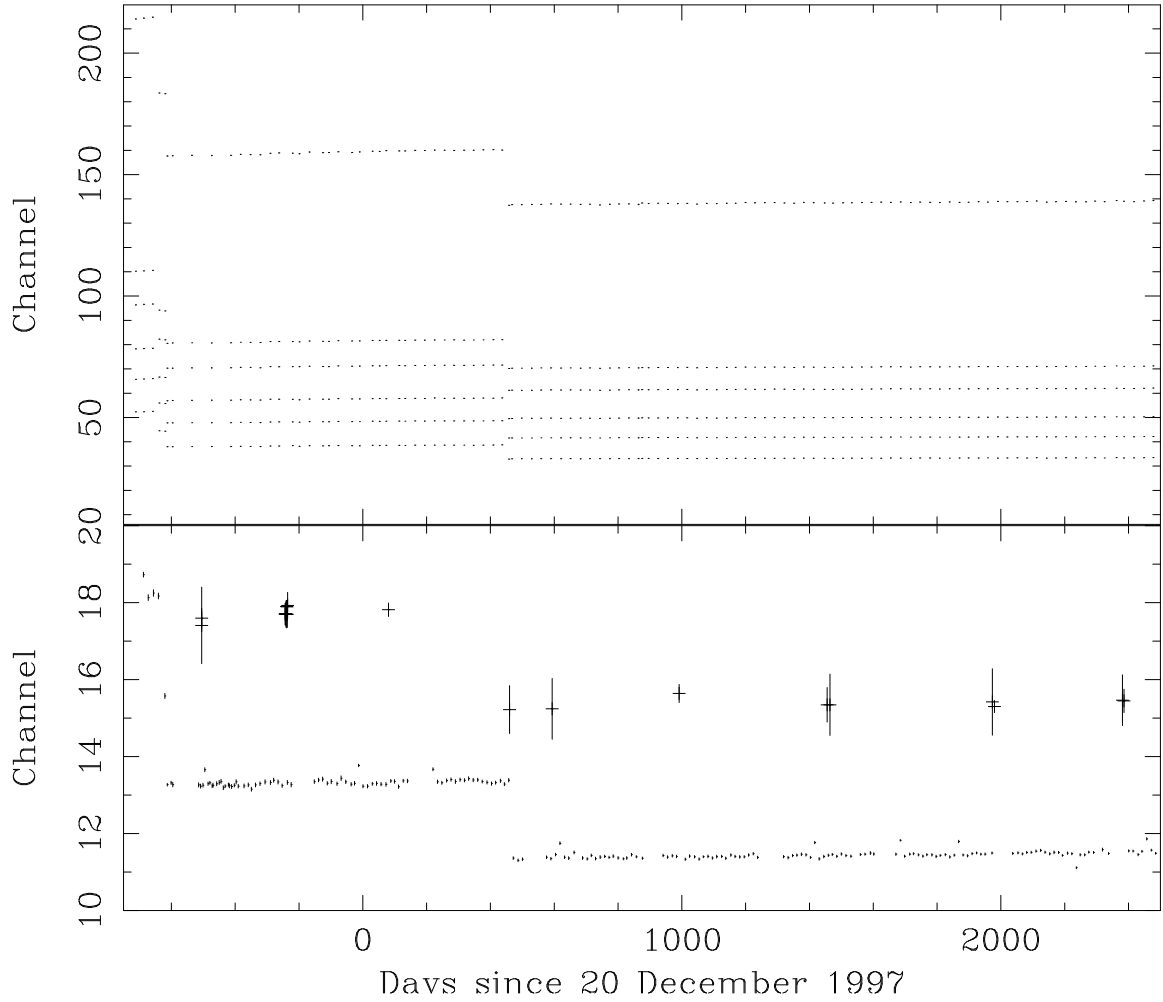}
\figcaption{
The time and channel pairs used to determine the energy to channel relationship
for PCU 2.  The discontinuities related to discrete high voltage changes
are readily apparent.  The upper panel shows the 6 lines from the $Am^{241}$
source while the lower panel show the fits to the Cas A Iron line and the
Xenon L edge.  The small time dependence can be discerned in the Xenon L
data.
\label{e2cfig}}
\end{figure}

We fit the energy to channel relationship separately for each detector, layer, and epoch 
using a $\chi^2$ minimization.  
Figure \ref{e2cfig} shows the data that goes into this fit for PCU 2.  The upper
panel shows the centroids fit to the 6 peaks in the calibration spectra on a 
monthly basis;  the
lower panel shows the channels fit to the Xenon L-edge from Crab monitoring
observations and the Iron line fit to approximately annual observations of 
Cas A.  The gaps in the Xenon L edge data correspond to the annual period when
the Crab is too close to the sun to be observed.

\placefigure{e2cfig}

The precision of the input data is
highest at energies well above the PCA peak sensitivity.  The errors on 
the calibration line channels, as a fraction of the channel, range 
from $10^{-3}$ for the lines at 13 and 21 keV to less than $5 \times 10^{-4}$ 
for the lines at 17, 30 and 60 keV.  The fractional errors on the L-edge 
channel and Cas-A line are $\sim 5 \times 10^{-3}$.  The errors on the
Cas-A line energy include a statistical component and are correlated with
exposure time.

The mean photon energy observed by the PCA is always below 10 keV;  to apply the
maximum weight to the points nearest this mean, we reduce the error associated with
the L-edge points by a factor of 10.  To apply approximately equal weight 
to each of the Cas-A points and remove the exposure dependence,
we set all errors to 0.1 channel (about the mean).  We are not using the 
chi-square minimization to estimate errors on the channel to energy 
parameters (equation \ref{e2c}) and we validate the results
by examining the fits to the Crab nebula (section \ref{residualsf}).  
This ad-hoc procedure for adjusting the
errors is justified not by statistical rigor, but by reasonable results.
Our fits are consistently poor for the line near 26 keV;  we attribute
this to poor knowledge of the mean energy of this blended line, and exclude it 
from our fits.  

We produce energy to channel parameters for a range of values of
the parameter $f$  (eq. \ref{wf});  selection of the
best value of  $f$ comes after the process of adjusting the
parameters associated with quantum efficiency and redistribution.

\subsection{Quantum efficiency and redistribution}
The response matrix contains many correlated parameters;  we have 
used our frequent observations of the Crab nebula to make numerous estimates of
the best values of the parameters, and then made a response matrix using 
averages over time, or over time and PCU.

For each observation of the Crab, we have collected the data and estimated
the background separately for each layer of each detector.  We use the
channel to energy law to select data from each layer with $E_{min}(layer) \le
E_p \le E_{max}(layer)$.  For the first layer we use $E_{min}$ = 3 keV and
$E_{max}$ = 50 keV;  for the second and third layers we accept data from
8 to 50 keV.  

These data are fit, via a chi-square minimization technique, to a model
which parameterizes the response matrix and the Crab input spectrum.  As
parameters are highly correlated (i.e. the amount of Xenon in the first
layer and the power law index for the Crab) we have adopted a procedure
that minimizes a few parameters at a time, and revisits earlier steps as needed.
The Crab photon index is fixed at $-2.1$ and interstellar column 
density is fixed at $3 \times 10^{21}$ H atoms ${\rm cm}^{-2}$
using the approximation to absorption given \citet{zombeck} (page 200).

\begin{figure}[t]
\includegraphics[width=0.8\columnwidth,angle=270.]{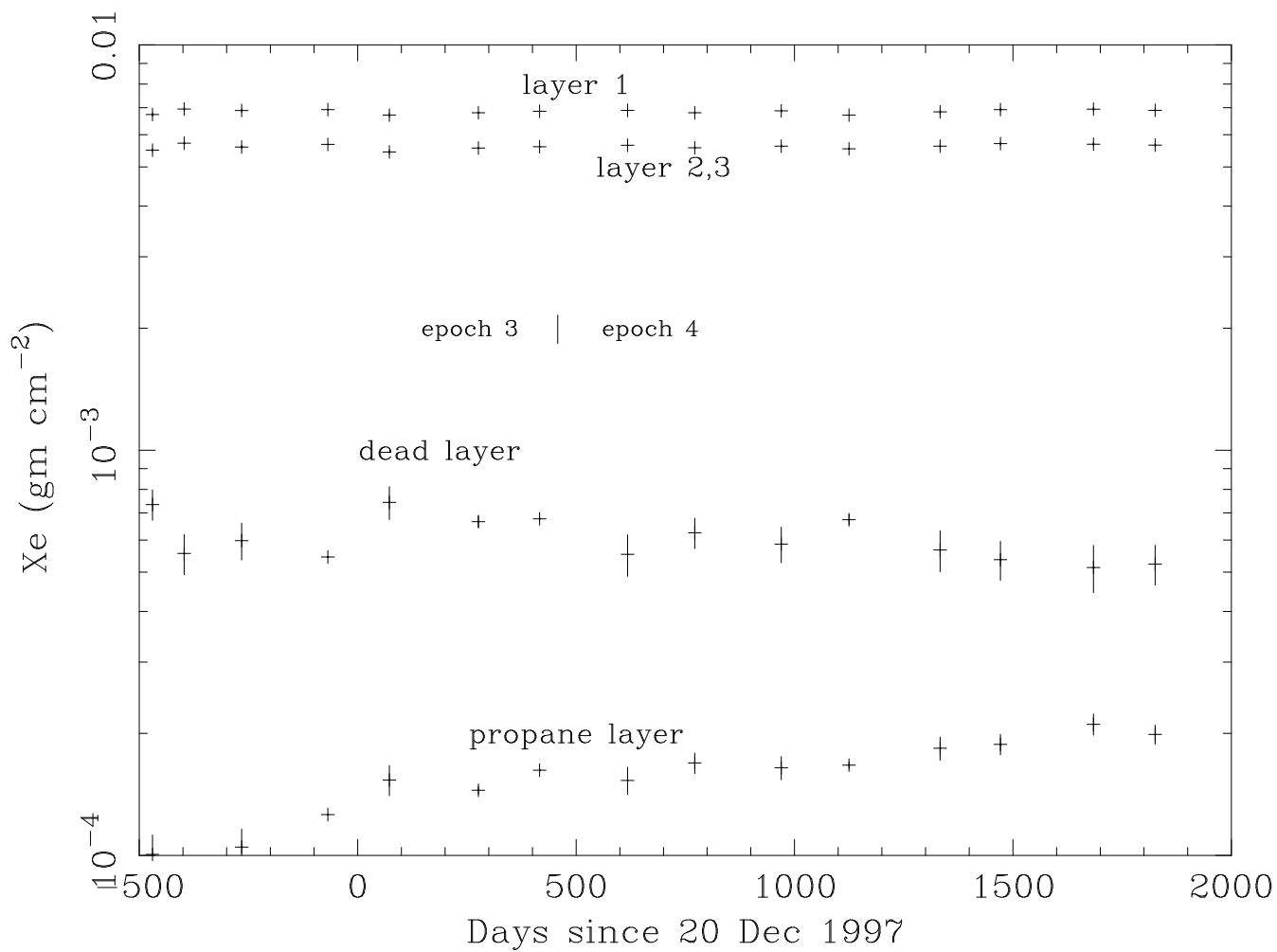}
\figcaption{Best fit values for the areal density of Xenon in the top layer, second and third
layers, dead layers, and propane layer and 1 $\sigma$ error bars.  The dead layers between Xenon layers
1 and 2 and Xenon layers 2 and 3 is assumed to be the same.
Our fits are consistent across the Epoch 3/4
boundary (day 457 on this scale).  Only the amount of Xenon in the propane
layer shows a time dependence;  this is approximated as a linear trend in the
response matrix generator.
\label{xe_fits}}
\end{figure}

Figure \ref{xe_fits}
shows the resulting fit values for the Xenon thicknesses of the first and
second signal layers, the propane volume, and the dead layer for PCU 2. 
We require the Xenon thickness of the third layer to be equal to the second layer.
The time scale is in days relative to 20 Dec 1997.  Note that on this plot
there is no discernable break in March 1999 (near day 500) when the high
voltage was changed.

\placefigure{xe_fits}

\begin{figure}[t]
\includegraphics[width=0.8\columnwidth,angle=270.]{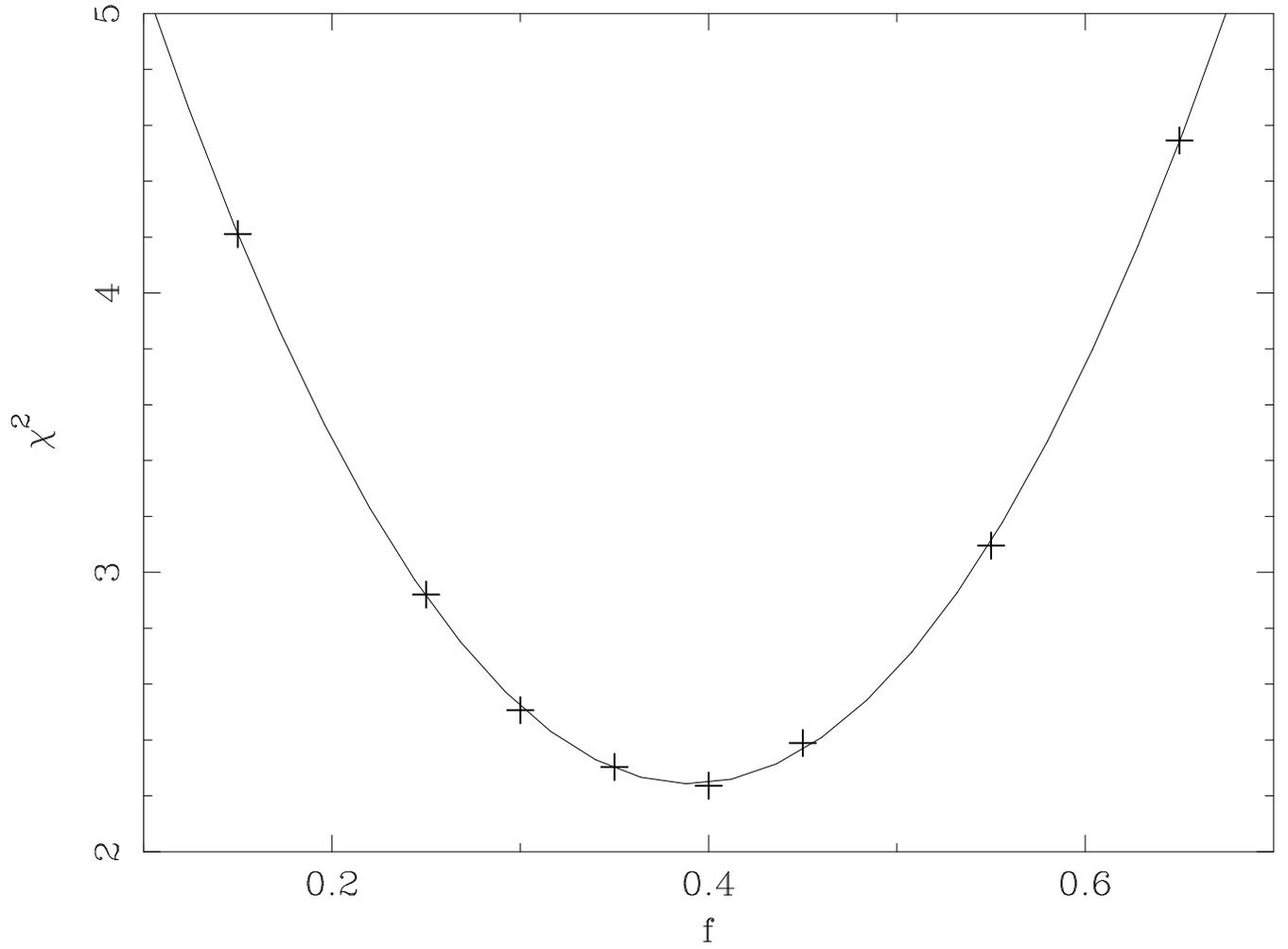}
\figcaption{Reduced $\chi^2$ for PCU 2 observations 
of the Crab nebula on 15 March 2000 as a function of $f$.  
The broad minimum near $f$ = 0.4 is typical.  The parabolic fit
guides the eye but has no physical significance.
The reduced $\chi^2$ is dominated by remaining systematic errors near the
Xenon L edges and in the lowest energy channels.
\label{best_w}}
\end{figure}

This procedure was repeated for several values of the parameter $f$.  We 
selected the best value of $f$ by comparing the values of $\chi^2_{red}$ 
for power law fits to the Crab data in all three layers.  By design all 
of these fits returned $\Gamma \approx 2.1$.  Using $\chi^2_{red}$ as a 
discriminating statistic, there is smooth variation with $f$, as shown in
fig \ref{best_w}.  For almost all observations of the Crab with 
$0.35 \le f_{min} \le 0.45$, so we adopted $f = 0.40$
as the best fit value.  The calibration of the {\it Ginga} Large Area Counters 
reports a total jump of 70 eV (equivalent to $f \approx 0.7$) summed across 
the three Xenon edges \citep{turner89}.  The discrepancy in $f$, which 
represents the properties of Xenon rather than detector details, indicates 
that systematics remain in the {\it Ginga} and/or PCA models.

Table \ref{v5.3_pars} gives the best fit values for the parameters.  
The parameters are more similar from detector to detector than for 
previous calibrations which were performed with a more ad-hoc approach.

\placefigure{best_w}

\subsection{Effective Area}
We adjusted the areas in {\tt xpcaarf} so that we match the canonical Crab flux
\citep{zombeck}.
Individual spectra from the Crab were corrected for both instrumental and
source induced deadtime (typically $\sim 6\%$;  see fig \ref{crab_std1_rates}.)  
Following this, the area parameters in {\tt xpcaarf} were adjusted so that
the best fit flux matches the literature values.  The net geometric areas were
subjected to an upwards correction of $\sim 12\%$.  The absolute error
in flux scale is believed to be smaller than this;  in situations where
absolute measurements are attempted, it is thus important to correct for
deadtime.
Our correction removes a substantial fraction of the discrepancy
noted by, e.g., \citet{kuulkers03} and references therein, which was performed 
with a previous
version of the response generator, which also systematically reported higher
values of the photon index for the Crab.

Our procedure is to fit an absorbed power-law to the many observations of the 
Crab nebula.  We use data from the first layer, and adjust the peak open area 
of the detector so that the average 2-10 keV flux is $2.4 \times 10^{-8} 
\ergscm$ \citep{zombeck}.  The derived geometric areas, which are inputs 
to {\tt xpcaarf}, are documented in table \ref{v5.3_pars}.  For PCU 0 we use 
data only from epoch 3 and 4 as the epoch 5 calibration remains poor (see 
the step function in the best fit index in fig \ref{crab_indices0}).

The PCA flux scale remains higher than for other instruments as the adopted
flux scale ignores interstellar absorption and because the the \citet{zombeck}
parameterization of the Crab nebula plus pulsar provides the highest
inferred 2-10 keV flux amongst measurements of the absolute flux.  More details
about the flux scale are provided on the PCA web page (footnote \ref{cal_page}).




\subsection{Checks on the Quality of the Redistribution Matrix}
\label{resultsf}

\subsubsection{Energy scale}
\begin{figure}[t]
\includegraphics[width=0.8\columnwidth,angle=270.]{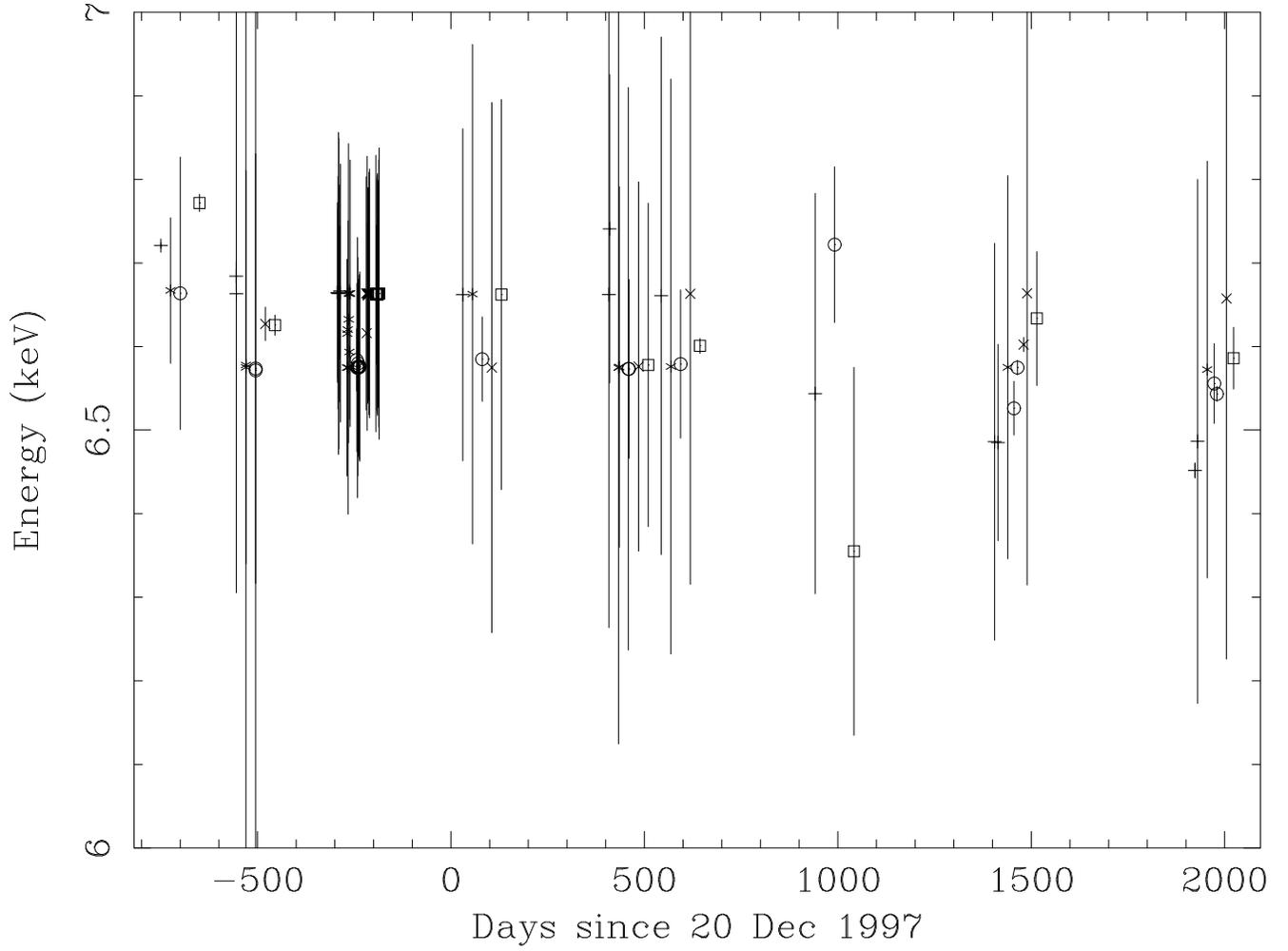}
\figcaption{Energy fit (with 1 $\sigma$ error bars) to the Fe line in Cas-A.  The model is described in section 
\ref{casa}.  Data from the different PCUs have been slightly offset in time 
for clarity.  The best fit energy is quite close to the
expected result of 6.59 keV.
\label{casa_line}}
\end{figure}

We re-fit the Cas-A line with our best fit matrix and plot the results in fig 
\ref{casa_line}.  The fits return a narrow line.  The centroid position is 
fit only to $\sim 0.15$ keV due to a combination of the intrinsic resolution 
of the detector and the steepness of the underlying continuum  ($\Gamma \approx 3$).
These errors are representative of the accuracy that can be expected from
fits to strong lines.

\placefigure{casa_line}

We have examined the stability of the energy scale on short time scales by collecting
the Am$^{241}$ calibration spectra on 1000 second intervals, taking data from all periods
when the detectors were on.  Such spectra allow fits for the amplitude and centroid
of the six lines although it is necessary to fix the width of the peaks at values
fit to spectra collected over one day.  We find no evidence for gain variation on
any timescale longer than 1000 seconds, and no effect on gain with respect to source
counting rate. (up to 2 Crab in the days we examined).

\subsubsection{Spectrum of the Crab Nebula}
\label{residualsf}
The Crab offers three checks on the quality of the response matrix.  First, it allows us to establish
consistency with previous results since the integrated emission is believed to be time stationary.
Second, it allows us to establish whether
our matrix properly accounts for the time dependencies in the detectors, and provides the same results
over the nearly nine years (and counting) lifetime of the mission.  Third, we can examine the residuals
to the fits and thus estimate how large residuals in fits to other sources must be before simple
continuum models are deemed inadequate.

\begin{figure}[t]
\includegraphics[width=0.75\columnwidth,angle=270.]{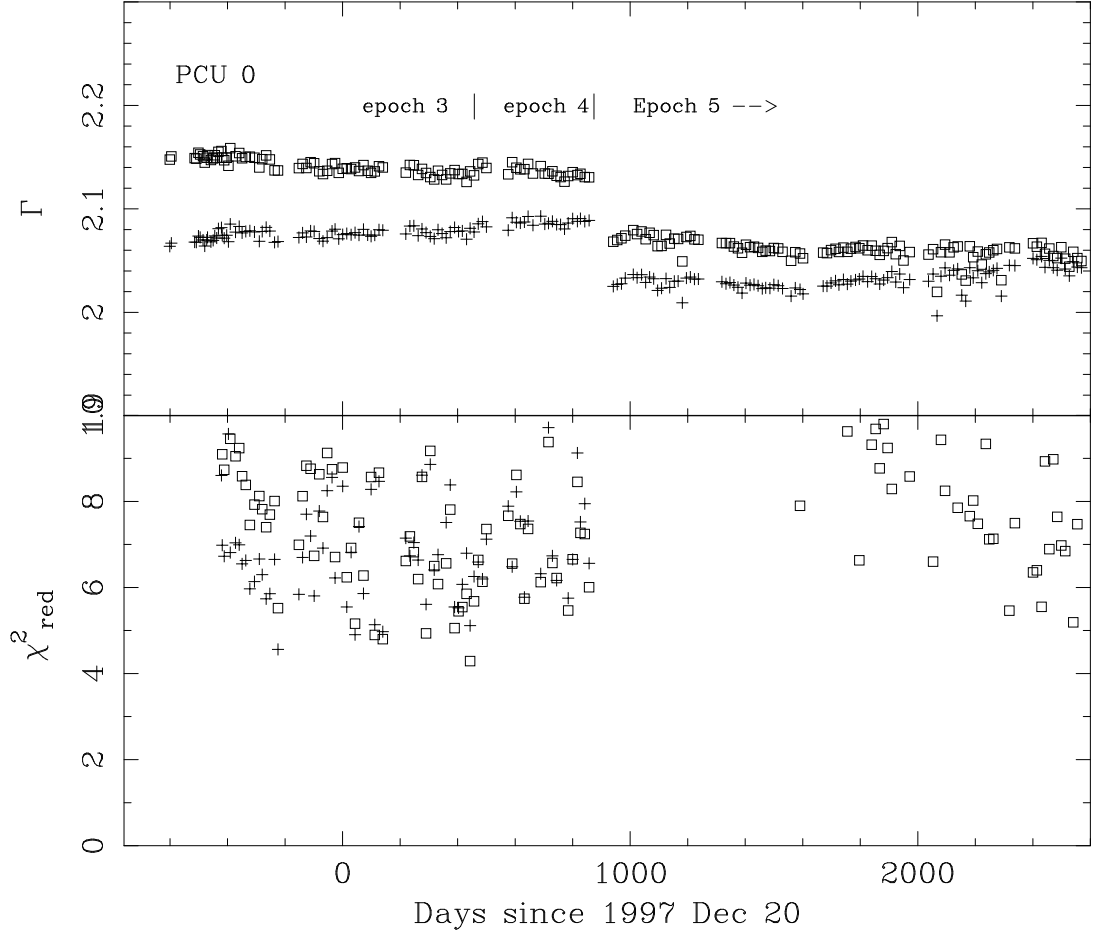}
\figcaption{The upper panel shows the photon index fit to data from the Crab for 
PCU 0.  The crosses represent the Ftools v5.3 response generator with the
default parameters (table \ref{v5.3_pars}) 
while the squares represent the previous calibration (Ftools v5.2).
The lower panel shows the reduced $\chi^2$ for each observation;  the symbols
have the same meaning.   The success of the current calibration is demonstrated 
by the consistent measure of $\Gamma$ with respect to time, and the good 
agreement between detectors.  The previous parameterization gave
slightly lower reduced $\chi^2$ for some detectors, at the cost of 
greater detector to detector variability.
For PCU 0, fits remain poor after the loss of the propane layer.
\label{crab_indices0}}
\end{figure}

\placefigure{crab_indices0}

\begin{figure}[t]
\includegraphics[width=0.75\columnwidth,angle=270.]{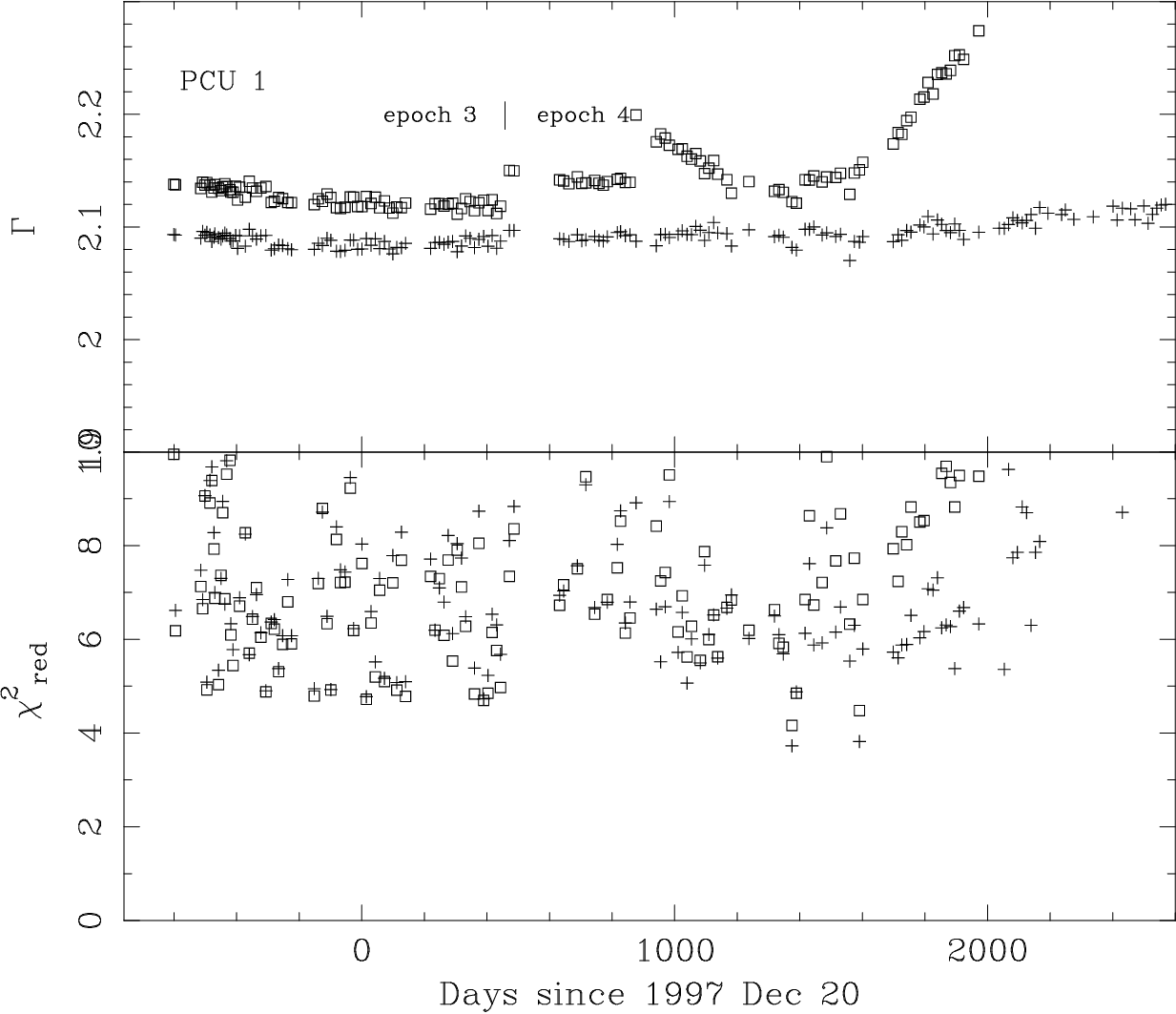}
\figcaption{Same as figure \ref{crab_indices0}, but for PCU 1.  
\label{crab_indices1}}
\end{figure}

\begin{figure}[h]
\includegraphics[width=0.75\columnwidth,angle=270.]{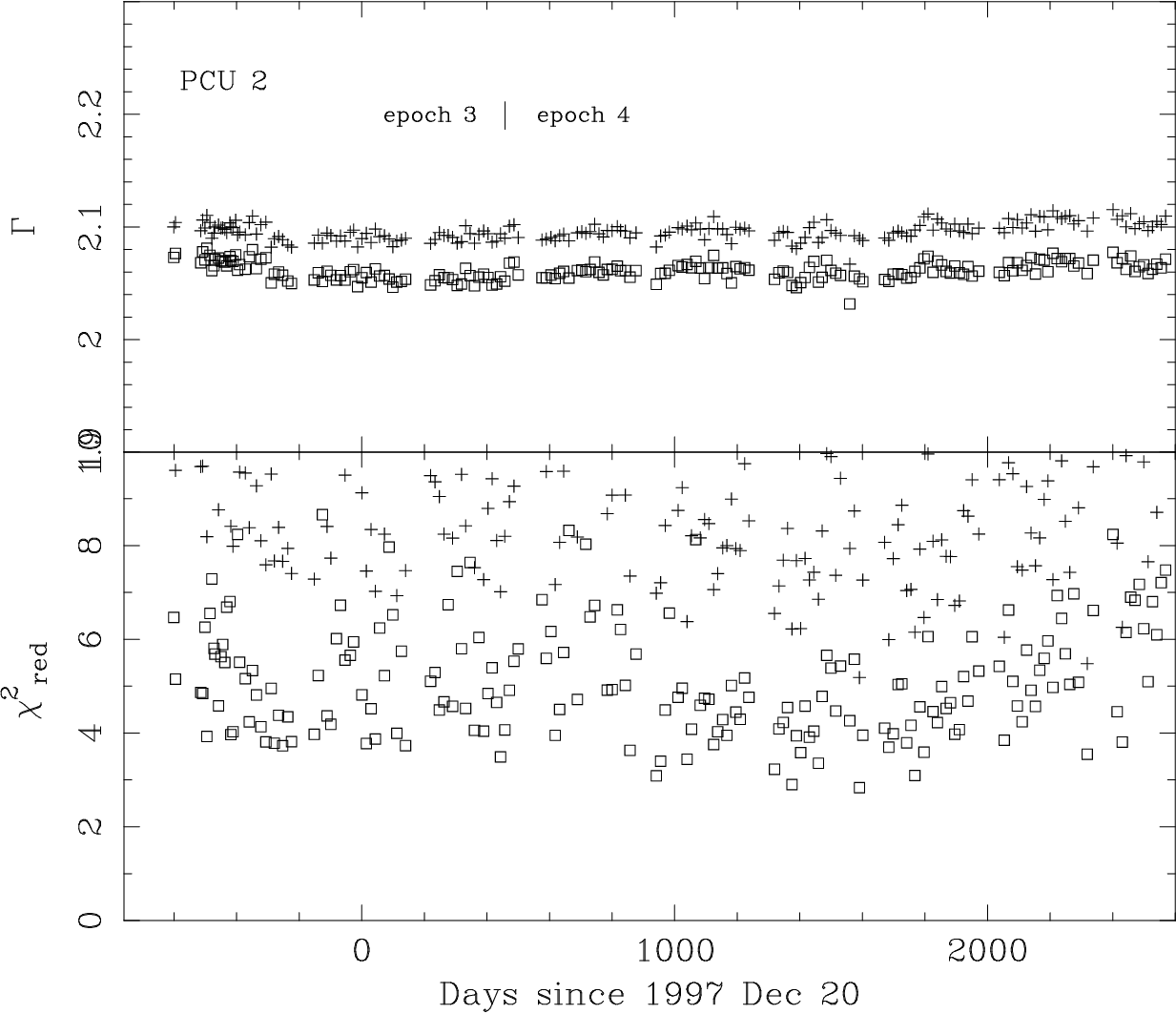}
\figcaption{Same as figure \ref{crab_indices0}, but for PCU 2.  
\label{crab_indices2}}
\end{figure}

\begin{figure}[h]
\includegraphics[width=0.75\columnwidth,angle=270.]{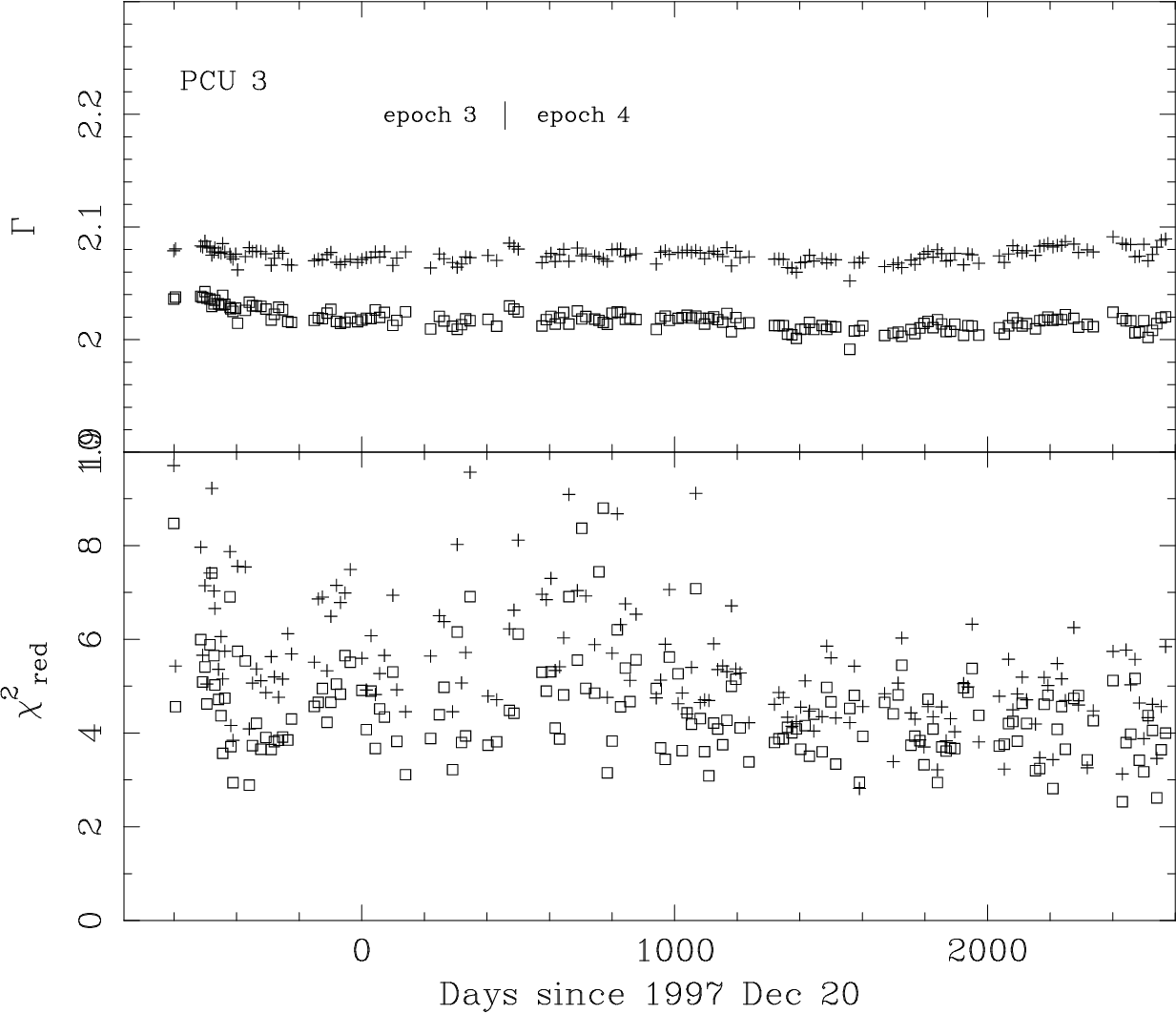}
\figcaption{Same as figure \ref{crab_indices0}, but for PCU 3.  
\label{crab_indices3}}
\end{figure}

\begin{figure}[h]
\includegraphics[width=0.75\columnwidth,angle=270.]{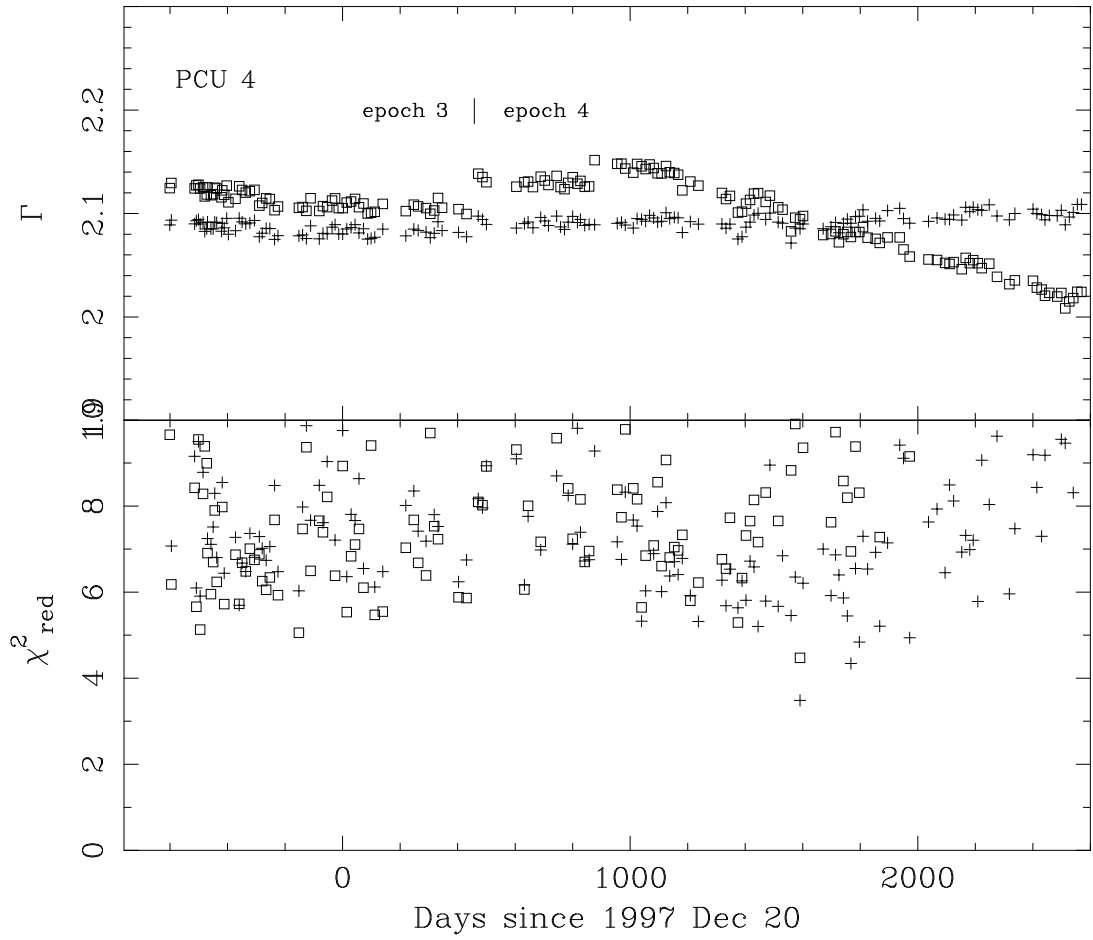}
\figcaption{Same as figure \ref{crab_indices4}, but for PCU 4.  
\label{crab_indices4}}
\end{figure}

The first two points are addressed in figures \ref{crab_indices0}  through 
\ref{crab_indices4} which present the best fit power-law index
for about one-quarter of the Crab monitoring observations using the 
matrix generator in the FTOOLS v5.3 release (crosses), as well as the 
previous calibration (open squares).  
The substantive differences between the two releases are
\begin{itemize}
\item More Crab and background data is included, thus better specifying
the slow drifts in the channel to energy relationship;
\item the parameters for each detector are minimized with the constraint
that the power law index of the Crab be -2.1;  this improves the detector
to detector agreement with a modest increase in chi-squared.
\item an error in the calculation of the instrument resolution (equation
\ref{energy_res}) was repaired.
\end{itemize}
For all fits the interstellar
absoprtion was fixed at a column density of $3 \times 10^{21}$ H atoms
${\rm cm}^{-2}$ using the {\tt XSPEC} model {\tt wabs}.\footnote{This is a more
detailed treatment than used in determination of the quantum efficiency
parameters.  The {\tt wabs} model has a discontinuity at each atomic
absorption edge rather than the \citet{zombeck} parameterization which
has an edge only at the Oxygen edge.  This slight inconsistency has a
minor effect}
 The lower panel in each figure shows 
the reduced $\chi^2$ associated with the FTOOLS v5.3 and v5.2.  
By construction, 
the best fit power-law index is now more consistent from PCU to PCU 
(particularly at late times where the previous calibration provided
an estimate based on extrapolation).  These figures show fits to layer 1 
only, which provides the large majority of the detected photons and the 
largest bandwidth.  That the reduced $\chi^2$ is typically greater than 1 
indicates that unmodelled systematics remain.

\placefigure{crab_indices1}
\placefigure{crab_indices2}
\placefigure{crab_indices3}
\placefigure{crab_indices4}

\begin{figure}[h]
\includegraphics[width=0.75\columnwidth,angle=270.]{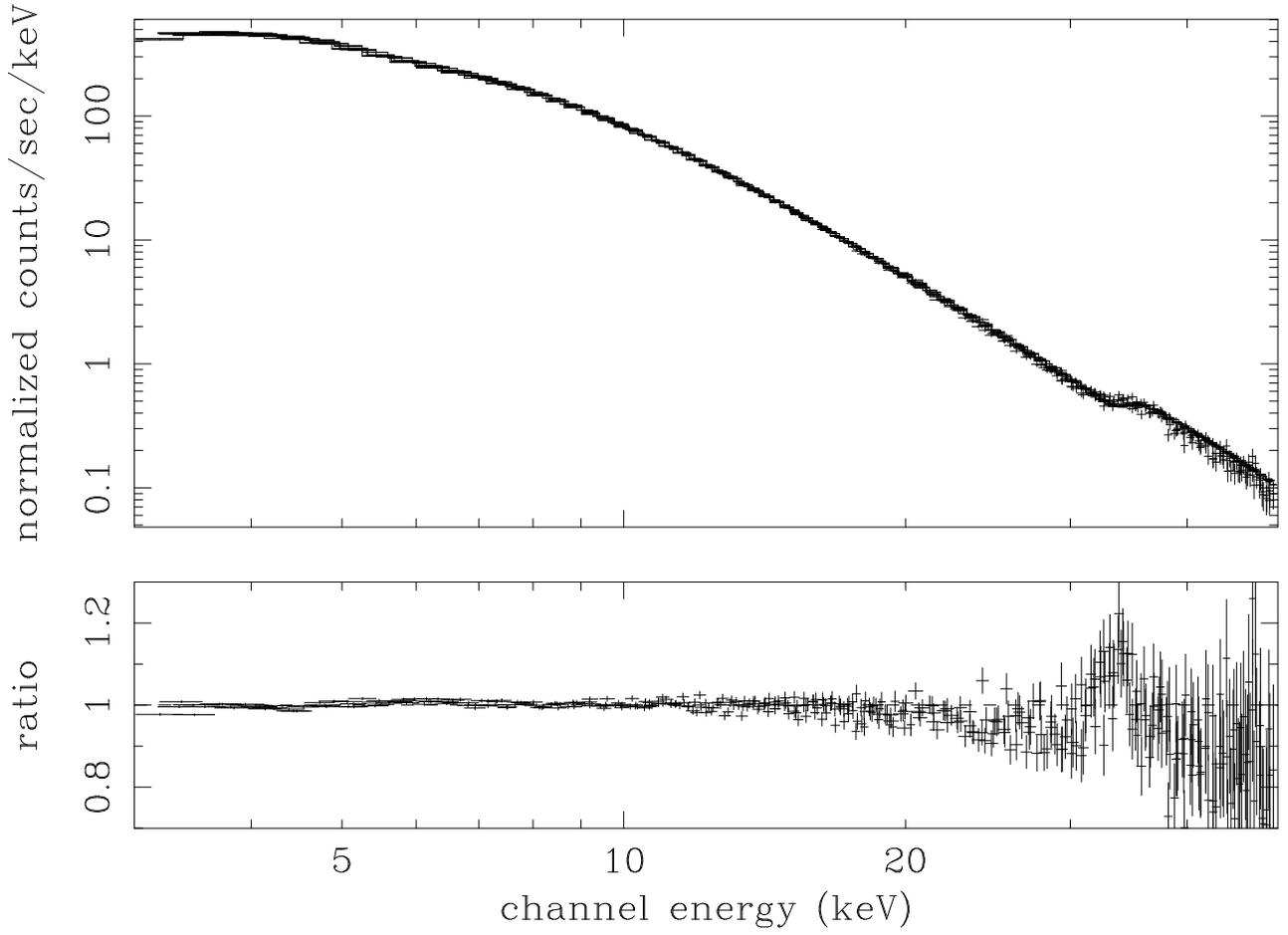}
\figcaption{Power-law fits to a Crab monitoring observation for
the first layers of the 5 PCUs.
The lower panel shows the ratio of the data to the model.
\label{res_5pcu}}
\end{figure}

\begin{figure}[h]
\includegraphics[width=0.75\columnwidth,angle=270.]{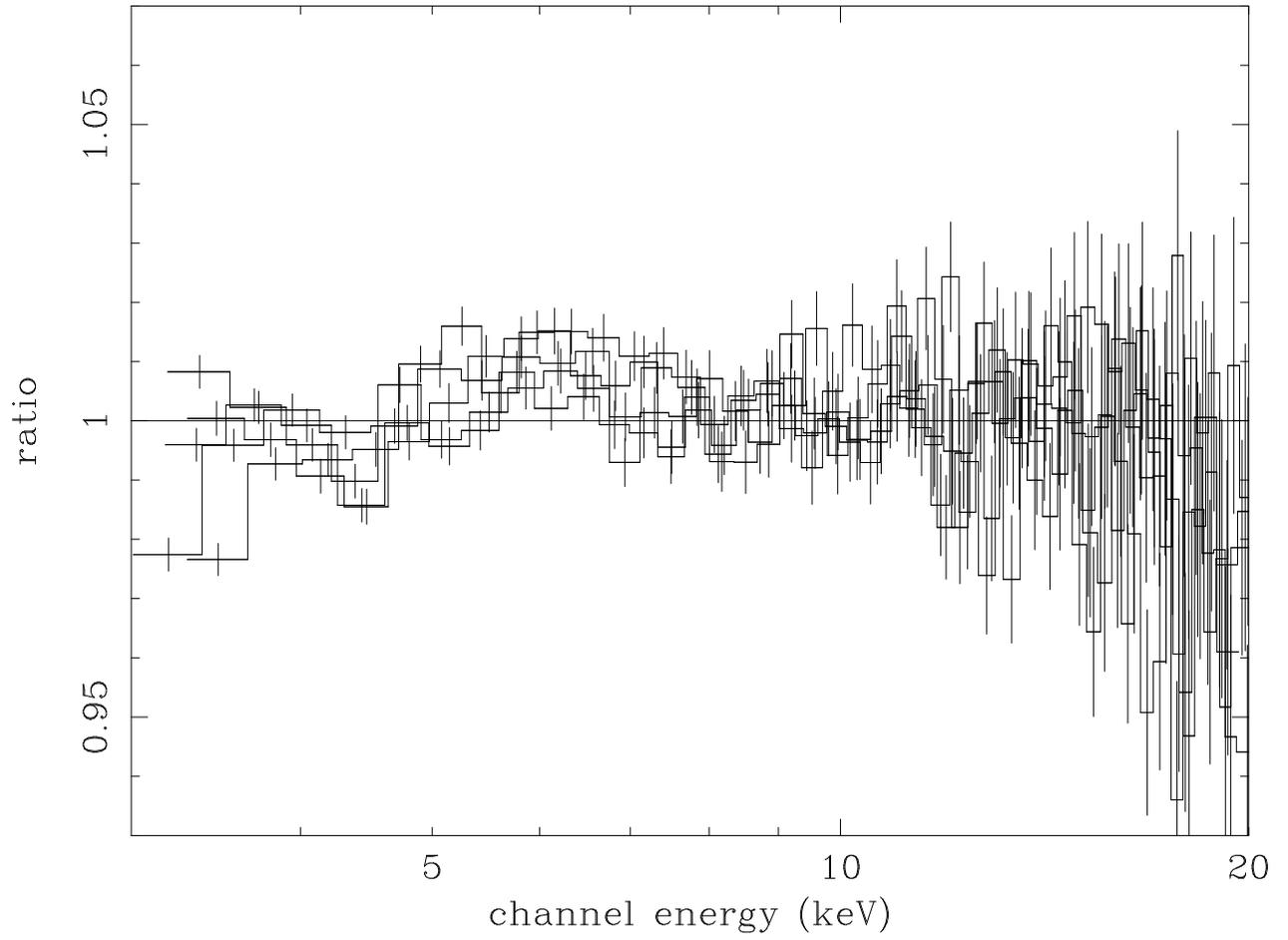}
\figcaption{Expanded view of the ratio in the lower panel of figure \ref{res_5pcu}.
\label{res_5pcub}}
\end{figure}

\begin{figure}[h]
\includegraphics[width=0.75\columnwidth,angle=270.]{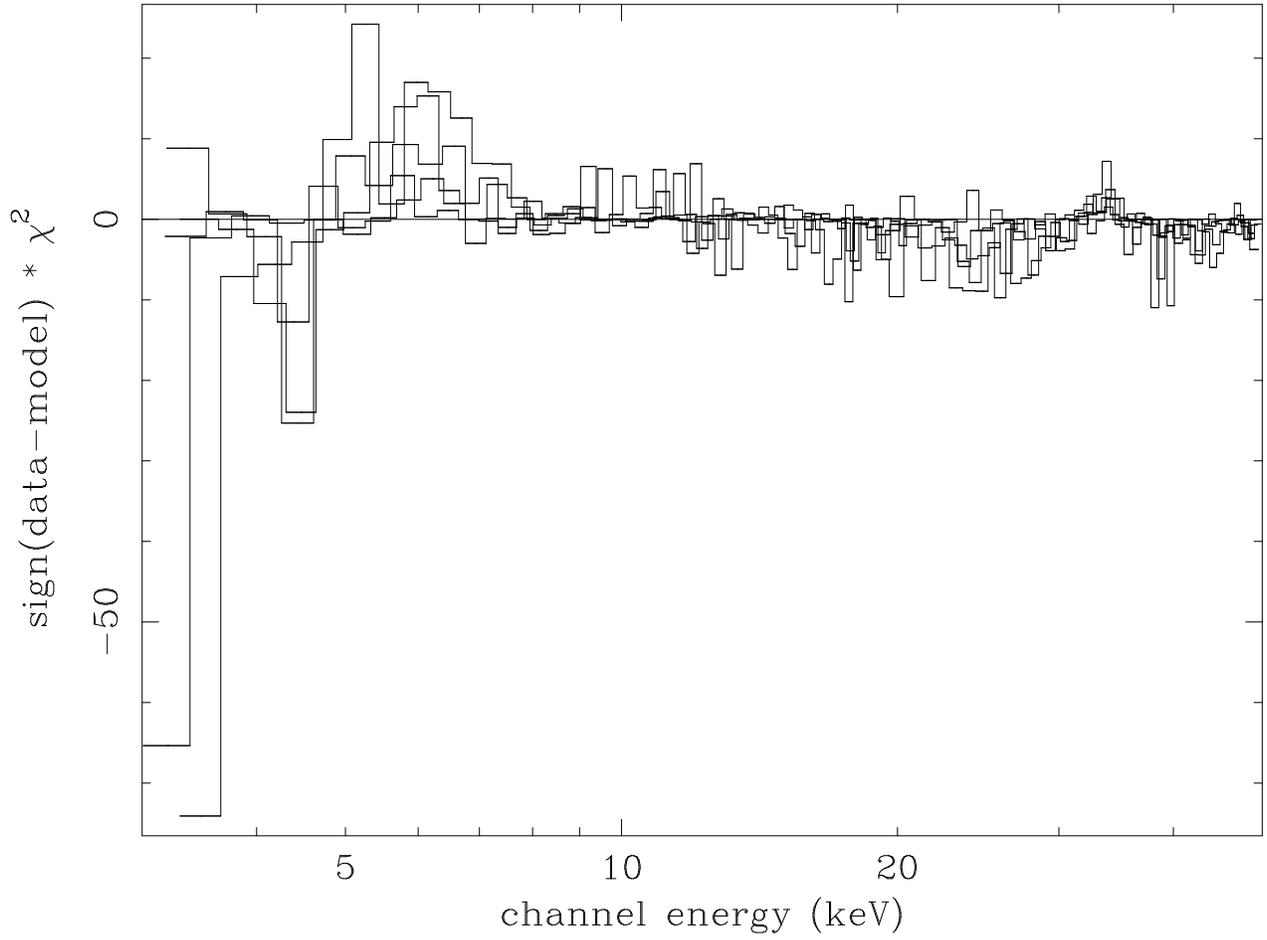}
\figcaption{Contributions to $\chi^2$ for the fits in figure \ref{res_5pcu}.
The reduced $\chi^2$ is dominated by contributions from the lowest channels and the region
near the Xenon L-edge.
\label{res_5chi}}
\end{figure}

\begin{figure}[h]
\includegraphics[width=0.75\columnwidth,angle=270.]{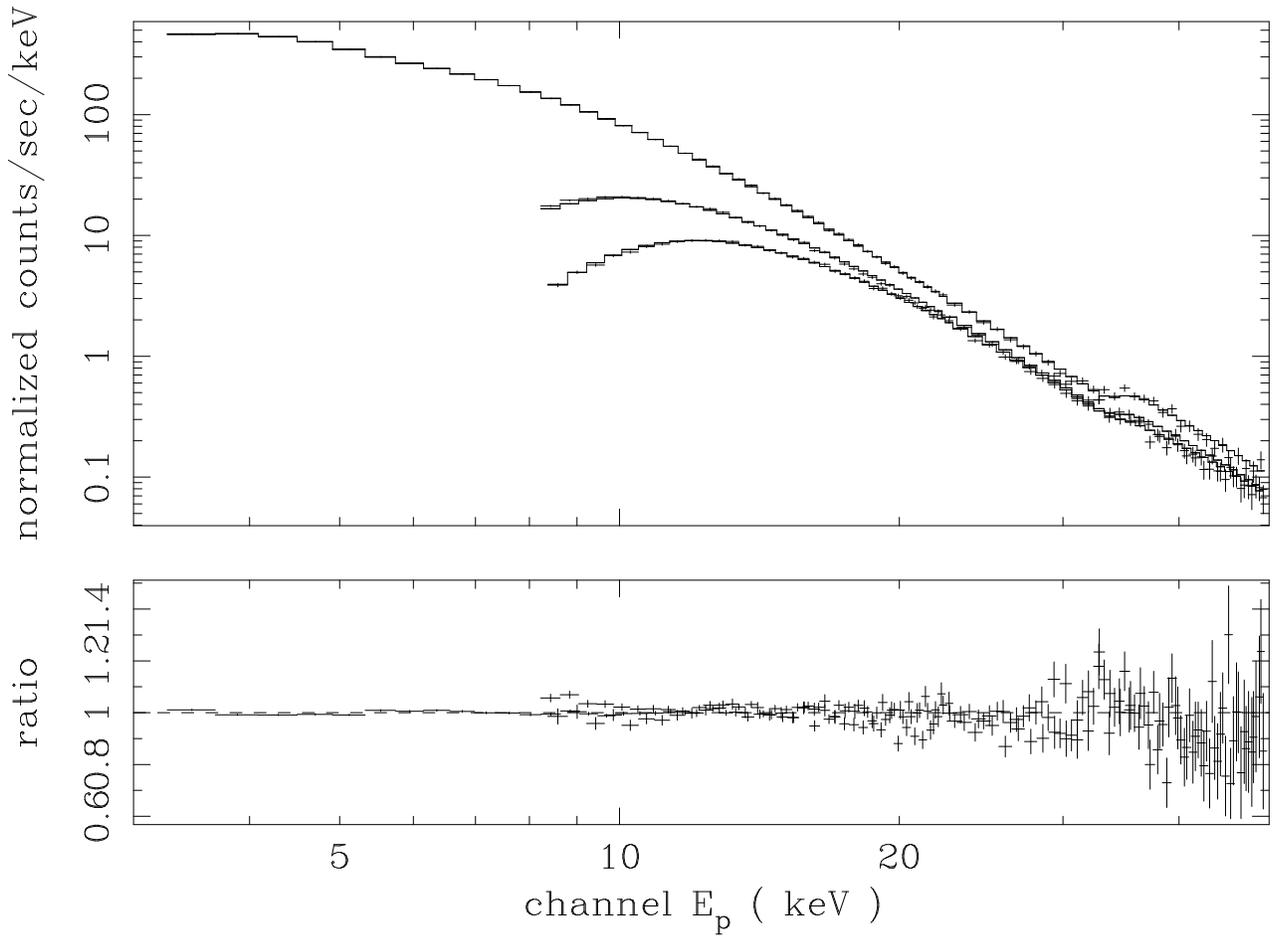}
\figcaption{Fits to three layers from
PCU 2.
\label{res_3layer}}
\end{figure}

The data and model, along with the ratio of data to model, fit to a representative Crab monitoring observation
are demonstrated in figures \ref{res_5pcu} and \ref{res_3layer}.
The data are from 1999-02-24 and the best fit power law indices are
2.088, 2.098, 2.105, 2.088, and 2.096.
The deviations from the power law are quite similar from one detector to the next.  Figure
\ref{res_5pcub} shows an enlargement of the 3-20 keV region;  below 10 keV the deviations are
less than $1\%$.  Between 10 and 20 keV the data begins to exceed the model slightly, though
only by $\sim 2\%$.  There is a clear underprediction just below the Xenon K edge;  although this
is quite obvious in the ratio presentation, the number of counts per channel (the convolution
of the intrinsic spectrum with the detector quantum efficiency, both of which decline rapidly
with energy) is nearly three orders of magnitude below the peak.  Figure \ref{res_5chi} shows
the contributions to $\chi^2$, which are dominated by deviations at the lowest energies and
near the Xenon L edge, where the counting rates are high.

Figure \ref{res_3layer} shows fits to data from all three layers.
The data are from 2003-02-26, PCU 2; the power law indices are constrained to be the same (best fit 2.123)
The normalizations are allowed to float; the variation (max to min) is less than 3\%.  
Requiring the normalizations to be the same does not change the best fit index.

\placefigure{res_5pcu}
\placefigure{res_5pcub}
\placefigure{res_5chi}
\placefigure{res_3layer}

\section{Background model Overview}
\label{bkgd}

The RXTE PCA is a non-imaging instrument; for both spectroscopy and light 
curve analysis the background must be subtracted based on an a 
priori model.  ``Background" is defined broadly to include anything that 
contributes non-source counts to the PCA instrument in orbit, including but 
not limited to:
\begin{itemize}
\item local particle environment;
\item induced radioactivity of the spacecraft; and
\item the cosmic X-ray background.
\end{itemize}
In general, these components vary as a function of time, and must be 
parameterized.  The parameterized model is adjusted to fit a set of dedicated 
observations by the PCA of blank sky regions. Once a good fit is achieved, 
the same parameterization can be applied to other observations.
The cosmic X-ray background, which in the PCA band is due to the integrated effect of unresolved
sources in the field of view, has spatial fluctuations of approximately $7\% (1 \sigma)$
\citep{rev03} in the 2-10 keV band.  These fluctuations are not (and cannot)
be predicted by the background model;  the magnitude of these fluctuations sets
the limit below which fluxes cannot be determined by the PCA. This confusion
limit is at $4 \times 10^{-12} \ergscm$.

\begin{figure}[t]
\includegraphics[width=0.8\columnwidth,angle=270.]{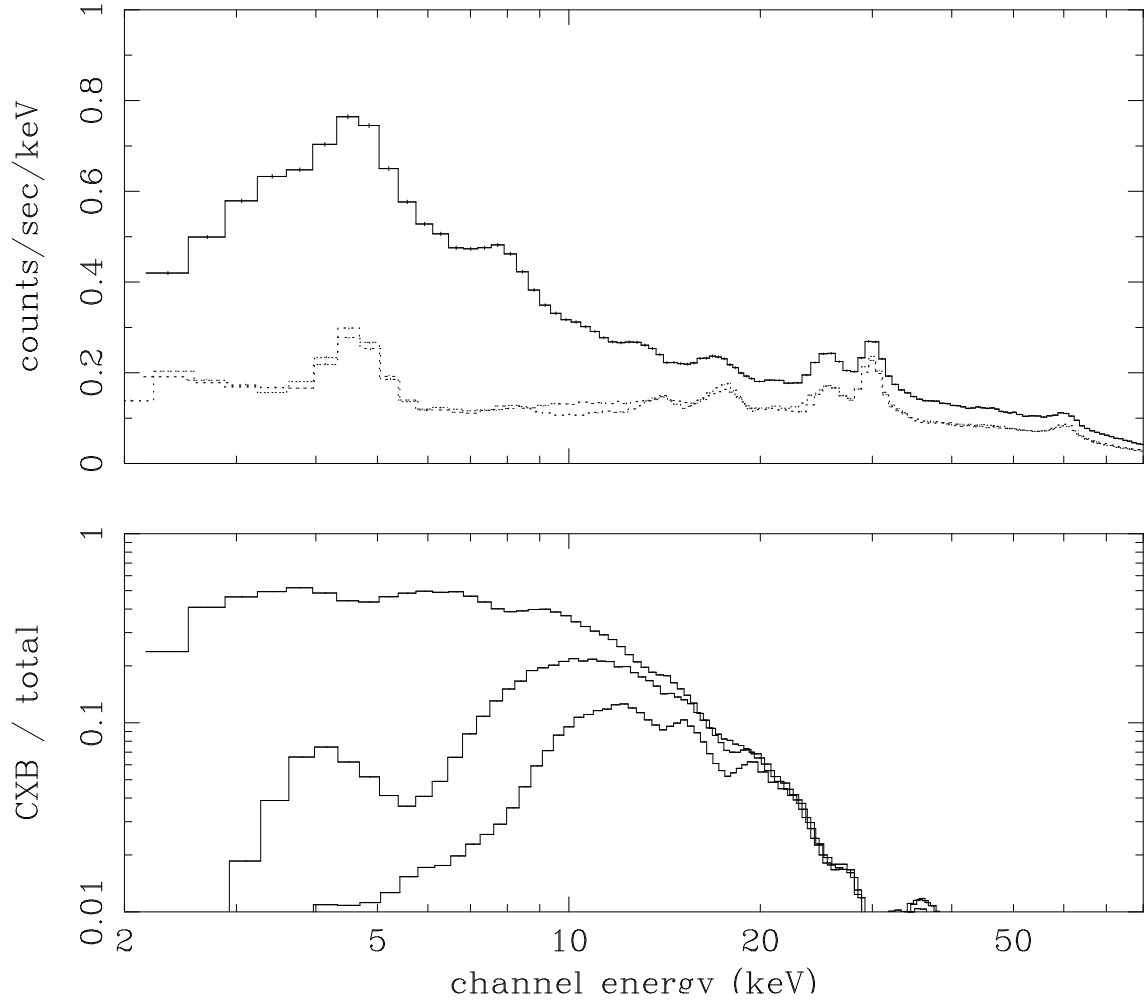}
\figcaption{Pulse height spectra from PCU 2 obtained during observations of
``blank" sky during May 1998.  These spectra are the sum of Cosmic
X-ray Background and unrejected instrument background.  The spectra are
collected separately for each layer;  the solid line represents the first
layer.  The lower panel shows the fraction of the total rate due to the
Cosmic X-ray background determined by differencing PCA observations of 
``blank" sky and dark earth \citep{rev03}.
\label{bkg_pha2}}
\end{figure}

\begin{figure}[h]
\includegraphics[width=0.8\columnwidth,angle=270.]{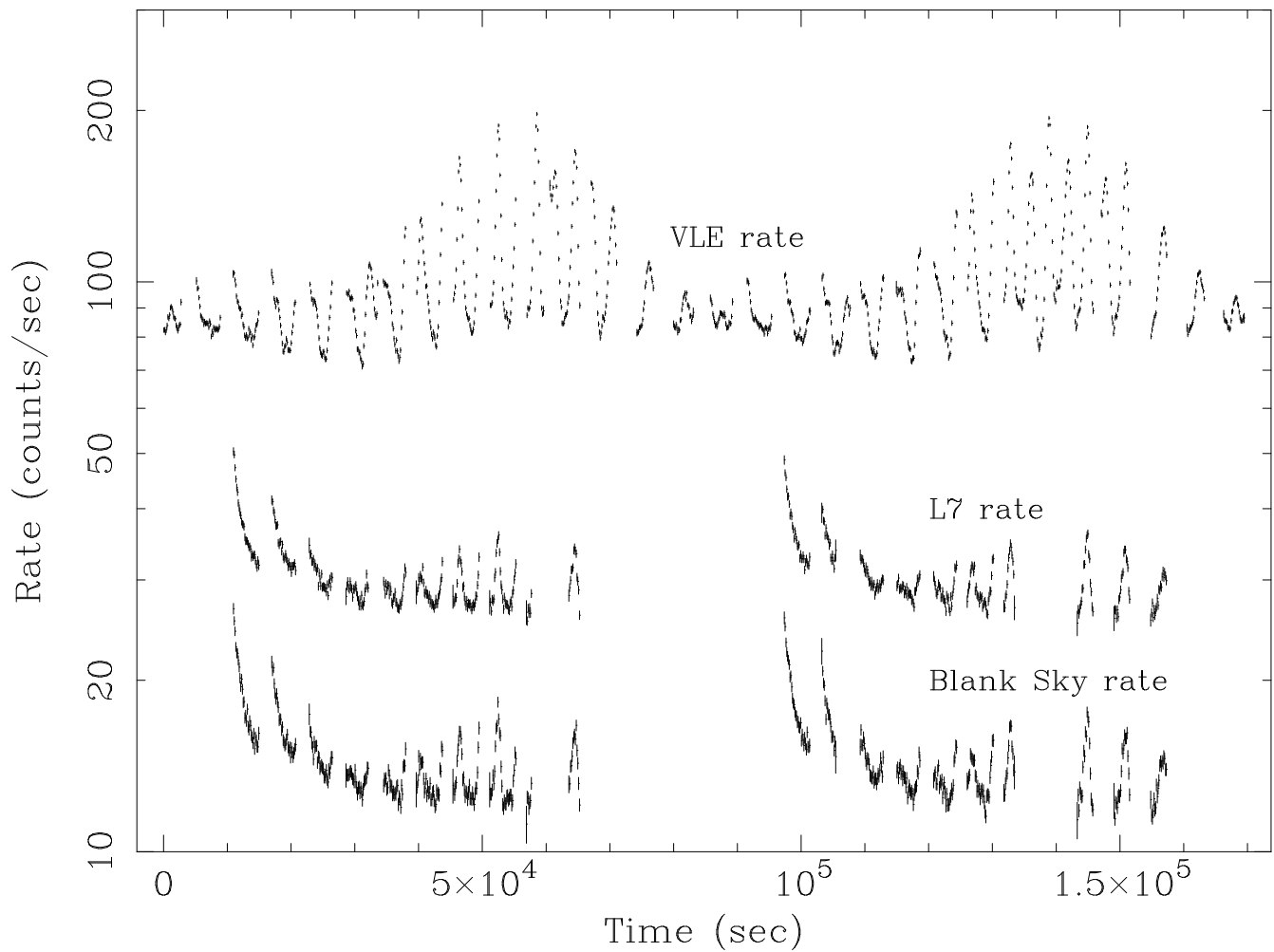}
\figcaption{The total background counting rate for one detector during a background
monitoring campaign.  The data are shown for all layers of PCU2.  Large orbital 
variations are easily visible, as well as the effect of passages through the 
South Atlantic Anomaly.  The large gaps are interruptions caused by observations
of other sources.  Also shown are the L7 and VLE rates, which are highly correlated
with the blank sky rate and which are used to parameterize estimates of the
instantaneous background.
\label{bkg_lc}}
\end{figure}

Figure \ref{bkg_pha2} shows the spectra of the ``good" count rate during
observations of blank sky;  this is the total (i.e. instrument plus
sky) background.
Spectra are shown separately for
the first, second, and third layers.  
The peaks near channels 26 and 30 keV are due to unflagged
events from the Am$^{241}$ calibration source.  
The fractional contribution of
sky background to each layer is shown in the lower panel.
The sky background is approximately 1 mCrab per beam.
The lower light curve in figure \ref{bkg_lc} shows
the total background rate over a two day period of nearly uninterrupted
observations of blank sky.  Variations by more than a factor of 2 in a 
day, and by up to a factor of 1.5 in an orbit are clearly visible.

\placefigure{bkg_pha2}
\placefigure{bkg_lc}

The background models for the PCA have components quite similar to those
for the {\it Ginga} Large Area Counters \citep{hayashida89}.  
The details of the background spectra and the amplitude of the variability
are different, as expected, due to differences in the orbit (RXTE has a lower
inclination and smaller eccentricity than {\it Ginga}) and details of
the detector construction.  We use data collected over a period
of years from repeated observations of several blank sky positions
to measure and model the background and its variation.

The most successful background model to date for faint sources is the 
``L7/240" model.  Here, ``L7" is the name of a housekeeping rate which 
is well correlated with most of the variation in the PCA background rate.
The L7 rate is the sum of all pairwise and adjacent coincidence rates in each PCU. 
The ``240" component refers to a radiactive decay timescale of approximately 240 minutes.  
This timescale may describe the combined effect of several radioactive elements.  
The L7/240 model is not appropriate for bright sources because the L7 rate 
can be modified by the source itself.

The ``Very Large Event" housekeeping rate is also correlated with the observed
background rate, although with more unmodelled residuals than the L7 rate.  
Both the L7 and VLE rates are shown in figure \ref{bkg_lc}.  Gaps in the blank sky 
and L7 rates exist due to observations of other sources while the VLE rate
is shown throughout, and is virtually unaffected by even bright X-ray sources.
Since the VLE rate is largely unaffected 
by the source rate, it can be used to parameterize a model
suitable for ``bright" sources.  Operationally, bright
is defined to be 40 source counts sec$^{-1}$ PCU$^{-1}$ or about 15 mCrab.

The background model is determined by fitting high latitude blank sky 
observations which are regularly obtained in 6 directions.  The output 
of the background model is an estimate of the combined spectrum of 
instrument background and the Cosmic X-ray Background observed at
high latitudes.  The model makes no attempt to predict the diffuse 
emission associated with the galaxy \citep{valinia98}.

When background observations were begun in November 1996, one day was 
devoted to this project every three or four weeks.  Eventually this 
was supplemented by short, twice daily observations to better sample 
the variations that are correlated with the apogee precession of $\sim 30$ days.
Because the orbit apogee and perigee differ by about 20 km and because the particle
flux within the South Atlantic Anomaly varies with altitude, the average daily 
particle fluence, and the radioactive decay component of the background, show 
a long term variation with this period as well.

The blank sky observations were originally divided among 5 distinct pointing 
directions.  Daily observations of a  sixth point were added in order to test the
success of the background modelling for the lengthy NGC 3516 monitoring campaign 
\citep{edelson99}.  The pointings are summarized in table \ref{bkg_pts}.

The current background models
\footnote{http://heasarc.gsfc.nasa.gov/docs/xte/pca\_news.html\#quick\_table
provides links to the current models, and details of the special requirements 
and limitations of the model for PCU 0 after the loss of the propane layer.
\label{pca_news}}
have explicitly linear dependences on L7 (or VLE), the radioactive decay term, 
and mission elapsed time.

In the construction of these models, we
\begin{itemize}
\item accounted for the variance in L7 (or VLE) itself;
\item derived coefficients for each Standard2 pulse height channel independently;
\item used a modified chi-square approach for low-statistics Poissonian data 
\citep{mighell99}
\item included data from immediately after the SAA (i.e. included more data than in earlier efforts)
\item selected data with a horizon angle of at least 10 deg and with the rates VPX1L and VPX1R (contained
within the Standard2 data) less than 100 sec$^{-1}$
\end{itemize}

The fitted background model for each channel,i, is:
\begin{equation} 
  BKG_i = A_i + B_i * L7 + C_i * DOSE + D_i * (t - t0)
\label{bkgd_mdl}
\end{equation}
  where $A_i$, $B_i$, $C_i$ and $D_i$ are the fit coefficients, L7 is the L7 rate in a PCU, 
DOSE is the SAA particle dosage summed over individual passages through the SAA
and decayed by a 240 minute folding timescale, and t is the epoch time.  
The SAA particle dosage is measured by the particle monitor on the 
High Energy X-ray Timing Experiment \citep{roth98} also aboard the RXTE;
each passage is defined as the period when the HEXTE high voltage is reduced.
Equation \ref{bkgd_mdl} is linear in all its terms, and allows for a secular 
drift over time.  The predicted model is also stretched (or compressed)
to account for the time dependence in the energy to channel relationship
(equations \ref{e2c} and \ref{e2c2}).  The correction contains only a linear
term derived from the peak shifts in the $Am^{241}$ spectra (and is thus
slightly inconsistent with the treatment in the response matrix generator).

\begin{figure}[t]
\includegraphics[width=0.8\columnwidth,angle=0.]{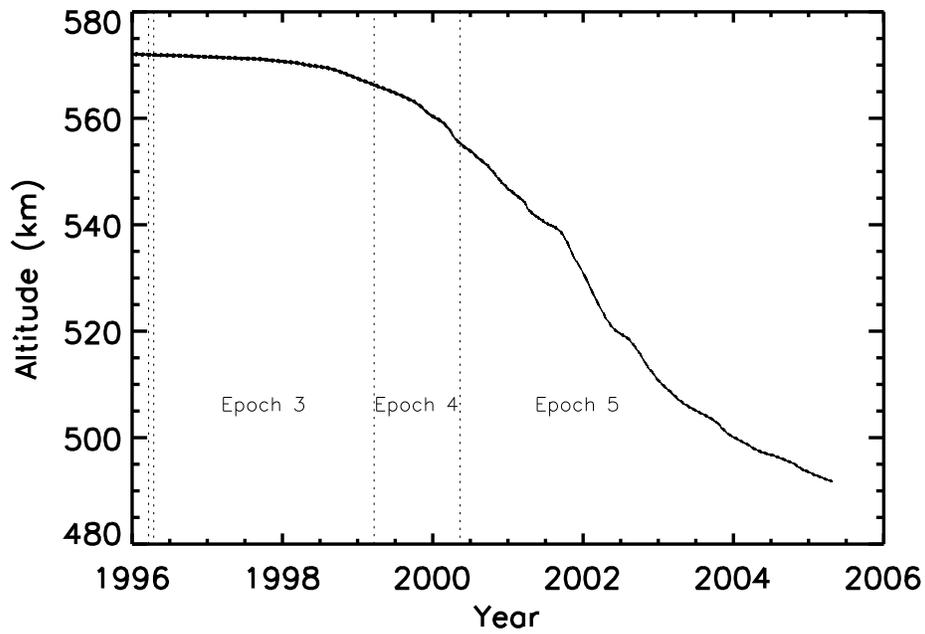}
\figcaption{The altitude of the RXTE orbit as a function of time.  The orbit
began to decay noticeably midway through epoch 3.  The small time dependent
background term, uncorrelated with the L7 rate, appears correlated and is
probably physically associated with the orbit altitude.
\label{xte_alt}}
\end{figure}

The secular drift in time has been $\sim 0.07$ ct s$^{-1}$ 
PCU$^{-1}$ yr$^{-1}$ from epoch 3B through the present, and is well correlated
with satellite altitude.  This term became significant when the RXTE orbit 
began to decay.  We parameterize the time dependent term as a linear trend 
within each epoch;  there is a discontinuity in the slope at epoch boundaries, 
with the significant discontinuity at the beginning of epoch 3B corellated 
with the time when the satellite orbit began to decay noticeably 
(fig \ref{xte_alt}).
The orbit decay is understood to be due to increased drag as the earth's 
atmosphere expanded in response to increased solar activity associated with 
the solar cycle.  The satellite altitude has changed from 580 km at the 
beginning of epoch 3B to $\sim 500$ km in fall 2003.
The secular drift term is not currently included in the VLE models.
\footnote{
An analysis of long observations (over 80 ksec in both March 2003
and Jan 2004) of Centaurus A \citep{roth05} which used the VLE model
found it necessary to reduce the predicted background spectrum by $\sim 4\%$
(i.e. maintaining the overall spectral shape).  These observations
were made at times {\it after} all of the blank sky observations that
went into the background model, which maximizes the error introduced
by ignoring the time dependence.  The apparent gain shift reported 
in \citep{roth05} is a consequence of both the extrapolation and the fact
that the gain shift used in the VLE model was determined over a period
even shorter than the span of blank sky observations.  These observations,
combined with the decrease in RXTE altitude, require a new
background model expected to be released in early 2006, which will be documented and
linked from the URLs in footnotes \ref{pca_news} and \ref{cal_page}.}
An effort is underway (in late 2005) to divide epoch 5 into three distinct
sub-epochs for the purpose of background estimation.  This will allow
a more precise estimate of the secular term which is apparently correlated
with the change in altitude, and will result in background models constructed
from more nearly contemporaneous data, minimizing effects due to the slowly
drifting energy to channel relationship.  These models will be available from
the RXTE web site.

VLE models require a second activation time
scale;  the second DOSE term is decayed by a 24 minute time scale.  Attempts to
fit the time scales are poorly constrained, probably due to the fact that both 
timescales are a mixture of several radioactive half-lives;  additionally, 
for the shorter timescale the DOSE term, which sums the fluence 
of particles on orbital timescales, may not contain enough time resolution.  
The quality of the 
resulting background spectra has been high enough that a more careful 
parameterization has not been needed, and we have not succeeded in 
identifying particular radioactive decays that are responsible.  
All of the identified variations were also observed by the Large Area 
Counters on {\it Ginga}, although with different amplitudes related 
to the details of the orbit and detector construction
\citep{turner89,hayashida89}

\placefigure{xte_alt}

\begin{figure}[h]
\includegraphics[width=\columnwidth,angle=0.]{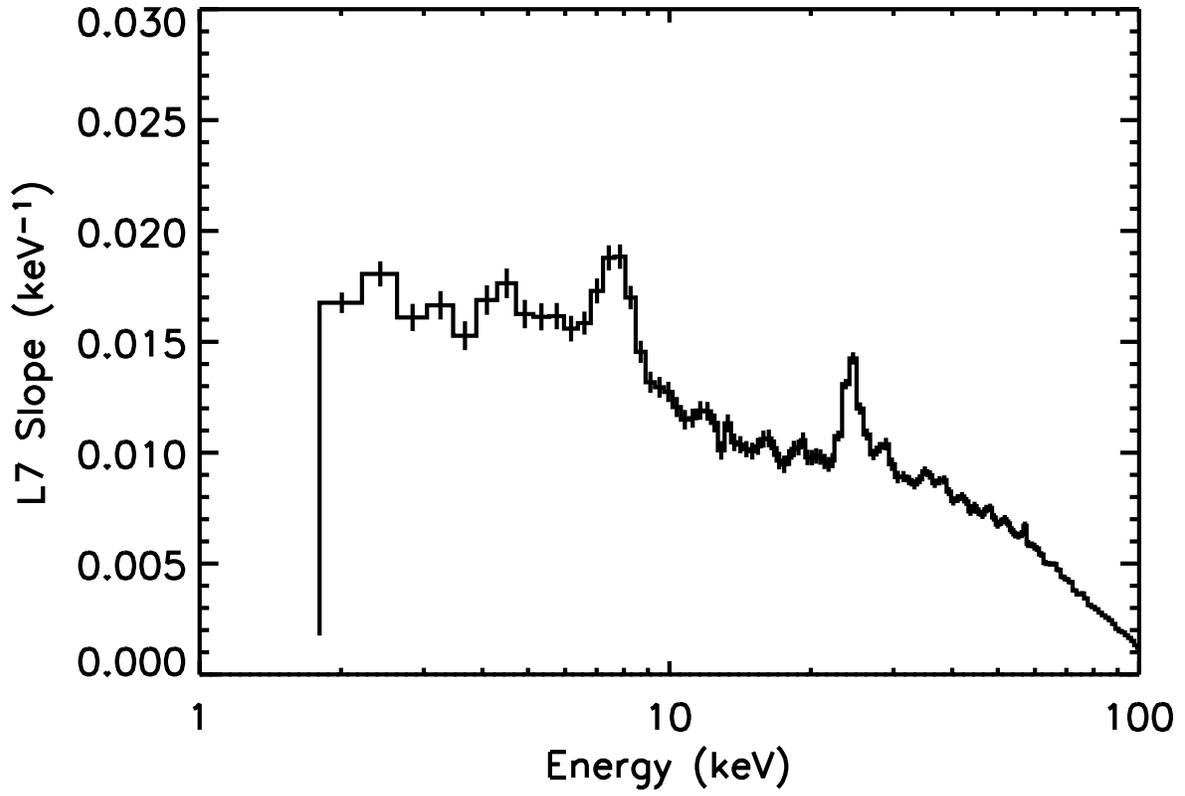}
\figcaption{Coincidence rate coefficients ($B_i$) of the background spectrum 
for PCU 2, layer 1, epoch 5.  These coefficients are derived for the L7 model.
\label{bkgd_B}}
\end{figure}

\begin{figure}[h]
\includegraphics[width=\columnwidth,angle=0.]{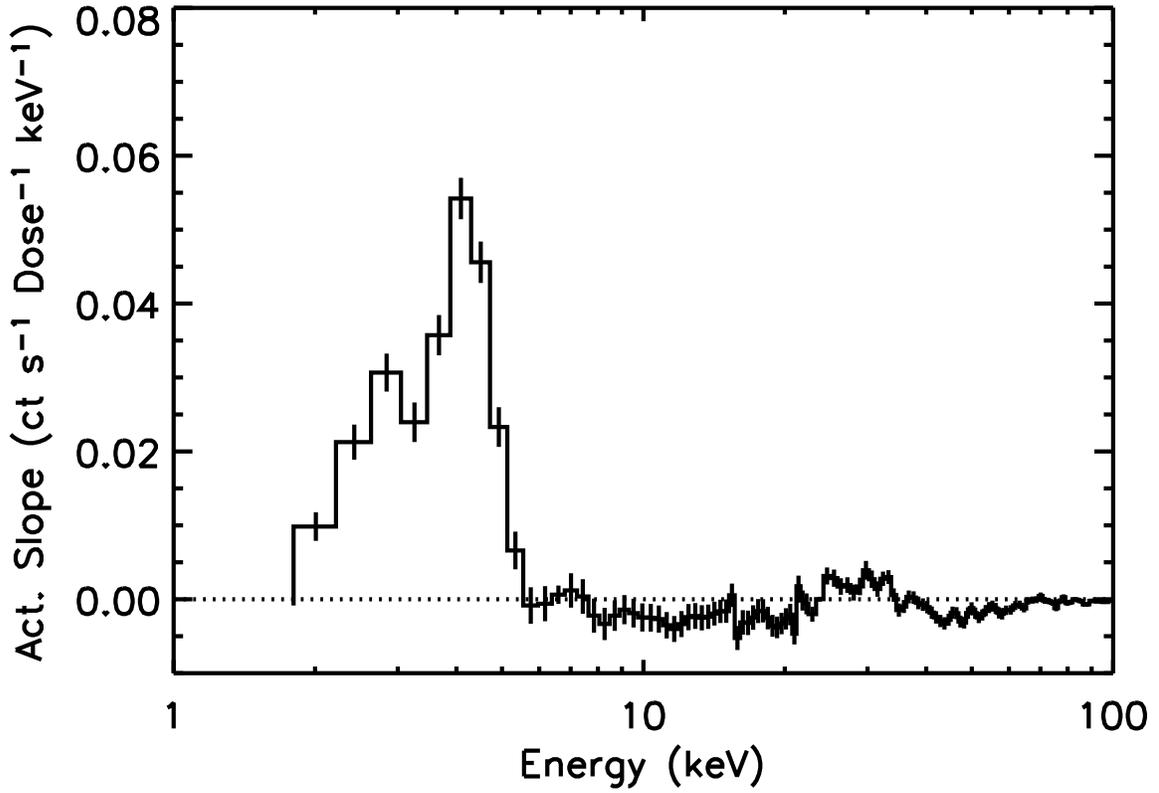}
\figcaption{Accumulated radioactive dosage coefficients ($C_i$) of the background 
spectrum  for PCU 2, layer 1, epoch 5.  These coefficients are derived for the 
L7 model.
\label{bkgd_C}}
\end{figure}

Figures \ref{bkgd_B} and \ref{bkgd_C} show the coefficients $B_i$ and 
$C_i$ for PCU 2, layer 1, during epoch 5 and show that the background 
varies in both amplitude and spectrum.

\placefigure{bkgd_B}
\placefigure{bkgd_C}

The fitting was applied to multiple, dedicated, 
PCA background pointings.  Each of the points on the sky has a slightly different 
sky background.  Our approach is to assign a different set of coefficients $A_i$
for each background pointing.  The production background model is determined by 
taking the weighted average of $A_i$'s for different pointings.  Thus, the 
background model represents an ``average" patch of high latitude sky.

\begin{figure}[t]
\includegraphics[width=0.8\columnwidth,angle=90.]{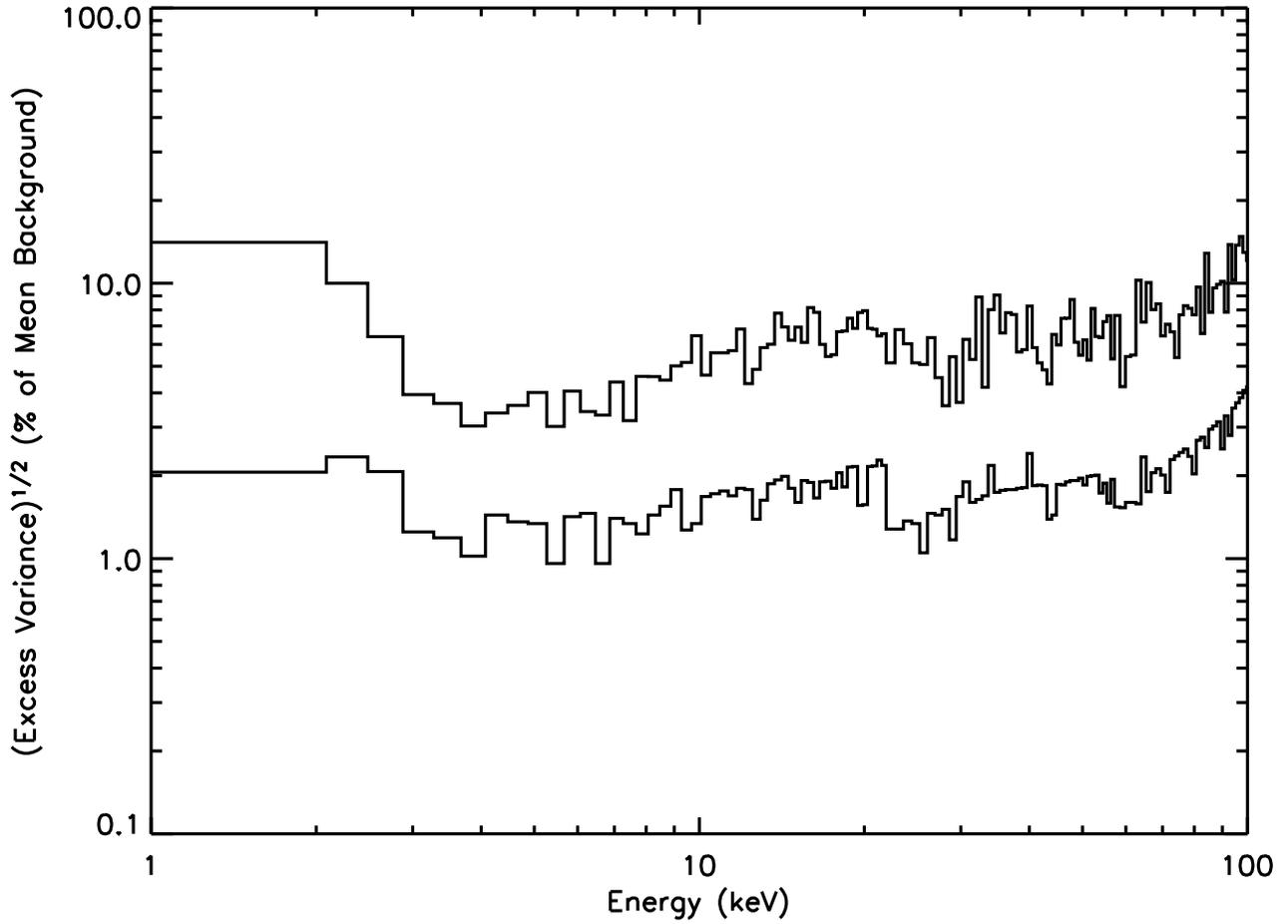}
\figcaption{Unmodelled background variations.  The upper line is plotted for
16 sec intervals, and is dominated by Poisson noise;  the lower line is 
plotted for 1600 sec intervals, and is representative of the systematic 
uncertainties.
\label{bkgd_sys}}
\end{figure}

Remaining count-rate variations after background subtraction are a measure of
systematic error.   We calculated the variance of blank sky rate minus
predicted background rate and compared that variance to that expected
from counting statistics alone.   The difference between these two
variances includes unmodelled effects.  
Figure \ref{bkgd_sys} gives the systematic error per 
channel while Table \ref{bkgd_sysT} gives some band averaged values 
for individual layers of each detector.  
In each case we report the square root of the excess variance as
a fraction of the total rate;  the table also gives an absolute
value in counts per sec.  The upper line in figure \ref{bkgd_sys}
includes a contribution due to the limited precision with which
the L7 rate is measured in any single 16 sec bin.  For observations
of faint sources with the PCA, the systematic errors in the count
rates reported in table \ref{bkgd_sysT} become larger than the Poisson
uncertainties in the count rates for observations longer than a
few hundred seconds.

\placefigure{bkgd_sys}

\section{Timing Calibration}
\label{timing}

Every event detected within the PCA, whether a cosmic X-ray or an instrument background
event, produces 19 bits which are passed to the EDS which adds the time stamp and
performs event selection and rebinning.  

Each signal chain has its own analog electronics chain consisting
of Charge Sensitive Amplifier, Shaping Amplifier, Discriminator, and Peak 
Finder.  The deadtime associated with the analog chains is paralyzable.  The
six Xenon signal chains and the Propane signal chain share a single
Analog to Digital Converter which produces a 256 bit pulse height;  the Xenon
veto chain is separately analyzed and produces a 2 bit pulse height.
The Analog to Digital Conversion is non-paralyzable.  ``Good" events produce
charge on a single chain;  the resulting pulse height can be unambiguously
identified with that chain.  For events which produce more than one analog
signal, the pulse height cannot be unambiguously assigned to a particular
chain;  such events are generally not included in the telemetry except to
be counted in the rates present in the Standard data modes.

\subsection{Deadtime}
\label{deadtimeps}

A detailed description of the effects of deadtime on a particular observation
depends on both the source brightness and spectrum.  The spectrum affects the
ratio of events observed in the first layer and other layers, and therefore changes
the details of the interaction of a paralyzable deadtime process (associated with
each analog chain) and a non paralyzable process (the analog to digital conversion).
Fortunately, a complete description is not required for the most needed
corrections.  We present useful
approximations for a common statistical representation of the data (power spectra)
and for the construction of light curves.

\subsubsection{Power Spectra}

Even at relatively low count rates, the probability that events are missed
has a significant effect on the shape of the power spectrum \citep{zhang95}.

Periods of deadtime can are generated by X-rays or charged particles
which interact within the detector.  Both the desired astrophysical signal
and the instrument background contribute to deadtime.

An X-ray event causes the detector to be dead in
two ways. First, it disables {\em its own} analog electronic chain, i.e., the
Charge Sensitive Amplifier (CSA), Shaping Amplifier (SA), 
and the associated discriminators for a period of time
which depends on the amount of energy it has deposited in the
detector;  this is a paralyzable dead time. Each detector has 
7 analog electronics chains (6 Xenon half-layers and
the propane layer) which share an Analog to Digital
Converter (ADC).   Second, this event causes the entire analog chain
to be busy for $10.0 \musec$;
this is a non-paralyzable dead time which consists of $6.5 \musec$ of analog to
digital conversion time plus fixed delays to allow the analog signal to settle
(prior to the conversion) and for the data to be latched for transfer to the
EDS and reset (after the conversion).  For events with energy less than 
$\approx 20$ keV, the analog chain is again live before the end of the ADC
conversion.  The minimum time between sequential events on the same or
different chains is $10 \musec$, and to a good approximation, all events
can be considered to have this deadtime.

Some events trigger two or more of the analog electronic chains.  Most such
events are
due to a charged particle which leaves a track through the detector.  However,
a single X-ray can trigger two signal chains, either by creating a fluorescence
photon which is reabsorbed elsewhere in the detector, or an energetic photo-electron
which travels into an adjacent cell.  At high fluxes, it is also possible that
two X-rays from the source will be detected simultaneously by different signal
chains \citep{vdk96}.  In all of these cases, both signal chains are paralyzably dead and 
one of the signals is presented to the ADC.  No information is available to the
EDS about which signal was digitized.  The ADC deadtime is as above, and $10 \musec$
remains a good approximation to the deadtime; for large equivalent
energies, one (or more) analog chains may remain dead after the detector as
a whole can process another event, just as in the case of an event which 
triggers a single chain.

An event which triggers the Alpha chain and one additional anode is marked
as a calibration event, and produces the same deadtime as any other event of similar
energy.  An event which triggers only the Alpha chain diables event processing
for $6\mu s$ (any event which occurs during this interval will be interpretted
as a calibration event.)  Events which occur on the $V_x$ chain also disable
the detector for $6\mu s$ as any event which occurs during this period will
be marked in coincidence with the $V_x$ event.  For events which include the
$V_x$ chain and a single additional chain, the digitzed pulse height corresponds
to the non $V_x$ chain.  Strictly speaking, events which trigger only $V_x$ OR Alpha
should not be included in the total deadtime calculation since it is known that
no other event was presented to the detector electronics in this interval.  This
effect is described at greater length in section \ref{faint}.

Very Large Events (VLE) are operationally defined as  events
which exceed the dynamic range of the experiment and which saturate the amplifier.  
The equivalent energy depends on the high voltage setting, and was
$\sim 75 $ keV in Epoch 1 and $\sim 120 $ keV in Epoch 5.
The default VLE window is $170 \musec$ \citep{zhang96}.  See also section
\ref{routine}.

The Poisson random noise level is suppressed by the correlation between 
events caused by the deadtime. To a good approximation, the Poisson
noise level with deadtime correction can be computed as
follows:

\begin{equation}
P_d (f) = P_1 - P_2 cos( {{\pi f}\over f_{Nyq}})
\label{P_dead}
\end{equation}

$P_1$ and $P_2$ depend in principle on the details of the electronics.
For count rates less than $10^4$ ct s$^{-1}$ PCU$^{-1}$ (or $\sim 4$ Crab)
the dependence is small and both coefficients can be estimated as if
the deadtime is purely paralyzable with

\begin{equation}
P_1 = 2[1-2r_0t_d(1-{t_d\over{2t_b}})] {\rm and}
\end{equation}

\begin{equation}
P_2 =  2 r_0 t_d {{N-1}\over N}({t_d\over t_b})
\end{equation}


where $r_0$ is the output event rate of all events (i.e. without distinguishing
between ``good" events and coincident events), $t_b$ the bin size, $t_d$ the
deadtime taken to be $10 \musec$, $N$ is the number of frequencies in the 
power spectrum, and $f_{Nyq}$ is the Nyquist frequency \citep{zhang95}.
The dead time in $P_1$ and $P_2$ is calculated as $10 \musec$ times the
total number of (non VLE) events transferred to the EDS per detector plus
$170 \musec$ times the number of VLE events per detector.  These rates
can be estimated from the Standard 1 rates assuming that the summed rates
come equally from all detectors which are on. 

We need to include the contribution 
to the power spectrum by the VLE events. 
Since VLE events cause ``anti-shots'' in the data, its contribution 
can be written as 
\begin{equation} 
P_{vle}(f) = 2r_{vle} r_0 \tau^2 ( { {\sin{\pi\tau f}}\over{\pi\tau f} })^2, 
\label{P_VLE} 
\end{equation} 
where $r_{vle}$ is the VLE rate, $r_0$ the good event rate, and $\tau$
the VLE window size \citep{zhang96}.
In practice one needs to add up equations~\ref{P_dead} and
\ref{P_VLE}.  

A more rigorous approach which deals simultaneously with the deadtime produced by VLE events 
and all other events (rather than treating the VLE events as anti-shots and adding a
contribution) has been developed by \citet{wei06}\footnote{We anticipate providing a
link to this thesis from the PCA web page (footnote \ref{cal_page}) once it is complete}.  Fitting for values of the deadtime
and VLE window yields 8.84 and 138 $\musec$.  These values underestimate the deadtime
correction required to produce a uniform rate from the Crab nebula over long observations
(section \ref{faint}).  The discrepancy suggests that the detector deadtime is more
complicated than the models so far employed to calibrate the parameters.

\subsubsection{Light Curves}

\paragraph{Faint Sources}
\label{faint}

For the purposes of this discussion, a faint source is one where the deadtime
correction is less than $10\%$.  This includes the Crab.  Dead time is produced by all
events within the detector, and can be estimated from either Standard1 or Standard2
data as both modes count each event presented to the EDS
once.  We illustrate this with observations of the Crab.

\begin{figure}
\includegraphics[width=0.8\columnwidth,angle=270.]{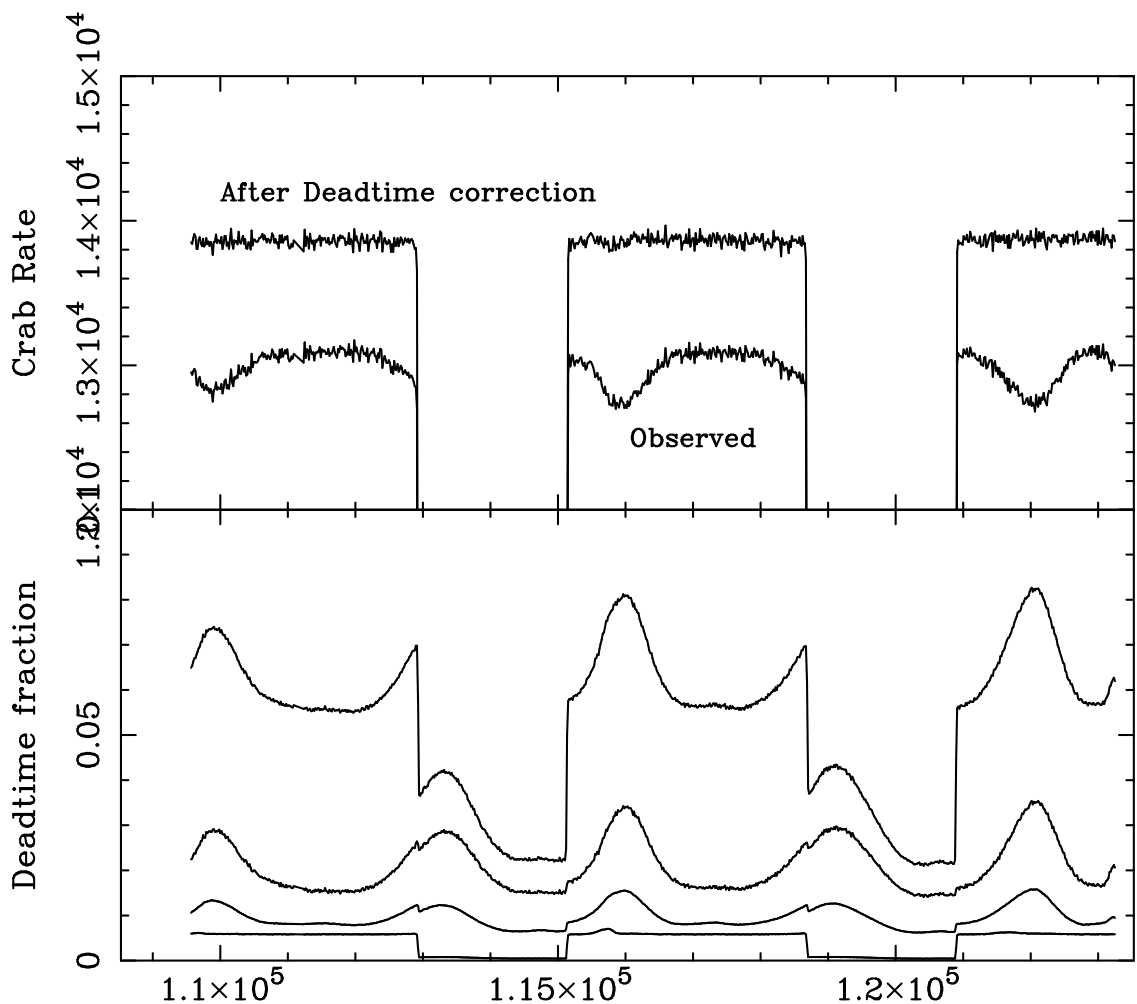}
\figcaption{Crab rates as observed and corrected for deadtime.  The lower panel 
shows the total deadtime from all sources.  Also shown in the deadtime calculated
from the ``Very Large Event", ``Remaining", and ``Propane" rates as
recorded in the Standard 1 data.  The Very Large events make the largest
contribution and the propane events the smallest.  The total deadtime
includes has a contribution from the source itself.
The figure shows 3 on source intervals separated
by observations of the earth.  The deadtime induced by the Crab is comparable to the
instrument background estimated from the occulted observations.
\label{crab_std1_rates}}
\end{figure}

Figure \ref{crab_std1_rates} shows the total good rate (sum of the 5 rates from the individual detectors) and
the deadtime corrected rate in the top panel;  the calculated deadtime is shown in the lower panel, along with the contributions to deadtime from instrumental background
rates.
The deadtime is calculated from the total rates.  Every event is recorded exactly
once in the eight rates present in the Standard 1 telemetry.  Five of these rates
record the frequency of ``good" events in each PCU (i.e. events which trigger a
single signal chain), one rate measures all events which have the Propane layer,
one rate measures all events which contain a VLE, and one rate (the ``Remaining"
rate) counts all other events.  Standard 1 data thus allows an evaluation of the
deadtime on a 0.125 second basis.  The deadtime can be evaluated on smaller timescales
if good event data is present on shorter timescales to the extent that the $V_x$, Propane,
and Remaining rates do not vary appreciably at frequencies above 8 Hz.  While the
Propane rate may have high frequency terms (due to source variability) this condition
is generally approximately true.

In calculating the dead time in figure \ref{crab_std1_rates}, we assume that each
event contributes $10 \musec$ to the deadtime except for Very Large Events
which contribute $170 \musec$.  For the rates
which are summed over all PCU, we assume that the contribution from each
PCU is the same.\footnote{This assumption becomes less true in Epoch 5, when
PCU 0 is no longer functionally identical to the others.  A more careful analysis
could estimate the instrumental contributions from Standard 2, which has
sufficient time resolution to capture variations in the instrument background
induced deadtime}
Estimating the incident rate with this proceedure is adequate in practice
although it ignores one detail of the PCA electronics.  Some of the events
included in the Standard 1 ``Remaining" rate trigger only the Xenon Veto
anode (i.e. and not any of the signal anodes) and are handled differently
than the other events.  These events do not initiate an analog to digital
conversion and the 8 pulse height bits transferred to the EDS are therefore
set to 0 as are the lower level discriminator bits.  
While there is a dead period where no good event can be recorded
due to the presence of the veto signal, it is known that no incident event 
was lost in these
dead periods.  (If there had been an event, the analog to digital conversion
would have been initiated, and a multiple anode - i.e. rejectable - event
would be transferred to the EDS.)  The dead time correction accounts for
periods where an incident event would go un-noticed.  Since the VX only events
do not prevent us from noticing incident events, they should not be included
in the sum of ``dead time".  In practice this is a small effect which can
be estimated on 16 second time scales from the Standard 2 data which records
the total rate of VLE (only) events.  For the interval shown in fig 
\ref{crab_std1_rates} about two-thirds of the instrument background induced
deadtime is due to Very Large Events.  The ``Remaining" rate contributes a
deadtime of $\sim 1\%$.  Approximately $ 14\%$ of the remaining
rate is due to $V_x$ only events.  Thus the deadtime fraction which is about $0.06$
on average is over estimated by $\sim 0.0014$ which is about $2\%$ of the total
deadtime.  The absolute value of this overestimate is likely to be nearly
constant over the mission;  the fractional contribution to the deadtime depends
on the source intensity.

The total observed rate is modulated by occulted periods,
and is $\sim 13,000$ count sec$^{-1}$ when on source.  The remaining count 
rate is modulated at twice the orbital period (and is correlated with 
earth latitude, McIlwain L, or rigidity);  in addition there is a 
contribution to the remaining count rate due to the source
itself.  This represents chance 2-fold coincidences of X-rays from the Crab.  
For the Crab, peak deadtime from 
all sources amounts to $\le 7\%$.

\placefigure{crab_std1_rates}

Deadtime corrections similar to this example will need to be performed for all 
observations attempting to measure relative flux variations to better than a few per
cent.  The time scales for background induced variation are about 45 minutes (half an
orbit) although source variability can cause variations in the deadtime on 
much faster time scales.

\paragraph{Bright Sources}

We operationally define bright sources as those sources for which it is
inappropriate to treat all incident events as independent.  The definition is
therefore dependent on how the data is used.  For instance, for power spectrum
analysis, deadtime must always be considered, as the chance that events are
miscounted or missed altogether changes the shape of the power spectrum.
For the construction of
light curves, on the other hand, the faint correction described above works well
for sources with net counting rates $\le 10,000$ count sec$^{-1}$ per PCU.  
At higher count rates it is important to correct for the chance that two cosmic
photons are simultaneously detected in different layers \citep{vdk96,jah99} or
in the same layer \citep{tomsick98}.

\section{Absolute Timing}
\label{absolute}

Accuracy of the RXTE absolute timing capability on scales longer than 1 second
has been verified by comparison of burst arrival times with BATSE.  We have used
both bursts from J1744-28 and a bright gamma-ray burst (960924) which produced a
large coincidence signal in the PCA to establish this agreement.  All information
on times finer
than 1 second is contained within individual EDS partitions.  The relationship
between EDS partitions and spacecraft time has been verified through ground testing
and correlation of time tagged muon data with PCA data containing signals from the same
muons.
The content of the RXTE telemetry, and the relation to absolute time, has been
thoroughly documented by \citet{rots98} and references therein.  The telemetry 
times in the RXTE mission data base give times that are accurate to $\le 100 
\musec$ and users who require times $\le 10 \musec$ can achieve this by 
applying correction terms\footnote{
http://heasarc.gsfc.nasa.gov/docs/xte/time\_news.html}
which are measured several times a day by RXTE mission operations personel.

The phase of the primary peak of PSR 1821-24 has been measured in 30 different 
satellite orbits over the course of 3 days \citep{saito01}.  The statistical 
accuracy of each measurement is $\sim 0.003$ in phase ($\sim 10 \musec$) 
which is virtually identical to the distribution of measured phases.  We 
can therefore conclude that the stability of the clock is better than $10 \musec$ 
over 3 days.  The scatter in the phases includes variation from clock variability,
pulsar timing noise, and statistical uncertainty.  The shape of the pulse is 
quite narrow \citep{rots98}.

\subsection{Error Budget}







%
%
%

In this section we examine the potential contributions to the
uncertainty in computing the photon arrival times at the solar system
barycenter.  We consider the effects attributable to the detector and
spacecraft, to the ground system, and to the barycentering
computation.  The effects are tabulated in table \ref{time_errors} and
discussed below.

X-ray photons enter the Xenon volume of the PCA where they interact
and are converted to photoelectrons, and subsequently to an ionization
cloud which drifts to the anode wires where an electron avalanche is
created.  The drift time is approximately $\sigma_{\rm drift} = 1 \musec$.  
Once collected at the anode wires, the electron pulse is amplified, 
shaped and converted to a digital pulse height.  This conversion 
process takes approximately $18.2  \musec$ for all PCUs, and is 
essentially independent of energy.\footnote{As discussed by \citet{rots98}, the time for the lower level discriminator threshold to be met is energy dependent, 
but the analog to digital process does not occur until the pulse {\it peak} is
reached.  Thus, the conversion time is largely energy independent.}

When an X-ray event is registered in the PCA electronics, its pulse
height is transferred to the EDS over a 4 MHz serial link, where a
time stamp is applied.  The resolution of the RXTE clock is
$\sigma_{\rm EDS} = 1 \musec$.  For some modes, including the ``GoodXenon"
mode which telemeters full timing information and all the pha and flag bits
presented to the EDS, the full time
precision of each event is kept.  For other event modes with coarser
time resolution, $\sigma_{\rm evt}$, the event time is rounded down
(i.e., truncated), so on average an event will appear ``early'' by a
time $\sigma_{\rm evt}/2$.  For binned modes, the time
tag is generally treated the same way, and refers to the 
beginning of the bin, as documented by the TIMEPIXR keyword in the
archived data files.

The RXTE clock, which is used to time-tag each X-ray event, is
calibrated using the User Spacecraft Clock Calibration System (USCCS)
\citep{rots98}.  In short, the White Sands Complex sends
a clock calibration signal via TDRSS to the spacecraft.  The
spacecraft immediately returns the signal to White Sands via TDRSS.
Both send and receive times from White Sands are recorded with $1
\musec$ precision for each calibration, and typically 50--150
calibrations are performed per day.  When the calibration signal is
received by the spacecraft, it also records the value of its clock.
When the spacecraft and White Sands time tags are later processed on
the ground in the RXTE Mission Operations Center (MOC), it is possible
to determine the time difference between the spacecraft clock and the
White Sands cesium atomic clock.  The effective frequency of the
spacecraft clock is routinely adjusted to keep time differences within
$\pm 70 \musec$ of the White Sands clock.

The USCCS calibration signals are embedded in the spacecraft
ranging measurements.  Individual signals are sent at intervals much
shorter than the light travel time to the spacecraft, so some
information on the orbit ephemeris is needed to pair up the downlink
time stamps with the correct uplink time stamps.  
The spacecraft ephemeris is
estimated by the Goddard Flight Dynamics Facility (FDF), using
spacecraft ranging data obtained through the TDRSS link.  The FDF
produces daily ``production'' solutions with approximately 8 hours of
overlap with the previous day.  We have performed a comparison of the
overlap regions between daily solutions in order to provide an
estimate of the ephemeris uncertainty.  Over the mission lifetime, the
overlap differences are less than $\sim$450 m with 99\% confidence,
or $<1.5  \musec$ light travel time.  However, before the increase in
solar activity starting around the year 2000, the uncertainty was
approximately a factor of ten smaller than this.

The White Sands clock is formally required to keep station time within
$\sigma_{\rm WS} = 5 \musec$ of UTC, as defined by the US Naval
Observatory master clock, but is also required to maintain knowledge
of this time difference at the $\pm 0.1 \musec$ level.  In practice,
over the time period 1996--2001, this difference has been kept to
within $\sigma_{\rm WS} = 1 \musec$ of UTC(USNO).  Station time
is actually compared against Global Positioning Satellite (GPS) time.
GPS time, in turn, is kept within $\sigma_{\rm GPS} = 100$ ns of
TT(BIPM), the international standard of ephemeris time, according to
published values in BIPM Circular T from 1996 to 2001.

The clock offsets derived from USCCS can be used to correct X-ray
event times to White Sands station time.  A piecewise continuous
quadratic function is fitted to segments of clock calibration offsets.
This function serves both to interpolate between gaps, but also to
smooth individual calibration points.  The function is constrained at
each endpoint to be continuous in value with surrounding segments.  In
addition, discontinuities in slope are known because spacecraft clock
frequency adjustments occur at known times and with known
magnitudes.  The granularity of clock frequency adjustments is
1/3072 Hz.  Thus, the actual function
is highly constrained, which is appropriate since we believe the clock
to be largely well behaved.  The clock model we have constructed
matches the data to within $\sigma_{\rm model} = 2.5 \musec$.

Before MJD 50,567 a software processing error in the MOC caused
individual clock calibration times to jitter at a level of $\pm 8
\musec$.  While the fitted quadratic model serves to smooth this
jitter significantly, the variance in model residuals is signficantly
larger before MJD 50,567.  We estimate that 99\% of the residuals lie
within $\sigma_{\rm MOC} = 4.4 \musec$ of zero.  After that date, the
variance of the calibration points is contained within the band
defined by $\sigma_{\rm model}$.

Assuming the effects in table \ref{time_errors} are uncorrelated, 
which we expect to be the
case, the total error will add the terms in quadrature.  Thus, the absolute
timing error for most of the mission is $< 3.4 \musec$ (99\%).

\section{Field of View}
\label{fov}

\subsection{Collimator model}

Each PCU has a collimated, approximately 
circular, field of view of radius $1^{\circ}$ from peak to zero throughput.
Each PCU has a collimator assembly made up of 5 individual collimator modules.
Each module contains a large number of identical hexagonal tubes which
provide the collimation and each module was aligned independently. 
The opening of each individual hexagon is 0.125 inch (flat to flat); the 
length of each collimator tube is 8 inches.

To model the on-orbit collimator efficiency we began with the theoretical
transmission function for a perfectly absorbing hexagonal tube with dimensions
equal to those comprising the PCA collimator modules. The fabrication, 
mounting and alignment of the collimator modules must introduce some level
of misalignment among all the individual hexagons making up the collimator
assembly. To produce a more realistic model 
we averaged the responses from a
large number of perfect hexagons but with the pointing direction of each
tube randomly displaced from the vertical. The random offsets were parameterized
with a single parameter, $\sigma$, which represents the width, in arcmin, of
a gaussian distribution centered on the vertical from which the random offsets
were sampled. Thus, the larger $\sigma$ the greater is the spread in pointing
directions of the individual hexagonal tubes. 
We calculate the model response by randomly
drawing an off axis angle, $\theta$, from a gaussian distribution centered on
$\theta = 0$ and with a standard deviation $\sigma$. A random value for the 
azimuthal angle $\phi$ is also selected. The theoretical transmission for this
orientation is calculated and added to the total. The process is repeated for
a large number of offsets, and the final total response is normalized to 1.0 at
its peak. We calculated a series of models for different values of $\sigma$,
ranging from 1 arcmin to 8 arcmin, and conclude that $\sigma = 6$ arcmin provides
the best description of the collimators.  The same value of $\sigma = 6$ is
used for all 5 detectors;  scans over the Crab nebula do not provide
evidence that this model requires separate values for each detector.

We used the Crab as an approximately constant and point-like source of X-rays
in order to determine the boresight direction for each PCU
as well as the value of $\sigma$ which gave the best fit to the 
scan data. The RXTE spacecraft attitude control system (ACS) computes an
estimate of the spacecraft attitude on a 0.25 second timescale. The attitude
information prescribes the orientation of the three spacecraft axes in 
Earth-centered inertial coordinates (epoch J2000). The attitude data is
provided by two on-board star trackers, and includes on board aberration
correction.  With this knowledge the
location of any X-ray source with respect to the spacecraft coordinates can be
calculated. We used the attitudes to determine the counting rate in each
PCU from the Crab as a function of position in the spacecraft frame. We then
minimized the function
\begin{equation}
\chi^2_j = \sum_i  \frac{\left ( O^j(Y_{S}^i,Z_{S}^i) - R^j 
M(Y_{S}^i - Y_{bore}^j, Z_{S}^i - Z_{bore}^j, \sigma ) \right
)^2}{O^j(Y_{S}^i, Z_{S}^i)} \; ,
\end{equation}
where $O$ is the observed countrate, $M$ is the model response, $i$ denotes the
individual rate and attitude samples, $j$ denotes the different PCUs,
$Y_{bore}^j$ and $Z_{bore}^j$ specify the
pointing direction (boresight) of each detector, $\sigma$ specifies the
smearing of the ideal hexagonal response discussed above, and $R^j$ denotes the
peak countrate for each detector, that is the counting rate at the peak of the 
response. Since sources are effectively at 
infinity only the direction is relevant. The direction to the source
with respect to the spacecraft frame can be uniquely 
specified with only two parameters. We elected to use the 
$Y_{S}$ and
$Z_{S}$ spacecraft coordinates as the independent variables. Since these form part
of a unit vector we have that $X_{S}^2
+ Y_{S}^2 + Z_{S}^2 = 1$.
Before performing the fit we first corrected the observed rates for detector
deadtime using the faint deadtime correction formula described above. 
We found that $\sigma = 5 - 6$ arcmin gave the smallest values of $\chi^2$.
Table \ref{alignments} summarizes the derived values 
of $Y_{bore}^j$ and $Z_{bore}^j$ as well as the definition of the
spacecraft ``science axis".  It is the science axis which the ACS points
towards a commanded position.

\subsection{Fidelity of the Collimator Model}

\begin{figure}[t]
\includegraphics[width=0.8\columnwidth,angle=0.]{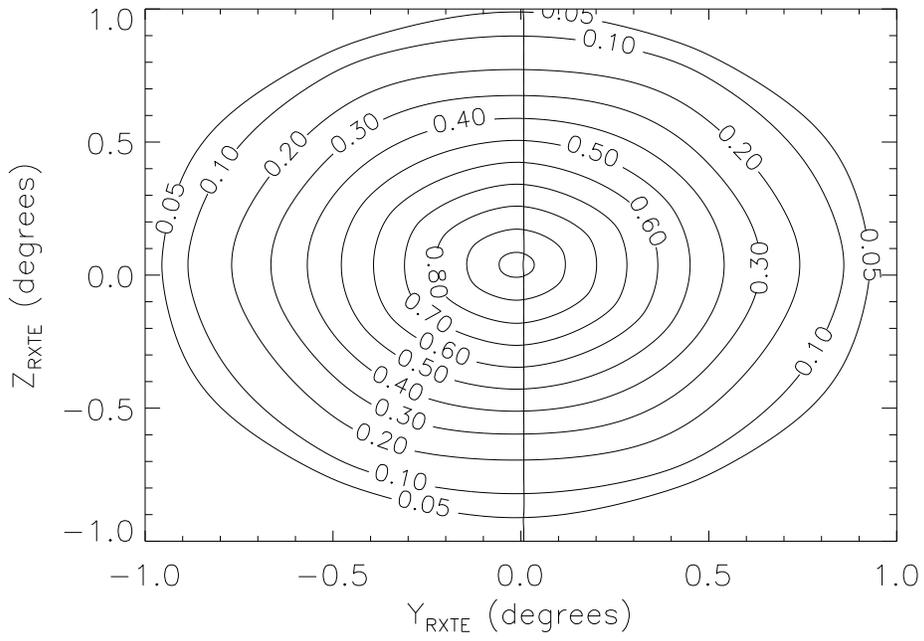}
\figcaption{The collimator efficiency model for PCU 0.   The vertical line
represents a scan trajectory discussed in the text.
\label{coll_eff}}
\end{figure}

\begin{figure}[h]
\includegraphics[width=0.8\columnwidth,angle=0.]{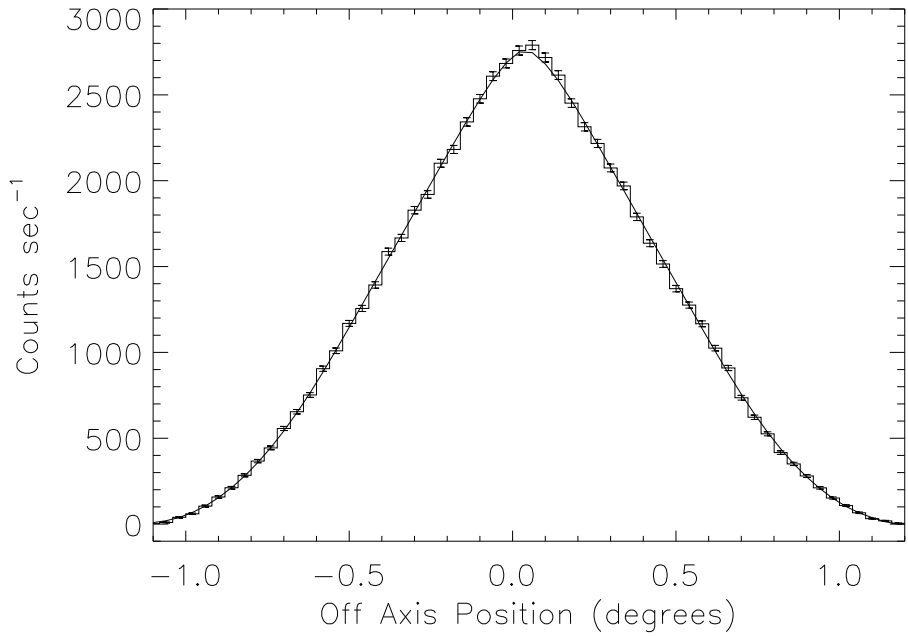}
\figcaption{Data and model along a scan trajectory which passes near the peak
of the response.
\label{coll_rate}}
\end{figure}

\begin{figure}[h]
\includegraphics[width=0.8\columnwidth,angle=0.]{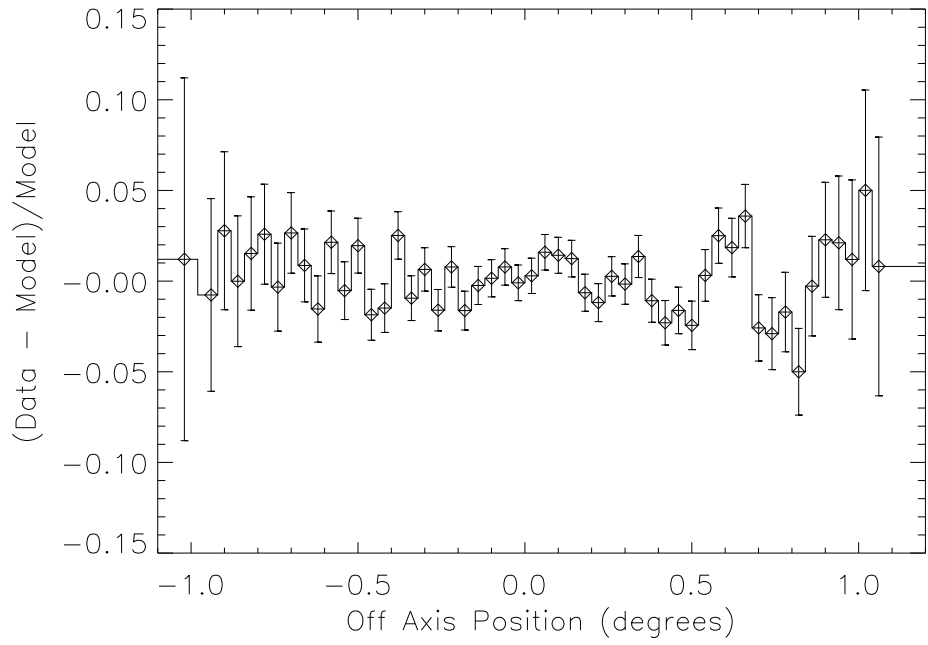}
\figcaption{Ratio of residuals to the model.  The collimator model is accurate
to better than a few percent along the entire scan.
\label{coll_ratio}}
\end{figure}

In order to assess the accuracy of the collimator model we have
carried out comparisons of the observed (background subtracted and
deadtime corrected) and predicted counting rate
from the Crab along various scan trajectories across the
collimators. 
Results from a characteristic scan are summarized in Figures 
\ref{coll_eff} to \ref{coll_ratio}.  Figure \ref{coll_eff} shows the 
collimator efficiency model (as a contour plot) for PCU 0;  the vertical 
line represents the scan trajectory of figures \ref{coll_rate} and 
\ref{coll_ratio}.  Figure \ref{coll_rate} shows the counting rate in 
PCU 0 along this scan trajectory (histogram with error bars) as well 
as the predicted countrate from the PCU 0 collimator model (solid curve). 
The observed rate tracks the predicted rate rather 
closely.  Finally, to better quantify the fidelity we show in Figure 
\ref{coll_ratio} the ratio of (Data - Model)/Model along the scan path.  
The collimator model is faithful to the data at the few percent level on this
scan, which is typical of all 5 PCUs.

\placefigure{coll_eff}
\placefigure{coll_rate}
\placefigure{coll_ratio}

\subsubsection{ Position accuracy using the collimator model}

An important capability of RXTE is its ability to respond quickly to
changes in the X-ray sky, for example, to observe the appearance of a
new X-ray source.  A crucial aspect of such observations is the
ability to rapidly localize the source so that observations in other
wavebands can be attempted.  Sky positions can be determined from
observations in which the PCA is scanned over the source of
interest. The observed lightcurve is compared to a model lightcurve
derived from the spacecraft attitude data, the collimator models, and
a source model which includes the celestial coordinates.  The
source model parameters are varied in order to minimize some goodness
of fit quantity (typcially a $\chi^2$ statistic). The collimator
models are an important component of this fitting procedure. 
Other factors which affect the precision of position determinations
are a statistical uncertainty related to the
brightness of the source, a systematic uncertainty introduced by
intrinsic source brightness variations, nearby and unmodelled sources,
and errors in the spacecraft
attitude determination.  Of these the most difficult to quantify for
any particular object is the systematic error due to source
variability.  Nevertheless, this suggests another way to probe the
quality of the collimator models by comparing the positions of known
sources to those determined from scanning observations with the PCA.

\begin{figure}[t]
\includegraphics[width=0.8\columnwidth,angle=0.]{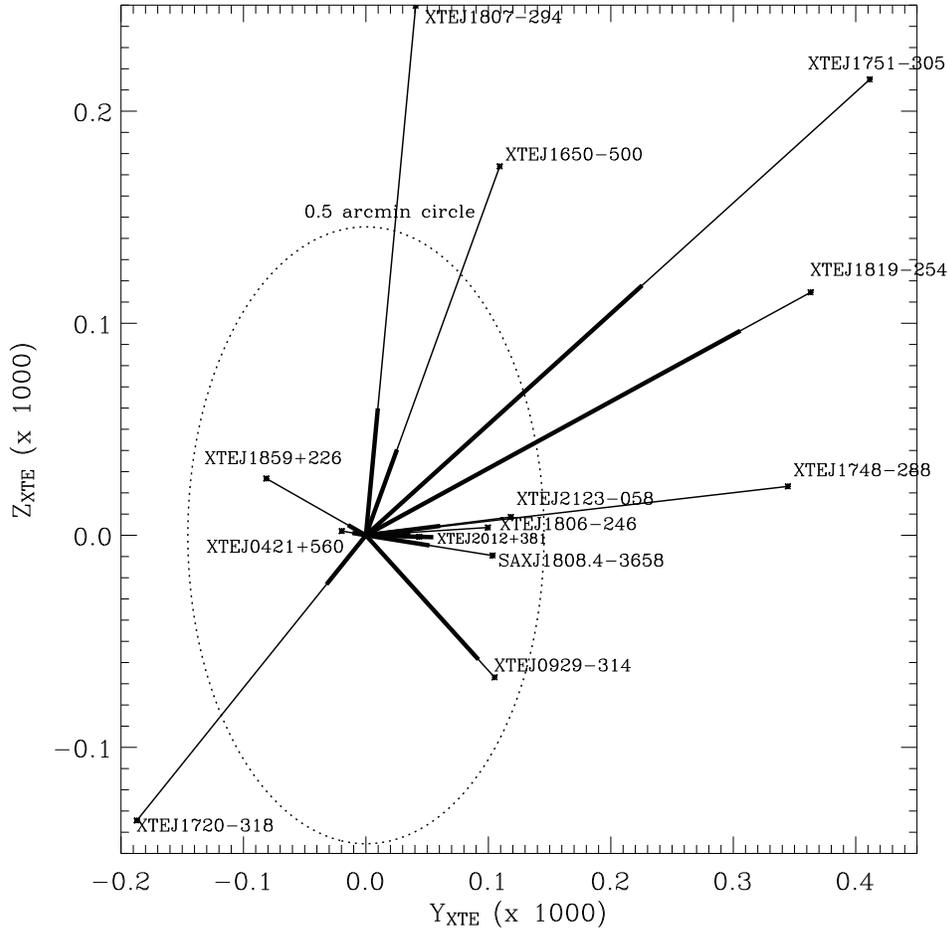}
\figcaption{Differences between PCA derived positions of known sources and
actual positions.  The dark part of the offset is attributable to counting
statistics;  the remainder comes from source variability, ACS systematics, and
inaccuracies in the collimator model.
\label{yzoff}}
\end{figure}

Over its $\sim 9$ yr mission RXTE has performed scanning
observations of many X-ray sources, and many of these either had or
now have positions accurate at the arc-second level. We analyzed a
sample of 13 bright sources and compared their known positions with
those derived from PCA scanning observations. The results are
summarized in Figure \ref{yzoff}, which shows the offsets between the known and
PCA-derived positions. For each source the offset is represented as a
vector in the spacecraft coordinate frame. The length of the vector
represents the angular separation between the known and fitted
position. The thick portion of each vector is an estimate of the
statistical uncertainty in the derived position. Thus, the thin
portion provides an indication of the scale of the systematic error
due to source variability and other effects.   The dotted
circle has a 1/2 arc-minute radius.  Most sources are localized to
better than 1 arc-minute, however, for a significant number of objects
with small statistical errors the systematic error dominates.

Since the component of error due to source variability should be
approximately random, one should be able to reduce it by averaging
position determinations from many independent scanning observations. A
bright source which is nearly ideal for such a study is the LMXB 4U
1820--30.  This object resides in the Galactic bulge and has been
observed hundreds of times during the course of PCA Galactic Bulge
monitoring \citep{swank01,markwardt2002} and its position is known to
sub-arcsecond accuracy \citep{sosin95}. We determined positions
using many such observations of 4U 1820--30 and for each PCU
separately. The results are summarized in Table 10. For each detector
we give the number of observations analyzed, the average derived right
ascension, $\alpha$, and declination, $\delta$, (both J2000), the
angular separation, $\Delta\theta$, from the known position, and an
estimate of the statistical error of the angular separation,
$\Delta\theta_{stat}$. These results suggest that in the absence of
source variability errors the position accuracy achievable with the
PCA collimator model is in the 2--10 arcsecond range. This is becoming
comparable to the pointing accuracy of RXTE. We note that the results
from PCUs 0--4 are all more or less consistent.

\placefigure{yzoff}

\subsection{Solid Angle}
The linear approximation for the 
response,
$f= 1 - r/r_0$  where $r_0$ is equal to $1 \deg$, overestimates the
solid angle.  Numerically integrating
$f(r) sin(r) d\Omega$ over $0 \le \theta \le 2\pi$ and $0 \le r \le r_0$
gives 0.000320578 sr.

Integrating the summed responses numerically for all the PCUs, using the 
model file pcacoll\_v100.2, we get 0.00029703 sr, or about $8 \%$ less. 
This is approximately the effective solid angle of a linearly falling 
response out to $r_0$ if $r_0 = 0.965 \deg$. Put another way, 
this is the solid angle you get by integrating a flat (unit) 
response from r= 0 to 0.55712 degrees.

\section{Summary}
The PCA is a large and versatile instrument with well understood and systematics
limited calibration.  PCA observers command only the data compression;  the appendix
describes how the 
effects of data compression can be exactly modelled.
The accuracy of the energy response function is 
limited by systematics below the Xe K edge;  
deviations from power-law fits to the Crab Nebula plus pulsar are
less than 1\% below 10 keV and gradually increase towards higher energies. 
Unmodelled variations in the instrument background are less than 2\% of the total
sky plus instrument background below 10 keV and less than 1\% between 10 and 20 keV.
The PCA has a dynamic range of 4.5 orders of magnitude:  it is confusion 
limited at fluxes below $\sim 4 \times 10^{-12} \ergscm$ and
deadtime limited at count rates greater than 20,000 ct sec$^{-1}$ PCU$^{-1}$
($\sim 2 \times 10^{-7} \ergscm$ for a Crab like spectrum).

\acknowledgements
The RXTE mission was made possible by the support of the Office of
Space Sciences at NASA Headquarters and by the hard and capable work
of scores of scientists, engineers, technicians, machinists, data
analysts, budget analysts, managers, administrative staff, and
reviewers.  We thank the anonymous referee for numerous constructive
suggestions which have improved the completeness and the presentation
of this work.

Facilities: \facility{RXTE(PCA)}, \facility{HEASARC}

\appendix
\section{Using PCA response matrix generator for user selected data modes}

Creating a PCA response matrix involves creating 256 channel matrices suitable for 
each detector ({\tt pcarmf}), shifting the channels following the EDS gain and
offset description({\tt rddescr}, {\tt pcagainset}, and {\tt gcorrmf}), rebinning
the channels to match the telemetered pulse height bins ({\tt rbnrmf}), estimating
the effective area of each detector after accounting for spacecraft pointing 
({\tt xpcaarf}), combining the area and redistribution matrices ({\tt marfrmf}), 
and adding the matrices from the 5 PCU ({\tt addrmf}).  The perl script {\tt pcarsp} 
takes care of these tasks.  Section \ref{sec3} described the contents and 
construction of {\tt pcarmf}.  The function and usage of the other tools
are described in the help files.  We describe the on-board gain shifting
here.

The gain of each of the PCU detectors is similar, but not identical.  Because
it is often desirable to co-add data from the 5 detectors, to maximize
time resolution within a given telemetry budget, the EDS does some channel
shifting in order to add more nearly equivalent energies.  The EDS does
integer arithmetic, parameterized by a gain and offset, to shift the 256
input pulse height channels to 256 corrected pulse height channels.  The 
gain term always results in an expansion, with the result that most channels
are shifted upwards, but many pairs of adjacent original channels are
shifted into corrected channels separated by two channels.  
The gain and offset parameters are not contained in
the telemetry, but can be associated with the pulse height files with the
ftool {\tt pcagainset}.  
The channel shifting algorithm is
\begin{equation}
{\rm I}_{corr} = { { ({\rm I}_{orig} \times (256 + gain) + 128)} \over {256} } + {\rm offset}
\end{equation}
where I$_{orig}$ is the pulse height produced by the PCU and I$_{corr}$ is the corrected pulse
height \citep{jah96}.
Table \ref{gainoff} has the values of gain and offset parameters used throughout
the RXTE mission.

\clearpage

\begin{deluxetable}{llr}
\tablewidth{0pt}
\tablecaption{Nominal PCU Dimensions  \label{dimensions}}
\tablehead{ \colhead{component} & \colhead{Material}	& \colhead{Dimension} }
\startdata
Thermal shield    &	polyimide	&   $76 \mu$  \\
Collimator sheet  &   BeCu		&   0.0069 cm \\
Collimator cell	(height)  &		&   $20$ cm \\
Collimator cell	(flat to flat)  &	&   $0.32$ cm \\
Entrance window	  &   Mylar		&   $25 \mu$  \\
\hspace{0.5cm} Window coating per side	  &   Aluminum          &   $70$ nm \\
Anti-coincidence  &   Propane		&   1.2 cm  \\
\hspace{0.5cm} pressure at 22 deg C  &  &   798 torr \\
Separation window &   Mylar             &   $25 \mu$\\
\hspace{0.5cm} Window coating per side	  &   Aluminum          &   $70$ nm  \\
Main volume pressure at 22 deg C  &   Xenon (90\%)/$CH_4$ (10\%) & 836 torr \\
\hspace{0.5cm} layer 1 depth         &                     &   $ 1.35 $ cm \\
\hspace{0.5cm} layer 2 depth         &                     &   $ 1.20 $ cm \\
\hspace{0.5cm} layer 3 depth         &                     &   $ 1.20 $ cm \\

Inner shield         &   Tin               &   $0.051$ cm  \\
Outer shield         &   Tantalum          &   $0.152$ cm  \\
\enddata
\end{deluxetable}
\clearpage
\begin{deluxetable}{clrrrrr}
\tablewidth{0pt}
\tablecaption{High Voltage Settings\label{hv_epochs}}
\tablehead{\colhead{Epoch} & \colhead{Start Date (UT)} &
           \colhead{PCU 0} &
           \colhead{PCU 1} &
           \colhead{PCU 2} &
           \colhead{PCU 3} &
           \colhead{PCU 4}\\
	   \cline{3-7}\\
	   \colhead{} & \colhead{} &
	   \colhead{Volts} & \colhead{Volts} & \colhead{Volts} & 
	   \colhead{Volts} & \colhead{Volts} }
\startdata
 1 &	Launch                   & 2030 & 2030 & 2026 & 2027 & 2048 \\
 2 &	1996 March 21 @ 18:34    & 2010 & 2010 & 2006 & 2007 & 2007 \\
 3A&    1996 April 15 @ 23:06   & 1990 & 1990 & 1986 & 1987 & 1988 \\
 3B\tablenotemark{a} &   1998 February 9 @ 01:00  &  "   &  "   &  "   &  "   &  "   \\
 4 &    1999 March 22 @ 17:39    & 1970 & 1970 & 1966 & 1967 & 1968 \\
 5\tablenotemark{b} &    2000 May 12 @ 1:06   "   &  "   &  "   &  "   &  "   &  "   \\
\enddata
\tablenotetext{a}{Epochs 3A and 3B distinguish background models with
different time dependence}
\tablenotetext{b}{PCU 0 lost pressure in the propane volume at the beginning of Epoch 5}
\end{deluxetable}
\clearpage
%
\begin{deluxetable}{lccccc}
\tablecaption{Quantum efficiency and redistribution parameters - v5.3 \label{v5.3_pars}}
\tablehead{ \colhead{parameter} & \colhead{PCU 0} &
            \colhead{PCU 1}     & \colhead{PCU 2} &
            \colhead{PCU 3}     & \colhead{PCU 4} }
\startdata
$Xe_{l1}$ (gm cm$^{-2}$)             & 0.00663    & 0.00669   &   0.00692  &  0.00652   &  0.00689 \\
$Xe_{l2,3}$ (gm cm$^{-2}$)           &  0.00542    & 0.00556   &   0.00568  &  0.00526   &  0.00570 \\
$Xe_{pr}$ (gm cm$^{-2}$ on 1997dec20)&  0.00015    & 0.00009   &   0.00013  &  0.00020   &  0.00013  \\
$Xe_{dl}$  (gm cm$^{-2}$)            &  0.00057    & 0.00062   &   0.00063  &  0.00071   &  0.00059  \\
Mylar  (gm cm$^{-2}$ in 2 windows)   &  0.00699    & 0.00696   &   0.00695  &  0.00696   &  0.00696 \\
$ d(Xe_{pr})/dt$ (gm cm$^{-2}$ day$^{-1}$) & 8.9E-08    & 5.2E-08   &   5.1E-08  &  4.7E-08   &  4.9E-08 \\
$Xe_{pr}$ (gm cm$^{-2}$ on 2000may13) & 7.1E-05   &            &            &          &     \\
$ d(Xe_{pr})/dt$ (gm cm$^{-2}$ day$^{-1}$) &  0.0E+00   &            &            &          &  \\
\cutinhead{PCA Universal parameters}
$E_{point}$  (keV)                    &             &            & 1.3E+01 &           &   \\
$A_o$                                 &             &            & 0.01677 &           &   \\
$\delta$                              &             &            & 2.90017 &           &   \\
$\kappa(5 keV)$                         &             &            & 0.043 &           &  \\
$a$ (equation \ref{energy_res})       &             &            & 0.12100 &           &    \\
$b$ (equation \ref{energy_res})       &             &            & 0.44200 & &    \\
$f$ (equation \ref{wf})              &             &            & 0.40000 &           &    \\
$T_0$ (equation \ref{e2c2})              &             &            & "1997-12-20" &           &    \\
K edge fration not self vetoed (section \ref{redistribution})   &   &    & 0.910 &     & \\
L edge fraction not self vetoed  (section \ref{redistribution}) &   &    & 1.000 &     & \\
Pr (gm cm$^{-2}$)                    &                  &          & 0.00261 &        &   \\
Al (gm cm$^{-2}$) total metal on 2 windows &            &          & 0.00008 &        &   \\
$K_{\alpha}$ escape fraction                &           &          & 0.545 &        &   \\
$K_{\beta}$ escape fraction                &            &          & 0.155 &        &   \\
$L$ escape fraction (layer 1)               &           &          & 0.010 &        &   \\
electron offset, $L$ escape             &               &          & 3.900 &        &   \\
electron offset, $K_{\alpha}$ escape            &               &          &-2.260 &        &   \\
electron offset, $K_{\beta}$ escape             &               &          & 3.840 &        &   \\
\cutinhead{Geometric Areas (from {\tt xpcaarf.par}) }
&   1567.0     &   1536.0      &  1563.0   &           1631.0   &    1598.0   \
\enddata
\end{deluxetable}
\clearpage
\begin{deluxetable}{rrl}
\tablewidth{0pt}
\tablecaption{In Flight calibration energies  \label{line_energies}}
\tablehead{ \colhead{Energy} & \colhead{Source} }
\startdata
13.930 & Np - L \\
17.530 & Np - L \\
21.130 & Np - L \\
26.350 & Am$^{241}$ \\
29.8 & Xe $K_{\alpha}$ escape\tablenotemark{a}\\ 
59.54 & Am$^{241}$ \\
\enddata
\tablenotetext{a}{Blend of Escape peak and Escape photon.  The 
escape photon is occasionally observed by itself associated with
events where the 59.54 keV photon is 
keV photon is absorbed in non-instrumented volume.}
\end{deluxetable}
\clearpage
\begin{deluxetable}{crr}
\tablewidth{0pt}
\tablecaption{Dedicated Blank Sky Pointing Directions \label{bkg_pts}}
\tablehead{\colhead{Target number\tablenotemark{a}} & \colhead{$\alpha$} & \colhead{$\delta$} } 
\startdata
 $N$0801-01 &	5.00  & -67.00  \\
 $N$0801-02 &	60.00  & 2.00   \\
 $N$0801-03 &    138.00 & 15.00  \\
 $N$0801-04 &    235.00 & 10.00  \\
 $N$0801-05 &    345.00 & -18.00   \\
 $N$0801-06 &    160.00 & 72.57 \\
\enddata
\tablenotetext{a}{ The index $N$ covers Announcement of Opportunity periods;  complete description of the RXTE observation ids are given at 
http://heasarc.gsfc.nasa.gov/docs/xte/start\_guide.html\#directories\_obid}
\end{deluxetable}
\clearpage
\begin{deluxetable}{lcrrrrrrrrrr}
\tablecaption{Background Systematics\tablenotemark{a} \label{bkgd_sysT}}
\tablehead{\colhead{Epoch} & \colhead{Layer} & \multicolumn{2}{c}{PCU 0} & \multicolumn{2}{c}{PCU 1}  & 
           \multicolumn{2}{c}{PCU 2}  & \multicolumn{2}{c}{PCU 3}  & \multicolumn{2}{c}{PCU 4} \\
\cline{3-4} \cline{5-6} \cline{7-8} \cline{9-10} \cline{11-12} \\
 & & \colhead{ct s$^{-1}$} & \colhead{\%} & \colhead{ct s$^{-1}$} & \colhead{\%} & \colhead{ct s$^{-1}$} & \colhead{\%} & \colhead{ct s$^{-1}$} & \colhead{\%} & \colhead{ct s$^{-1}$} & \colhead{\%} }
\startdata
\cutinhead{2-10 keV}
3A & 1 & 0.0289 & 0.82\% & 0.0300 & 0.82\% & 0.0250 & 0.68\% & 0.0254 & 0.76\% & 0.0332 & 0.88\% \\
3A & 2 & 0.0101 & 0.91\% & 0.0170 & 1.56\% & 0.0168 & 1.51\% & 0.0161 & 1.50\% & 0.0191 & 1.60\% \\
3A & 3 & 0.0173 & 1.56\% & 0.0215 & 1.99\% & 0.0439 & 3.83\% & 0.0163 & 1.58\% & 0.0211 & 1.77\% \\
3B & 1 & 0.0277 & 0.79\% & 0.0357 & 0.98\% & 0.0285 & 0.78\% & 0.0349 & 1.06\% & 0.0420 & 1.11\% \\
3B & 2 & 0.0185 & 1.71\% & 0.0160 & 1.52\% & 0.0172 & 1.59\% & 0.0177 & 1.69\% & 0.0209 & 1.75\% \\
3B & 3 & 0.0179 & 1.65\% & 0.0185 & 1.77\% & 0.0360 & 3.27\% & 0.0193 & 1.91\% & 0.0270 & 2.29\% \\
4 & 1 & 0.0315 & 0.85\% & 0.0223 & 0.59\% & 0.0357 & 0.92\% & 0.0312 & 0.88\% & 0.0237 & 0.60\% \\
4 & 2 & 0.0127 & 1.10\% & 0.0091 & 0.83\% & 0.0149 & 1.30\% & 0.0112 & 1.02\% & 0.0116 & 0.95\% \\
4 & 3 & 0.0185 & 1.63\% & 0.0108 & 1.03\% & 0.0253 & 2.23\% & 0.0164 & 1.58\% & 0.0110 & 0.93\% \\
5 & 1 & 0.1096 & 1.52\% & 0.0392 & 1.12\% & 0.0365 & 1.03\% & 0.0370 & 1.15\% & 0.0480 & 1.35\% \\
5 & 2 & 0.0123 & 1.30\% & 0.0085 & 0.94\% & 0.0108 & 1.15\% & 0.0098 & 1.11\% & 0.0137 & 1.39\% \\
5 & 3 & 0.0150 & 1.64\% & 0.0119 & 1.38\% & 0.0192 & 2.09\% & 0.0127 & 1.55\% & 0.0133 & 1.42\% \\
\cutinhead{10-20 keV}
3A & 1 & 0.0109 & 0.49\% & 0.0148 & 0.67\% & 0.0130 & 0.59\% & 0.0119 & 0.56\% & 0.0137 & 0.61\% \\
3A & 2 & 0.0090 & 0.76\% & 0.0081 & 0.68\% & 0.0131 & 1.13\% & 0.0081 & 0.78\% & 0.0092 & 0.72\% \\
3A & 3 & 0.0114 & 1.04\% & 0.0139 & 1.24\% & 0.0127 & 1.10\% & 0.0114 & 1.20\% & 0.0130 & 1.07\% \\
3B & 1 & 0.0084 & 0.38\% & 0.0076 & 0.35\% & 0.0125 & 0.57\% & 0.0120 & 0.57\% & 0.0085 & 0.38\% \\
3B & 2 & 0.0074 & 0.63\% & 0.0065 & 0.55\% & 0.0122 & 1.06\% & 0.0074 & 0.71\% & 0.0127 & 1.01\% \\
3B & 3 & 0.0098 & 0.90\% & 0.0051 & 0.46\% & 0.0128 & 1.14\% & 0.0100 & 1.07\% & 0.0122 & 1.01\% \\
4 & 1 & 0.0132 & 0.60\% & 0.0171 & 0.76\% & 0.0113 & 0.51\% & 0.0204 & 0.97\% & 0.0160 & 0.70\% \\
4 & 2 & 0.0097 & 0.74\% & 0.0106 & 0.77\% & 0.0080 & 0.61\% & 0.0115 & 0.96\% & 0.0125 & 0.89\% \\
4 & 3 & 0.0079 & 0.63\% & 0.0136 & 0.96\% & 0.0125 & 0.95\% & 0.0141 & 1.26\% & 0.0127 & 0.92\% \\
5 & 1 & 0.0428 & 1.16\% & 0.0108 & 0.52\% & 0.0137 & 0.67\% & 0.0139 & 0.71\% & 0.0194 & 0.92\% \\
5 & 2 & 0.0124 & 1.02\% & 0.0122 & 0.93\% & 0.0083 & 0.67\% & 0.0096 & 0.87\% & 0.0127 & 0.98\% \\
5 & 3 & 0.0080 & 0.70\% & 0.0100 & 0.76\% & 0.0057 & 0.47\% & 0.0089 & 0.87\% & 0.0076 & 0.60\% \\
\enddata
\tablenotetext{a}{The Epoch 5 systematics are determined from data prior to November 2003}
\end{deluxetable}
\clearpage
\begin{deluxetable}{lcc}
\tablewidth{0pt}
\tablecaption{Timing Error Budget \label{time_errors}}
\tablehead { \colhead{Description} & & \colhead{Uncertainty} \\
                                   & & \colhead{($\sim$99\%; $\musec$)} }
\startdata 
Electron Drift            & $\sigma_{\rm drift}$ & $<$1 \\
GoodXenon EDS Timestamp   & $\sigma_{\rm EDS}$   & 1    \\ 
RXTE Clock Model          & $\sigma_{\rm model}$ & 2.5  \\
MOC Variance\tablenotemark{a}
                          & $\sigma_{\rm var}$   & 4.4  \\
White Sands Station       & $\sigma_{\rm WS}$    &  1 \\
UTC(GPS) - TT(BIPM)       & $\sigma_{\rm GPS}$   & 0.1  \\
TT - TDB                  & $\sigma_{\rm TDB}$   & 0.023\\
Satellite Ephemeris       & $\sigma_{\rm eph}$   & 1.5  \\
\enddata
\tablenotetext{a}{Applies to observations preceding MJD 50,567}
\end{deluxetable}
\clearpage
%
\begin{deluxetable}{lrr}
\tablewidth{0pt}
\tablecaption{PCU alignements  \label{alignments}}
\tablehead{ \colhead{PCU} & \colhead{$Y_{bore}$} &
	    \colhead{$Z_{bore}$}  }
\startdata

   0 &  -0.0000385 &  0.000629  \\
   1 &   0.0001046 &  0.000529  \\
   2 &  -0.0000500 &  0.000746 \\
   3 &   0.0002940 &  0.001340 \\
   4 &   0.0002900 &  0.001970 \\
\tableline
RXTE\tablenotemark{a} &   0.0000000 &  0.000700 \\
\enddata
\tablenotetext{a}{The science axis is controlled by pointing commands to the spacecraft}
\end{deluxetable}

\begin{deluxetable}{cccccc}
\tablewidth{0pt}
\tablecaption{PCA Position Measurements for 4U 1820-30}
\tablehead{ \colhead{PCU} & \colhead{observations} & \colhead{$<\alpha >$ (J2000)} &
            \colhead{ $<\delta >$ (J2000)} & 
	    \colhead{ $\Delta\theta$ (arcsec)} & 
	    \colhead{$\Delta\theta_{stat}$ (arcsec)} \\
	\colhead{} & \colhead{} & \colhead{(J2000)} & \colhead{(J2000)} &
	\colhead{(arcsec)} & \colhead{(arcsec)} }
\startdata
0 & 208 & $275.9196^{\circ}$ & $-30.3617^{\circ}$ & 3.83 & 1.63 \\
1 & 22  & $275.9205^{\circ}$ & $-30.3638^{\circ}$ & 11.34 & 5.46 \\
2 & 212 & $275.9196^{\circ}$ & $-30.3601^{\circ}$ & 4.67 & 1.65 \\
3 & 186 & $275.9181^{\circ}$ & $-30.3602^{\circ}$ & 2.20 & 1.75 \\
4 & 45  & $275.9219^{\circ}$ & $-30.3616^{\circ}$ & 10.38 & 3.90 \\
\enddata
\end{deluxetable}
\clearpage
\begin{deluxetable}{lrrrrrrrrrr}
\tablewidth{0pt}
\tablecaption{Gain and offset values for the PCA  \label{gainoff}}
\tablehead{ \colhead{Date} & \multicolumn{2}{c}{PCU 0} & 
                                              \multicolumn{2}{c}{PCU 1} & 
                                              \multicolumn{2}{c}{PCU 2} & 
                                              \multicolumn{2}{c}{PCU 3} & 
                                              \multicolumn{2}{c}{PCU 4} \\
\cline{2-3} \cline{4-5} \cline{6-7} \cline{8-9} \cline{10-11} \\
\colhead{} & \colhead{gain} & \colhead{offset} & \colhead{gain} & \colhead{offset} & 
           \colhead{gain} & \colhead{offset} & \colhead{gain} & \colhead{offset} &
           \colhead{gain} & \colhead{offset} }
\startdata
12/30/95:00:00:00\tablenotemark{a} & 24 & 1 & 22 & 1 & 29 & 1 & 10 & 0 &  2 &  0 \\
01/18/96:18:31:45                  &  0 & 0 &  0 & 0 &  0 & 0 &  0 & 0 &  0 &  0 \\
02/28/96:19:24:40                  & 29 &-1 & 20 &-1 & 27 &-1 & 14 &-1 & -1 & -1 \\
03/25/96:21:02:50                  & 29 &-1 & 20 &-1 & 27 &-1 & 14 &-1 & 33 & -1 \\
\enddata
\tablenotetext{a}{first data and time for which these settings apply}
\end{deluxetable}

\clearpage

\clearpage

\clearpage

\clearpage

\clearpage

\clearpage

\clearpage

\clearpage

\clearpage

\end{document}